# Geometry of nonequilibrium reaction networks

Sara Dal Cengio[*] and Vivien Lecomte
*Université Grenoble Alpes, CNRS, LIPhy, FR-38000 Grenoble, France*

Matteo Polettini
*Complex Systems and Statistical Mechanics, Department of Physics and Materials Science,
University of Luxembourg, L-1511 Luxembourg, Luxembourg*

(Dated: March 31, 2023)

The modern thermodynamics of discrete systems is based on graph theory, which provides both algebraic methods to define observables and a geometric intuition of their meaning and role. However, because chemical reactions are usually many-to-many, chemical networks are rather described by hypergraphs, which lack a systematized algebraic treatment and a clear geometric intuition. Here we fill this gap by building fundamental bases of chemical cycles (encoding stationary behavior) and cocycles (encoding finite-time relaxation). We interpret them in terms of circulations and gradients on the hypergraph, and use them to properly identify nonequilibrium observables. As application, we unveil hidden symmetries in linear response and, within this regime, propose a reconstruction algorithm for large metabolic networks consistent with Kirchhoff's Voltage and Current Laws.

### CONTENTS

---

[*] sara.dal-cengio@univ-grenoble-alpes.fr

## I. INTRODUCTION

### A. Context and motivations

One main task of nonequilibrium physics is identifying the nontrivial forces which drive a system out of equilibrium and the currents which develop both inside the system and in its interface with the outside environment.

Perhaps the simplest example is that of a Brownian particle moving on a ring. The system is described by the Langevin equation $\dot{x}(t) = F(x(t)) + \eta(t)$ for the particle's position $x(t)$, where $F(x)$ is the deterministic force and $\eta(t)$ is the thermal noise (Gaussian and white). The one-dimensional nature of the problem makes it easy to decompose the force into conservative and non-conservative contributions: $F(x) = -V'(x) + f$, where, for a ring of length $L$, $V(x) = V(x + L)$ is a periodic potential and $f = \int_0^L dx\, F(x)$. The source of the drive is identified in the scalar parameter $f$: for $f = 0$ the system relaxes with a vanishing current to an equilibrium steady state governed by the Boltzmann distribution associated to the potential $V(x)$; while for $f \neq 0$ the system is driven to a nonequilibrium steady state characterized by a non-vanishing current [1]. Notice that it is the geometry of the ring that allows for a nonzero steady-state current in this latter case.

The aforementioned model has arguably little relevance in real-world settings, but there are alternative scenarios where cycles are encountered naturally in relation to nonequilibrium behavior. For instance, molecular motors perform in cycles [2, 3] and cycles appear in most biochemical reactions, such as those involved in gene regulation and metabolic functions of living systems. This motivates us to consider the framework of chemical reaction networks (CRNs) [4–8], describing sets of reactions involving chemical species. Each reaction has its inherent chemical activity: it transforms (a combination of) reactants into products giving rise to a net flux of matter, the current, in response to an intrinsic chemical force, the affinity. At equilibrium currents and affinities vanish. Thereupon, external currents can be injected in the system through external chemostats which then foster nonequilibrium behavior. Consider for example a minimal model of glycolysis [9] for the consumption of ATP in the cell:

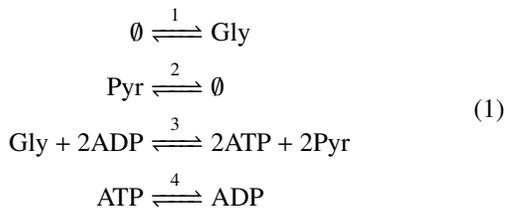

$$\begin{aligned}
\emptyset &\xrightleftharpoons{1} \text{Gly} \\
\text{Pyr} &\xrightleftharpoons{2} \emptyset \\
\text{Gly} + 2\text{ADP} &\xrightleftharpoons{3} 2\text{ATP} + 2\text{Pyr} \\
\text{ATP} &\xrightleftharpoons{4} \text{ADP}
\end{aligned} \quad (1)$$

The first two reactions stand for the couplings with external chemostats (depicted by $\emptyset$): the cell imports and expels glucose (Gly) and pyruvate (Pyr), effectively fixing their concentrations. In reaction 3, a molecule of Gly is used to convert two molecules of low-energy adenosine diphosphate (ADP) into two molecules of high-energy adenosine triphosphate (ATP). The chemical energy stored in ATP is then released during the spontaneous dephosphorylation (reaction 4), and used to fuel the physiological activity of the cell. One sees that whenever the four reactions are performed respectively once, twice, once, twice the number of molecules of each species is preserved. This is an example of chemical cycle [8, 10], that is, a sequence of reactions which does not alter the overall state of the system. The cell is maintained in a nonequilibrium steady state, dissipating energy for its metabolic activity, since chemostats sustain net currents of Gly (consumed) and Pyr (produced). Interestingly, the chemical cycle plays a role similar to periodicity for the Brownian particle on a ring.

In this example and in generic CRNs, the full analogy is hindered by the fact that interactions are inherently discrete and dependent on the topology of the reaction network. It is thus natural to ask: How can one identify the conservative and non-conservative contributions to the chemical force? How do they rise transient and steady currents? Based on works by Kirchhoff on electrical circuits and Kolmogorov on Markov chains, Hill and Schnakenberg [4, 6, 11] – among others – proposed a framework to describe, in steady state, the source of irreversibility as stemming from chemical cycles. Using the graph-theory notion of spanning tree, they identified a fundamental set of cycles defined on the population graph and explained their physical relevance to identify the chemical driving forces. This approach has been successfully exploited in several applications [12–16] and recently extended, by one of us, by introducing graph cocycles [17], a notion dual to that of cycles. Worthily, the notion of cocycle was the missing piece in Schnakenberg's analysis to understand the finite-time structure of chemical forces, beyond steady state. Albeit elegant, such settings in practice apply only to noninteracting (linear) networks which can be represented as simple graphs [18]. It is the case for instance of resistor (or flow) networks [19], minimal biochemical models [16, 20–22], or unimolecular CRNs [23] where each reaction involves one reactant and one product [e.g. reaction 4 in Eq. (1)].

However, real-world networks involve interactions among several species [e.g. reaction 3 in Eq. (1)] making them best represented as hypergraphs [24], that is, generalized graphs where hyper-edges connect more than two nodes (see Fig. 1). Hypergraphs have recently emerged as a new challenge in network science [25–27] and lack a comprehensive theoretical understanding. A key point is that no notion of spanning tree exists for hypergraphs, precluding the Hill–Schnakenberg analysis. Interactions are of course fundamental in inorganic chemistry where heterogeneous catalysis increases the efficiency of reaction pathways [28], as well as in intracellular processes, where autocatalytic interactions are at the core of the capability of living systems to self-replicate [29]. Interactions give rise to nonlinearities at the level of chemical concentrations, resulting in a spectrum of dynamical behaviors not displayed by noninteracting networks [30]. It thus appears crucial, in interacting CRNs, to build a geometry of hypergraphs aimed at identifying a decomposition of nonequilibrium physical observables such as currents and affinities. This is the objective of the present work.

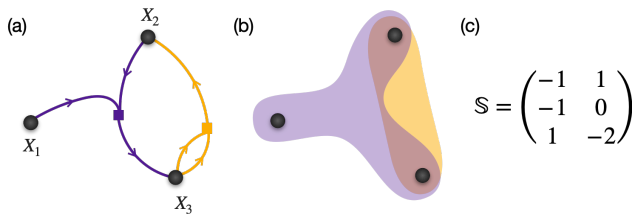

FIG. 1. Reactions $X_1 + X_2 \overset{1}{\rightleftharpoons} X_3$ , $2X_3 \overset{2}{\rightleftharpoons} X_1$ involves an interspecific interaction between species $X_1$ and $X_2$ and an intraspecific interaction between two molecules of $X_3$. **(a)** The hypergraph representation is composed of $N = 3$ nodes corresponding to the species $\{X_1, X_2, X_3\}$ and two hyperedges corresponding to reaction 1 (purple) and reaction 2 (orange). The hyperedges differ from simple edges as they present branches which can connect to different nodes (purple) or to the same node (orange). **(b)** The first reaction corresponds to a non-pairwise interaction as it involves all three species. The second reaction is still pairwise at the level of nodes, but the multiplicity of its edge is different from one, due to the branching. **(c)** As a consequence, the stoichiometric matrix $\mathbb{S}$ associated with the network is not an incidence matrix (see Sec. I B).

### B. Main results and structure of the paper

Here we establish a framework that extends the Hill–Schnakenberg analysis to the case of interacting CRNs. Crucially, we follow a novel algebraic approach to build generalized notions of cycles and cocycles (see Fig. 2). This allows us to bypass the conceptual bottleneck of previous analyses of non-interacting CRNs, that relied on graph theory (specifically, on spanning trees). To do so, a conceptual shift is required: from graph-theoretical objects to vector spaces. The newly defined cycles and cocycles reduce to the Hill–Schnakenberg ones for non-interacting CRNs and, in the fully interacting case, allow one to build geometrical notions that (*i*) generalize the graph-theoretical ones, (*ii*) provide a physical decomposition of observables (currents and forces) and (*iii*) reveal hidden structures in the underlying network exchanges.

A generic reaction network is described by a stoichiometric matrix $\mathbb{S}$, encoding its topology (see Fig. 1). For noninteracting networks, $\mathbb{S}$ coincides with an incidence matrix relating the nodes and edges of an oriented graph, with $\mathbb{S}^\top$ (resp. $\mathbb{S}$) representing a discretized gradient (resp. divergence). The non-invertibility of $\mathbb{S}$, due to the interdependence of its degrees of freedom, is a known issue [31, 32]. A key step in our approach is to construct geometrically a Green-function integrator $\mathbb{G}^\top$ which allows for a partial inversion of $\mathbb{S}^\top$ along a spanning tree. The procedure is explicit and naturally generalizes to the case of interacting CRNs, enabling us to extend the notion of integration and differentiation along a hypergraph. Doing so, we establish a potential condition for the forces and connect it to the notion of reversibility for the dynamics, and to that of (chemical) potential for the thermodynamics.

Equipped with such geometric interpretation, we put forward a decomposition of chemical forces into conservative and non-conservative contributions, akin to the Helmholtz–Hodge decomposition of vector calculus in $\mathbf{R}^3$. For the unfamiliar reader, such a decomposition of a vector field, $\mathbf{F} = -\boldsymbol{\nabla} V + \boldsymbol{\nabla} \times \mathbf{A} \in \mathbf{R}^3$ provides a separation of the force into two components: (*i*) a gradient force that is conservative, (*ii*) a non-gradient force with zero divergence (i.e. of zero total flux through any closed surface) that drives irreversible stationary behavior. We discuss the analogy with our decomposition of chemical forces and its implication for nonequilibrium physics [33, 34]. Physically, conservative and non-conservative forces generate currents of different geometric types. On one side, 'tidal' currents control the transient relaxation to steady state and are due to conservative forces; on the other side cyclic currents characterize the steady state in presence of non-conservative forces. Although easily pictured on graphs, such features survive also on hypergraphs (see Fig. 2). We provide a guideline to identify the different sets of currents in this case, based on the notions of algebraic cycles and cocycles. Such concepts have direct consequences for the dynamics of interacting CRNs: we show for instance that the slow modes of non-linear relaxation are controlled by the cocycles when a timescale separation occurs (and this, arbitrarily far from steady state). Close to equilibrium, we show that the linear responses to external field and to initial condition turn out to presenting a hidden spectral symmetry.

As a practical application, we consider the problem of thermodynamically consistent reconstruction of CRNs involved in various cellular functions, e.g. metabolism [35–42] (but other multi-omics datasets could also be considered). The problem is, roughly, the following: DNA sequencing grants knowledge of the enzymes possibly present in a cell, and enzyme specificity identifies the substrates (metabolites) that bind and interact. Thus the stoichiometry of the metabolite network is known. However, the currents of the reactions are not, and one would like to have some principles to make an informed guess about the overall currents expressed by the cell. Such currents, and thus the (phenotypic) state of the network, are subject to myriad of constraints. In particular, fundamental physical constraints are (*i*) mass balance, and (*ii*) thermodynamic feasibility. While the first is a simple linear constraint, the second is non-linear, and it has proven difficult to implement in reconstruction algorithms [43–47]. Given the network topology, these algorithms aim at navigating in the landscape of possible metabolite currents compatible with some known value of uptake and secretion rates. On one hand, our framework grants a simple linear-regime approximation of such landscape that allows one to explore it. The best feature of this reconstruction approach is that the only free parameters are some positive real numbers, one per internal reaction involved in a cycle. Once these parameters are given and some intuition about which are the independent external currents is built, the reconstruction is just a simple linear formula that allows one to explore a landscape of solutions. On the other hand, the geometric tools that we introduce enable us to identify a non-trivial set of exact linear relations between combinations of internal and external currents, that are valid arbitrarily far from the linear regime, and thus provide strong constraints on the landscape of solutions.

The paper is organized as follows. For the sake of clarity, we dedicate Part II to noninteracting networks, whose configuration space is a graph, and review the analysis of Hill–Schnakenberg, extending it to include finite-time relaxation (in the spirit of Ref. [17]). In this context, we discuss the complete mapping between the graph-theoretical analysis and the algebraic framework, which sets the stage for Part III. Therein, we show how to fully extend the theory to interacting CRNs, by defining generalized cycles and cocycles that give a geometrical meaning to integration and differentiation on hypergraphs. In Part IV, we put forward the announced decompositions of chemical forces (affinities) and currents and explain how these tools provide a rationale behind the notions of chemical affinity and chemical potential, as well as stochastic aspects such as local detailed balance and entropy production rate. Finally, as applications of the formalism, we study in Part V the linear responses of CRNs disclosing a spectral symmetry between the equilibrium relaxation and the driven steady-state perturbations. In conclusion, we propose an algorithmic-like procedure for feasible reconstructions of realistic metabolic networks.

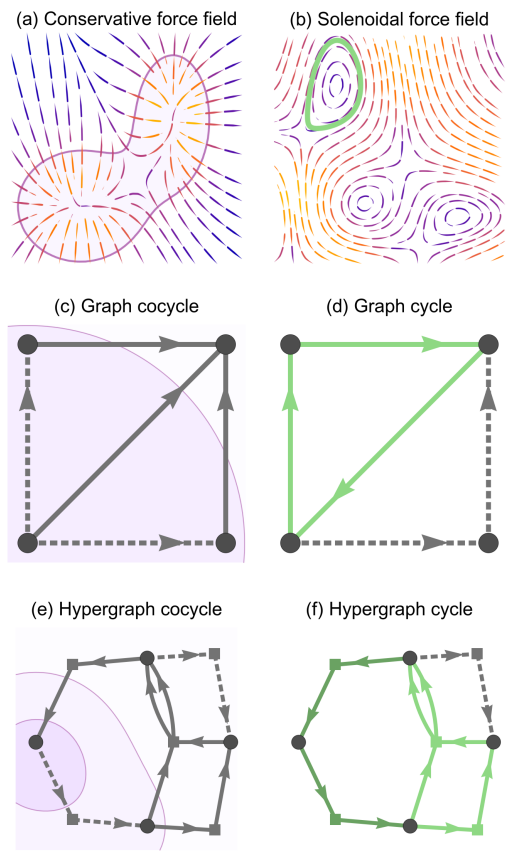

FIG. 2. Graphical summary of core geometrical concepts introduced in the paper. **(a)** A gradient field, deriving from a potential, in continuum space. A level line of the potential splits space in two components, the bounded one being represented as an "island" (purple). **(b)** A zero-divergence non-gradient field in continuum space, which generates forces along cycles (green). **(c)** On the graph representing a non-interacting network, an island (purple) splits the species in two sets and defines a characteristic potential (1 on the island, 0 elsewhere). Its gradient is localized at the island boundary, on the set of outward edges (non-dashed) that defines a cocycle. We identify a core set of islands/cocycles which form a basis of conservative forces. Such forces drive transient ("tidal") currents, that flow through the island boundaries, and control the overall relaxation of currents. **(d)** A graph cycle (green) supports stationary currents, driven by non-equilibrium forces (quantified by summing forces along cycles). **(e)** On the hypergraph of an interacting network, we define a potential landscape that generalizes the above notions of islands, which are now non-flat (shades of purple representing the "altitude" map). The boundary of an island defines a cocycle, as a weighted set of hyperedges (non-dashed). **(f)** Similarly, hypergraph cycles possess a complex topology and involve reactions with different weights (levels of green). The physical decompositions of forces and currents extend from graph to hypergraph.

### C. Setup and notation

In this paper we work with CRNs with mass-action kinetics. More precisely, we consider a dilute, well-stirred mixture of $N$ chemical species interacting through $R$ reactions. Each reaction is uniquely specified by two (non-negative) integer valued vectors, $\boldsymbol{\nu}^{+\rho}$ and $\boldsymbol{\nu}^{-\rho}$, which give the number of molecules per species being produced (+) and consumed (−) by reaction $\rho$. Grouping the species in a vector $\boldsymbol{X} = (X_1, ..., X_N)^\top$, reaction $\rho$ can be written as

$$\boldsymbol{\nu}^{+\rho} \cdot \boldsymbol{X} \underset{k_\rho^-}{\overset{k_\rho^+}{\rightleftharpoons}} \boldsymbol{\nu}^{-\rho} \cdot \boldsymbol{X}, \qquad (2)$$

where $\cdot$ is the scalar product (see Eq. (1) for an example). Each reaction is strictly reversible, that is, can occur both in the forward and backward direction with reaction rate constants $k_\rho^\pm > 0$. Thus, for each reaction we introduce a pair of velocities $\lambda_\rho^\pm(\boldsymbol{x})$ describing the rate of change of the chemical concentrations $\boldsymbol{x} = (x_1, ..., x_N)^\top$ in the corresponding direction. For large number of particles (i.e. negligible fluctuations), the velocities are proportional to the concentrations of the species partaking in the reaction $\pm\rho$,

$$\lambda_\rho^\pm(\boldsymbol{x}) = k_\rho^\pm \boldsymbol{x}^{\boldsymbol{\nu}^{\pm\rho}} \quad \forall \rho, \qquad (3)$$

with the notation convention $\boldsymbol{a}^{\boldsymbol{b}} = \prod_i a_i^{b_i}$. Then, we define the net current $J_\rho$ of reaction $\rho$ as the difference between the two reaction velocities

$$J_\rho(\boldsymbol{x}) = \lambda_\rho^+(\boldsymbol{x}) - \lambda_\rho^-(\boldsymbol{x}). \qquad (4)$$

The dynamical evolution of the concentration vector $\boldsymbol{x}(t)$ in time, given some initial concentrations at $t = 0$, is given by a deterministic rate equation:

$$\frac{d}{dt}\boldsymbol{x}(t) = \mathbb{S}\boldsymbol{J}(\boldsymbol{x}(t)), \qquad (5)$$

where $\mathbb{S}$ is the stoichiometric matrix, of dimensions $N \times R$, whose columns describe the net stoichiometry of each reaction $\mathbb{S}_\rho = \boldsymbol{\nu}^{-\rho} - \boldsymbol{\nu}^{+\rho}$. As such, $\mathbb{S}$ encodes the topology of the network, and acts as a discrete divergence in Eq. (5), which can be seen as a continuity equation. Finally, for each reaction $\rho$ we introduce the reaction affinity $A_\rho$ defined as:

$$A_\rho(\boldsymbol{x}) = \log \frac{\lambda_\rho^+(\boldsymbol{x})}{\lambda_\rho^-(\boldsymbol{x})} = \log\left(\frac{k_\rho^+}{k_\rho^-} \boldsymbol{x}^{-\mathbb{S}_\rho}\right). \qquad (6)$$

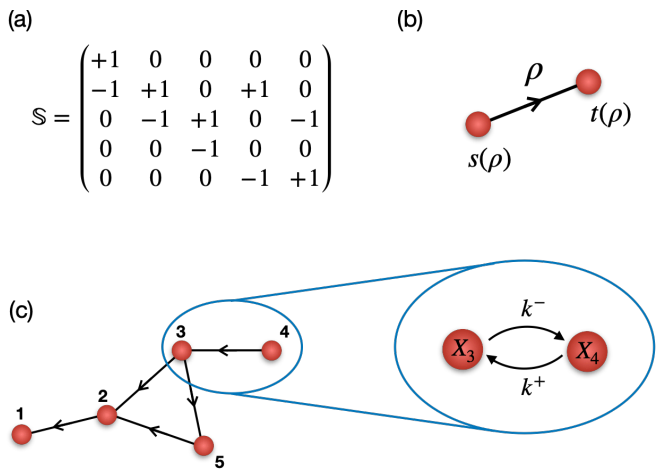

FIG. 3. Example of graphical representation for a noninteracting reaction network of the type of Eq. (8). **(a)** The stoichiometric matrix of a noninteracting network is an incidence matrix describing the relation between the species/nodes and the reactions/edges. Each column $\mathbb{S}_\rho$ has exactly a +1 entry corresponding to the species produced by the reaction $\rho$ and a −1 entry corresponding to the species consumed by the reaction $\rho$. **(b)** Accordingly, one can represent each column of $\mathbb{S}$ as an edge between a source node $s(\rho)$ and a target node $t(\rho)$. **(c)** By doing so for all the reactions in the network, one obtains a graph representation where each node is a species and each edge is a reaction. Since the reactions are reversible, the orientation of the edges is conventional.

Eq. (6) corresponds to the usual mass-action force, which implies the following constitutive relation between $J_\rho$ and $A_\rho$:

$$J_\rho(\mathbf{x}) = \lambda_\rho^+(\mathbf{x})\left[1 - \exp\left(-A_\rho(\mathbf{x})\right)\right], \quad (7)$$

and quantifies the chemical drive, i.e. how an imbalance in the concentrations of reactants and products results in a net reaction current.

Finally, while we adopt here the language of chemical reactions, the framework of Eqs. (2)-(7) describes generally the dynamics of interacting populations in the large system size limit. The only stringent assumption being that reactions (e.g. infection events in epidemic models or genetic mutations in evolutionary dynamics) are reversible, so as for Eq. (6) to be well-defined.

## II. NONINTERACTING REACTION NETWORKS

We dedicate this section to noninteracting networks where each reaction involves the transformation of a species into another:

$$X_i \underset{k_\rho^-}{\overset{k_\rho^+}{\rightleftharpoons}} X_j. \quad (8)$$

In this case, the stoichiometric matrix $\mathbb{S}$ takes the form of an incidence matrix, namely:

$$\mathbb{S}_{i\rho} = \begin{cases} -1, & \text{if } i \text{ is the species consumed by } \rho, \\ +1, & \text{if } i \text{ is the species produced by } \rho, \\ 0, & \text{otherwise}. \end{cases} \quad (9)$$

The objective of this part is to relate the algebraic and graph-theoretical pictures underlying such set of reactions, in view of extending these to the case of interacting CRNs (see Sec. III). To do so, we review rapidly the Hill–Schnakenberg approach in Sec. II A and translate it in the algebraic framework in Sec. II B. We then provide the first result of the paper: an integrator matrix $\mathbb{G}^\top$ inspired by geometry which allows one to integrate conservative forces on a graph and obtain the potentials from which they derive. Finally, notice that Eq. (8) also describes a Markov chain between states labeled by the $X_i$'s (this is one of Schnakenberg's standpoints).

### A. Cycles and cocycles in graph theory

Noninteracting CRNs like Eq. (8) admit a graphical representation in terms of nodes (or vertices) and edges (or links). The incidence matrix Eq. (9) describes the topology of an oriented graph $\mathcal{G}$ where each reaction $\rho$ is a directed edge pointing from a source node $s(\rho)$ to a target node $t(\rho)$, and each node represents a species (see Fig. 3 for an example). Notice that, since we assume reactions to be reversible, the orientation of the edge coincides with the orientation of the (chosen) forward direction in Eq. (2), as entailed in $\mathbb{S}$. Without loss of generality, we consider simply connected graphs.

Following the Hill–Schnakenberg approach [4, 6, 11], we introduce the notion of spanning tree, defined as a connected subgraph of $\mathcal{G}$, containing every node but no closed paths (Fig. 4b). Clearly in general there are several spanning trees and their number depends on the topology of $\mathcal{G}$. We fix one that we call $T_\mathcal{G}$. Choosing $T_\mathcal{G}$ corresponds to splitting the edges of $\mathcal{G}$ into edges that are excluded from the spanning tree and edges that belong to it (Fig. 4c). In graph theory [18], these distinct edges are respectively named chords and cochords and we associate them with two indices, $\alpha \notin T_\mathcal{G}$ spanning the set of chords and $\gamma \in T_\mathcal{G}$ spanning the set of cochords.

Adding a chord $\alpha$ back in $T_\mathcal{G}$ generates a closed path. Removing a cochord $\gamma$ from $T_\mathcal{G}$ generates a cut, i.e. a splitting of the nodes of $\mathcal{G}$ into two disconnected islands/components. Such closed paths and cuts can be given an orientation: that of the closed path prescribes the direction of going along it; and that of the cut is a choice of source and target among the two disconnected islands. We thus define the cycle $C(\alpha)$ as the closed path generated by restoring the chord $\alpha$ into $T_\mathcal{G}$, and oriented in the same direction as the generating chord $\alpha$ (Fig. 4d). We define the cocycle $C(\gamma)$ as the set of edges that sew the cut generated by the removal of the cochord $\gamma$. Conventionally, the source island is chosen to be the island containing the source node $s(\gamma)$ so that all the edges in $C(\gamma)$ are taken parallel to the generating cochord $\gamma$ (Fig. 4f). Notice that if the cochord $\gamma$ is a bridge, i.e. if it does not belong to any closed path, the corresponding cocycle contains only the cochord $\gamma$. On the other hand, if the cochord $\gamma$ belongs to one (or more) closed path in $\mathcal{G}$, the corresponding cocycle contains all the chords (possibly flipped) associated with those closed paths. Namely, for any pair of chord and cochord $(\alpha, \gamma)$ we have that $\gamma \in C(\alpha) \Leftrightarrow \alpha \in C(\gamma)$, with the pair of edges $(\alpha, \gamma)$ oriented parallel to each other in $C(\gamma)$ and antiparallel (head-tail orientation) in $C(\alpha)$. This encodes the fact that whenever $\gamma \in C(\alpha)$, the cycle $C(\alpha)$ possesses nodes both in the source and in the target of the cut corresponding to the cocycle $C(\gamma)$, and the edges of this cycle must present a zero flux in total between source and target. This 'duality' property between $C(\alpha)$ and $C(\gamma)$ is non-trivial, and its manifestation will be met in multiple forms in the following[1].

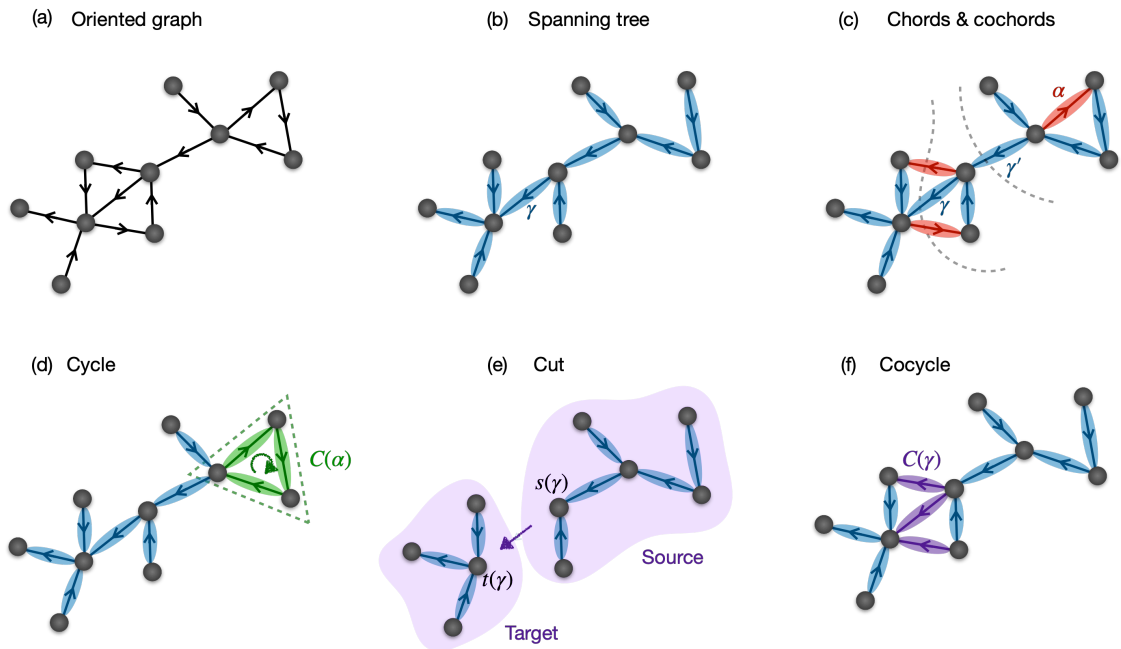

FIG. 4. A summary of the notions from graph theory which are discussed in the text. **(a)** Example of an oriented graph $\mathcal{G}$ obtained from the stoichiometric/incidence matrix $\mathbb{S}$ of a noninteracting CRN: each node represents a species and each edge a reaction. **(b)** A spanning tree $T_\mathcal{G}$, here represented with blue-marked edges, is obtained by pruning edges from the original graph so as to remove every closed path (while keeping a connected tree). **(c)** Picking $T_\mathcal{G}$ corresponds to (a choice of) splitting of the edges in $\mathcal{G}$ between the cochords (in blue), forming $T_\mathcal{G}$, and the chords (in red). The set of chords is spanned by the index $\alpha$ and the set of cochords by the index $\gamma$. Cochords like $\gamma'$ are named bridges as they do not belong to any closed path, in contrast for instance to the cochord $\gamma$ which belongs to two closed paths. **(d)** Reintroducing a chord $\alpha$ into the spanning tree generates a cycle $C(\alpha)$ which is oriented according to the orientation of the chord $\alpha$. By construction, each cycle only contains a single chord (the generating one). **(e)** The removal of a cochord $\gamma$ from the spanning tree generates a cut, i.e. a splitting of the full graph in two disconnected islands/components. In particular, we name 'source island' (resp. 'target island') the component containing the source node $s(\gamma)$ (resp. target node $t(\gamma)$) of the generating cochord. This allows one to establish an orientation of the cut. **(f)** The cocycle $C(\gamma)$ is defined as the set of edges that reconnect the source island to the target island after removing the cochord $\gamma$. In the example it has three elements: the cochord $\gamma$ and the two chords associated to cycles that contain $\gamma$. All the three edges are oriented parallel to the cochord $\gamma$, in order to connect the source to the target.

Cocycles and cycles are the central graph-theoretical ingredients of this paper. A cycle is a closed path in the space of reactions. As such, it corresponds to a sequence of transformations which connects a node back to itself. In the spirit of Hill–Schnakenberg's theory, we will use them to characterize nonequilibrium steady states. The interpretation of cocycles is less intuitive but will play a central role. Those are sets of reactions associated to a binary splitting of the graph into two separate sub-systems; as will become clear in the following sections, they are associated (*i*) to fluxes of matter with no circulation, relating to the modes of relaxation of the dynamics,

---
[1] A proof of this property can be found in §2-2 of Nakanishi's book [18]. Anticipating on algebraic notions, it encodes that a same matrix $\mathbb{T}$ describes both the dependencies between reactions and the cycles of CRNs; in this framework, a purely algebraic proof of this can be found at the end of our Appendix A 1.

and (*ii*) to conservative forces, that 'derive' from a potential.

### B. Algebraic definition of cycles and cocycles

We now detail how to algebraically relate cycles and cocycles to $\mathbb{S}$. Namely, we associate to cycles and cocycles two families of vectors which not only retain the same properties as on the graph but also, algebraically, occur to be bases of two complementary and orthogonal (real valued) vector spaces,

$$\operatorname{Ker} \mathbb{S} \perp \operatorname{Im} \mathbb{S}^\top, \quad (10)$$

namely the kernel $\operatorname{Ker} \mathbb{S}$ and the coimage $\operatorname{Im} \mathbb{S}^\top$ of the stoichiometric matrix.

To start with, we count the number of independent degrees of freedom. We know from the graph construction that the total number of cycles and cocycles is equal to the number of reactions/edges $R$ and, likewise, to the number of columns of $\mathbb{S}$. The latter is in turn related to the dimensions of the image and kernel of $\mathbb{S}$ via the rank-nullity theorem:

$$R = \operatorname{rank} \mathbb{S} + \dim \operatorname{Ker} \mathbb{S}. \quad (11)$$

It is known that the rank $M$ of the incidence matrix of any connected graph is $N - 1$ [2] with the matrix possessing a sole left nullvector $\ell_0 = (1, 1, \cdots, 1)$. This reflects the fact that the sum of the entries in any given column of $\mathbb{S}$ is zero. Physically, $\ell_0$ has the status of a mass conservation law and one verifies that the dynamics in Eq. (5) indeed conserves the quantity $\ell_0 \cdot \boldsymbol{x}(t)$: in a closed system, i.e. in the absence of fluxes in and out of the system, Lavoisier's law of mass conservation is satisfied. Algebraically, the $N$ species hence present $M = N - 1$ independent degrees of freedom. Accordingly, one can use Euler's formula together with Eq. (11) to relate the number of cycles and cocycles to the fundamental subspaces of $\mathbb{S}$: one finds that the cycles are in number equal to the dimension of the kernel, $\dim \operatorname{Ker} \mathbb{S} = R - M$; and that the cocycles are in number equal to $M = \operatorname{rank} \mathbb{S}$, the number of independent columns of $\mathbb{S}$ (i.e. independent reactions).

In the previous section we introduced two indices, $\alpha$ and $\gamma$, to span the chords and cochords sets respectively. The same labeling can be introduced for the columns of $\mathbb{S}$. From the definition of $T_{\mathcal{G}}$, the $M$ columns labeled with $\gamma$ are (a choice of) linearly independent columns of $\mathbb{S}$. For convenience, we order them in such a way that $1 \le \gamma \le M$ and $M + 1 \le \alpha \le R$. Then, we introduce two families of column vectors in $\mathbf{R}^R$, respectively denominated $\{\boldsymbol{c}^\alpha\}$ and $\{\boldsymbol{c}^\gamma\}$ and defined as:

$$(\ldots, \boldsymbol{c}^\alpha, \ldots) = \begin{pmatrix} -\mathbb{T} \\ \mathbb{1}_{R-M} \end{pmatrix}, \quad (\ldots, \boldsymbol{c}^\gamma, \ldots) = \begin{pmatrix} \mathbb{1}_M \\ \mathbb{T}^\top \end{pmatrix}. \quad (12)$$

Here $\mathbb{1}_n$ is the $n \times n$ identity matrix and $\mathbb{T}$ is a $M \times (R - M)$ rectangular matrix defined from the graph $\mathcal{G}$ as:

$$\mathbb{T}_{\gamma\alpha} = \begin{cases} +1, & \text{if cochord } \gamma \in C(\alpha) \text{ and } \| \text{ to it}, \\ -1, & \text{if cochord } \gamma \in C(\alpha) \text{ and } \nparallel \text{ to it}, \\ 0, & \text{otherwise}, \end{cases} \quad (13)$$

where $\|$ and $\nparallel$ refer to the orientation of the edge $\gamma$ (as prescribed by $\mathbb{S}$) matching or not the orientation of the cycle it belongs to. By construction, the vector $\boldsymbol{c}^\alpha$ thus specifies the composition (and orientation) of the edges entering in $C(\alpha)$, with $c^\alpha_{\alpha'} = \delta_{\alpha\alpha'}$ since in the chord set only the generating chord $\alpha$ belongs to $C(\alpha)$ and dictates its orientation. (Here and below, $\delta_{ij}$ denotes the Kronecker delta.) Analogously, the vector $\boldsymbol{c}^\gamma$ contains non-zero entries for any edge that belongs to the cocycle $C(\gamma)$, such that $c^\gamma_\rho \ne 0$ if and only if $\rho \in C(\gamma)$ and $c^\gamma_{\gamma'} = \delta_{\gamma\gamma'}$ since the only cochord contained in $C(\gamma)$ is the generating one. Albeit not obvious, the same matrix $\mathbb{T}$ (up to a sign) controls the composition of both cycles and cocycles, as expressed by Eq. (12). This is the algebraic encoding of the duality discussed in the previous section.

Due to the identity matrices in Eq. (12) all vectors $\boldsymbol{c}^\alpha$ and $\boldsymbol{c}^\gamma$ are linearly independent and one easily checks that they span orthogonal subspaces, since:

$$\boldsymbol{c}^\gamma \cdot \boldsymbol{c}^\alpha = 0 \quad \forall \gamma, \alpha. \quad (14)$$

Furthermore, the geometric construction ensures that the vectors $\boldsymbol{c}^\alpha$ belong to the kernel of $\mathbb{S}$, that is

$$\sum_\rho \mathbb{S}_{i\rho} c^\alpha_\rho = 0 \quad \forall \alpha, i. \quad (15)$$

This represents the fact that any node $i$ in a cycle has exactly one incoming and one outgoing edge. As a consequence, cycles and cocycles form a basis for, respectively, the kernel of $\mathbb{S}$ and its orthogonal complement, i.e. the coimage of $\mathbb{S}$, $\operatorname{Im} \mathbb{S}^\top$. This is the algebraic characterization of cycles and cocycles, which complements their definition from graph theory. Likewise, a vectorial representation holds for the chords and the cochords. Those are the canonical vectors $e^\gamma_\rho = \delta_{\gamma\rho}$ and $e^\alpha_\rho = \delta_{\alpha\rho}$ in $\mathbf{R}^R$. All in all, we have identified two alternative bases for $\mathbf{R}^R$, that we can merge in the following two matrices:

$$(\boldsymbol{e}^\gamma, \boldsymbol{e}^\alpha) = \begin{pmatrix} \mathbb{1}_M & \mathbb{0} \\ \mathbb{0} & \mathbb{1}_{R-M} \end{pmatrix}, \quad (\boldsymbol{c}^\gamma, \boldsymbol{c}^\alpha) = \begin{pmatrix} \mathbb{1}_M & -\mathbb{T} \\ \mathbb{T}^\top & \mathbb{1}_{R-M} \end{pmatrix}. \quad (16)$$

The left-hand matrix is the canonical basis in $\mathbf{R}^R$ obtained from the chords/cochords vectorial representation. The right-hand matrix is the non-orthogonal basis formed by the vectorial representation of cycles and cocycles. One easily verifies the following orthogonality relations:

$$\begin{aligned} \boldsymbol{e}^\alpha \cdot \boldsymbol{c}^{\alpha'} &= \delta_{\alpha\alpha'} \\ \boldsymbol{e}^\gamma \cdot \boldsymbol{c}^{\gamma'} &= \delta_{\gamma\gamma'}. \end{aligned} \quad (17)$$

From now on, we will call cycles and cocycles both the vectors defined in Eq. (12) and the graph-theoretical objects defined in the previous paragraph, as they are equivalent. We designate by chemical cycle any vector generated by a linear combination of the $\boldsymbol{c}^\alpha$'s.

### C. Integrating conservative forces on the graph

As anticipated in the introduction, the transposed incidence matrix $\mathbb{S}^\top$ is the discrete gradient operator prescribing the relation between the nodes and the edges of the graph. Accordingly, we define a force $\boldsymbol{F} \in \mathbf{R}^R$ to be conservative if it verifies a potential condition $\boldsymbol{F} = -\mathbb{S}^\top \boldsymbol{V}$, with $\boldsymbol{V} \in \mathbf{R}^N$ a potential defined on the nodes of $\mathcal{G}$ and $F_\rho = -(\mathbb{S}^\top \boldsymbol{V})_\rho = V_{s(\rho)} - V_{t(\rho)}$. Algebraically, it is equivalent to $\boldsymbol{F} \in \operatorname{Im} \mathbb{S}^\top = (\operatorname{Ker} \mathbb{S})^\perp$ which is the space spanned by the cocycles, so that we can write:

$$\boldsymbol{F} = \sum_\gamma F^c_\gamma \boldsymbol{c}^\gamma, \quad (18)$$

or equivalently $\boldsymbol{F} \cdot \boldsymbol{c}^\alpha = 0$, $\forall \alpha$. Notice that, from Eqs. (12) and (17), the coefficient $F^c_\gamma$ of the linear combination coincides with the entry $\gamma$ of $\boldsymbol{F}$, i.e. $F^c_\gamma = \boldsymbol{F} \cdot \boldsymbol{e}^\gamma = F_\gamma$.

Here the main difficulty to solve for the potential $\boldsymbol{V}$ in $\boldsymbol{F} = -\mathbb{S}^\top \boldsymbol{V}$ is the non-invertibility of $\mathbb{S}$, which prevents the identification of a discrete 'integrator' associated to $\mathbb{S}^\top$. Once again, graph theory comes in handy. Upon fixing arbitrarily a root node, let us orient all edges in $T_\mathcal{G}$ towards it and define $\forall i$ the subset $\mathcal{U}(i)$ containing all the nodes that are upstream node $i$ along the spanning tree, including $i$ itself; also, we number reactions starting from the root (see Appendix A for details). Then, we introduce the $N \times N$ square matrix $\mathbb{G}$ defined as

$$\mathbb{G}_{ij} = \begin{cases} 1, & \text{if node } j \in \mathcal{U}(i) \\ 0, & \text{otherwise.} \end{cases} \qquad (19)$$

which is invertible. See Fig. 5 for an example. In Appendix A we prove that the matrix $\mathbb{G}^\top$ then takes the form of a lower triangular integration operator on $T_\mathcal{G}$: namely, if $F$ is conservative, a solution to $F = -\mathbb{S}^\top V$ is given by the matrix $\mathbb{G}^\top$ in the following way. For an arbitrary vector $F$, let us define a potential $V_i$ on each node from the set of coefficients $F^c_\gamma$ in Eq. (18) as:

$$V_i[F^c_\gamma] = \sum_{1 \leq \gamma \leq M} \left(\mathbb{G}^\top\right)_{i\,s(\gamma)} F^c_\gamma = \sum_{1 \leq \gamma \leq M} F^c_\gamma \, \delta_{i \in \mathcal{U}(s(\gamma))}. \qquad (20)$$

---

[2] The rank-nullity theorem yields $\operatorname{rank} \mathbb{S} = \operatorname{rank} \mathbb{S}^\top = N - \dim \operatorname{Ker} \mathbb{S}^\top = N - 1$ since $\dim \operatorname{Ker} \mathbb{S}^\top = 1$, as indeed $\operatorname{Ker} \mathbb{S}^\top$ is spanned by $\ell_0^\top = (1, \cdots, 1)^\top$. It contains no other independent vector: *ad absurdum*, if such a vector would exist, one could build, by linear combination with $\ell_0^\top$, a non-zero vector $\boldsymbol{\ell}^\top \in \operatorname{Ker} \mathbb{S}^\top$ containing a 0 component $\ell_i = 0$; this is impossible, since using $\boldsymbol{\ell} \mathbb{S} = 0$ by recursion along the connected graph $\mathcal{G}$ starting from node $i$, we find $\ell_j = 0, \forall j$. See also for instance Ref. [48].

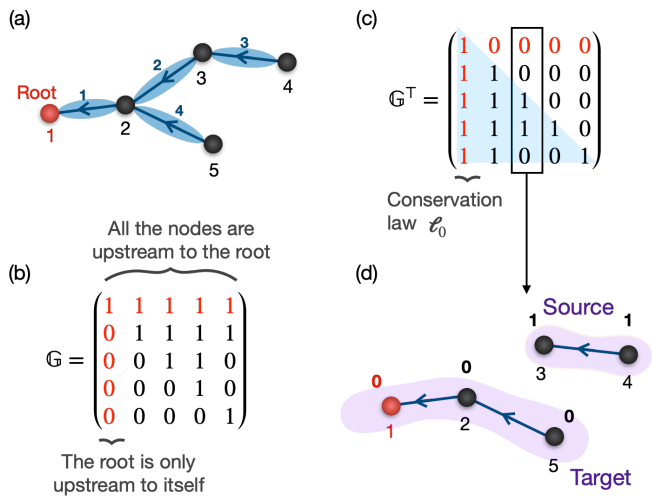

FIG. 5. **(a)** Spanning tree for the example CRN in Fig. 3. We picked node 1 (red) as the root and oriented all edges towards it. The enumeration of the nodes (in black) and the edges (bold blue) follows a natural convention detailed in Appendix A: It entails a simple one-to-one correspondence between the nodes $j \neq$ root (in black) and the cochords $\gamma \in T(\mathcal{G})$: namely, $\forall j \neq$ root $\exists \gamma$ s.t. $j = \gamma + 1 = s(\gamma)$. **(b)** The $N \times N$ matrix $\mathbb{G}$ for the aforementioned spanning tree, constructed as detailed in the main text. The first line and first column (in red) refer to the root and reflect the fact that we have oriented all the edges towards it. As a consequence, all the nodes $\neq$ root are upstream to it, hence the first line full of 1's. **(c)** The matrix $\mathbb{G}^\top$ presents the lower triangular structure of an integral operator on the spanning tree. The first column (in red) coincides to the mass conservation law $\ell_0$ and the 0's in the first line (red) fixes the root potential to $V_{\text{root}} = 0$ [see Eq. (20)]. **(d)** The last $M = N - 1$ columns of $\mathbb{G}^\top$ are in one-to-one correspondence with the $M$ cocycles. In particular, one reads from the +1 entries of the column $j = \gamma + 1$ the source island of the cocycle $C(\gamma)$. For example, if we remove the cochord $\gamma = 2$ from the spanning tree in panel (a), we generate a cut with the nodes 3 and 4 being disconnected from the rest of the graph. Algebraically, the appearance of a source island corresponds to the emergence of a new conservation law. Indeed the last $M$ columns of $\mathbb{G}^\top$ correspond to the conservation laws obtained by removing the cocycles.

The sum in Eq. (20) runs over all the edges in $T_\mathcal{G}$. Using the definition of $\mathbb{G}$ one sees that, for fixed $i$, $\sum_j (\mathbb{G}^\top)_{ij}$ runs over all the nodes which have $i$ among their upstreamers. Moreover, for every node $j$, except the root, there exists exactly one $\gamma$ such that $j = s(\gamma) = \gamma + 1$. Thus, the sum in Eq. (20) runs over the unique path on $T(\mathcal{G})$ between node $i$ and the root. Hence, it corresponds to a discrete integration of the entries $F^c_\gamma$ along the spanning tree. Notice that, from Eq. (20), the potential of the root is zero since the root, by convention, only has incoming edges. This means that the potential in Eq. (20) is uniquely defined up to a constant shift (fixed by $V_{\text{root}} = 0$), in analogy to the constant of integration in standard calculus.

Notably, the product $\mathbb{S}^\top \mathbb{G}^\top$ reads, as proved in Appendix A

$$-\mathbb{S}^\top \mathbb{G}^\top = \begin{pmatrix} 0 & \mathbb{1}_M \\ 0 & \mathbb{T}^\top \end{pmatrix} \qquad (21)$$

where the first column is full of zeroes and the matrix $\mathbb{T}$ is the same matrix as defined in Eq. (13). This special structure encodes the fact that the stoichiometric matrix is not full-rank, but it contains some built-in redundancy. The $M \times M$ square identity matrix $\mathbb{1}_M$ represents the inversion procedure between $\mathbb{S}^\top$ and $\mathbb{G}^\top$, illustrated in Eq. (20). The first column $(0 \ldots 0)^\top$ reflects the existence of the conservation law $\ell_0$ and the $\mathbb{T}$ matrix reflects the interdependence among reactions. Namely, only $M$ out of the $R$ columns of $\mathbb{S}$ are linearly independent while the remaining columns, labeled with $\alpha$ and associated to the cycles $c^\alpha$, can be obtained as a linear combination of the former. This is what the matrix $\mathbb{T}$ encodes: denoting by $\mathbb{S}_M$ the first $M$ (independent) columns of $\mathbb{S}$ and $\mathbb{S}_{\text{dep}}$ the last ones (corresponding to $R - M$ dependent reactions), one reads from Eq. (21) that $\mathbb{S}_{\text{dep}} = \mathbb{S}_M \mathbb{T}$. The relation in Eq. (21) is key in our analysis because it lies at the core of its extension from noninteracting to interacting CRNs, presented in Sec. III.

Excluding the first null-column, one recognizes on the right-hand side (r.h.s.) of Eq. (21) the cocycle basis of Eq. (12). This means that the columns of $\mathbb{G}^\top$, except the first one associated to the root, can be seen as $M = N - 1$ potential landscapes which, upon 'differentiation' via $\mathbb{S}^\top$, give the cocycle vectors $c^\gamma$'s. We thus define the potential vector $v^\gamma = (\mathbb{G}^\top)_{s(\gamma)}$, as the column $s(\gamma)$ of $\mathbb{G}^\top$. Then $c^\gamma = -\mathbb{S}^\top v^\gamma$ and $v^\gamma$ is a characteristic potential landscape (entries are 0 or 1) defined on the nodes of $\mathcal{G}$. It is in fact characteristic of the cut generated by the removal of the cocycle $c^\gamma$, as discussed previously. In particular, the target island, containing the root, is the sub-graph held at zero potential while the source island, corresponding to the +1 entries in $v^\gamma$, is held at unit potential. From a graphical viewpoint, the cocycle $c^\gamma$ is therefore the boundary of the source island $v^\gamma$ which, upon removing $c^\gamma$, remains isolated (Fig. 5).

## III. INTERACTING REACTION NETWORKS

We now show how the notions of cycles and cocycles, together with the geometrical picture of islands, can be extended to hypergraphs using linear algebra and discuss some important consequences for the macroscopic dynamics of $\boldsymbol{x}(t)$.

### A. Alternative construction of cycles and cocycles

Let us consider a generic interacting CRN whose topology is encoded in the stoichiometric matrix $\mathbb{S}$ with rank $M$. Algebraically, the rank of $\mathbb{S}$ quantifies the number of independent species and independent reactions, which are the same by rank-nullity theorem. Accordingly, one can pick $M$ independent reactions and, following the same convention as in Sec. II B, reorder the columns of $\mathbb{S}$ in a way such that they are placed first. Their label index is then $1 \leq \gamma \leq M$. These are the independent reactions which, in the case of noninteracting networks, constituted the cochords defining the spanning tree $T_\mathcal{G}$. The remaining $\alpha$-labeled reactions are in number $R - M = \dim \operatorname{Ker} \mathbb{S}$, so that $M + 1 \leq \alpha \leq R$. Contrary to noninteracting networks, the conservation laws are generally more than one, being in number $N - M = \dim \operatorname{Ker} \mathbb{S}^\top \geq 0$. By Eq. (5), each of them is associated to a physical quantity which is conserved.

We now show how the algebraic row reduction of $\mathbb{S}$ allows one to identify: (i) a choice of $N - M$ conserved quantities and (ii) a generalization of the cycle and cocycle bases for the (non-incidence) matrix $\mathbb{S}$. Using for instance the Gauss–Jordan elimination, a standard procedure in linear algebra [49], the stoichiometric matrix $\mathbb{S}$ is reduced to

$$-\mathbb{G}\,\mathbb{S} = \begin{pmatrix} \mathbb{0} & \mathbb{0} \\ \mathbb{1}_M & \mathbb{T} \end{pmatrix}. \quad (22)$$

Here the $N \times N$ matrix $\mathbb{G}$ is invertible and encodes the elementary operations performing the Gauss–Jordan elimination (see also Appendix A for another – explicit – construction of $\mathbb{G}$). Upon a permutation of rows, one recognizes in the r.h.s. of Eq. (22) the canonical reduced row echelon form [49], where the $M$ pivot elements constitute the bottom-left identity matrix $\mathbb{1}_M$ [3]. The first $N - M \geq 0$ rows filled with zeroes reflect the fact that $\mathbb{S}$ is not full row-rank in general, due to the possible existence of conservation laws. We stress that the reduced row echelon form, hence the matrix $\mathbb{T}$ in Eq. (22), is unique; it does not depend on the specific form of $\mathbb{G}$ (which is not unique). Applying the matrix $\mathbb{G}$ to Eq. (5) one obtains:

$$\frac{d}{dt}(\mathbb{G}\,\boldsymbol{x}(t))_i = 0, \quad \text{for } 0 \leq i \leq N - M. \quad (23)$$

Thus, the first $N - M$ elements of $\mathbb{G}\boldsymbol{x}$ are (a choice of) conserved quantities for the evolution of concentrations.

The interdependence among reactions is what the matrix $\mathbb{T}$ in Eq. (22) encodes: each dependent column $\mathbb{S}_\alpha$ is given by $\mathbb{S}_\alpha = \sum_\gamma \mathbb{S}_\gamma \mathbb{T}_{\alpha\gamma}$. Notably, we can use the notion of independence/dependence among reactions to restore the terms chords and cochords even in the absence of a spanning tree. In particular, we name chords (resp. cochords) the set of dependent (resp. independent) reactions in $\mathbb{S}$.

Let us now use the invertibility of $\mathbb{G}$ and take the transpose of Eq. (22), so that:

$$\mathbb{S}^\top = -\begin{pmatrix} \mathbb{0} & \mathbb{1}_M \\ \mathbb{0} & \mathbb{T}^\top \end{pmatrix} (\mathbb{G}^{-1})^\top. \quad (24)$$

Since $\mathbb{G}$ is full-rank, the image of $\mathbb{S}^\top$ is spanned by the $M$ non-zero columns of the reduced row echelon form. This explains the choice of the same notation $\mathbb{T}$ as for the matrix in Eq. (13) which was used to construct the cycles and cocycles for simple graphs. In that case, the matrix $\mathbb{T}$ was built from the spanning tree (see Sec. II A) while here it is obtained by algebraic row reduction. As a consequence, the entries of the new matrix $\mathbb{T}$ are no longer restricted to $\{0, \pm 1\}$ as in Eq. (13), but may take fractional entries (see Appendix A). In both cases, $\mathbb{T}$ allows one to define a basis for $\operatorname{Im} \mathbb{S}^\top$, which in the previous case was identified as the space spanned by the cocycles $\boldsymbol{c}^\gamma$'s. Accordingly, we interpret the column vectors in $(\mathbb{1}_M \; \mathbb{T})^\top$ as a family of generalized cocycles $\{\boldsymbol{c}^\gamma\}$. Following in the analogy, it is natural to use matrix $\mathbb{T}$ in Eq. (22) to construct a basis for the kernel of $\mathbb{S}$. In particular, we define a family of generalized cycles $\{\boldsymbol{c}^\alpha\}$ as the column vectors in $(-\mathbb{T}^\top \; \mathbb{1}_{R-M})^\top$. As in Sec. II B, the rank-nullity theorem ensures that these vectors constitute a basis for $\operatorname{Ker} \mathbb{S}$. In fact, it is possible to show that any basis of $\operatorname{Ker} \mathbb{S}$ can be reduced to that form, with $\mathbb{T}$ uniquely defined by Eq. (22) (see Appendix A).

We have shown how the row reduction of $\mathbb{S}$ allows one to identify bases for the kernel and the coimage of $\mathbb{S}$ which we connect to the previously defined cycles and cocycles. In fact, using this algebraic procedure for noninteracting CRNs yields the previous expressions Eq. (12) for the cycles and cocycles – provided the graph is oriented using the convention depicted in Fig. 5. In this sense, the newly defined vectors $\boldsymbol{c}^\alpha$'s and $\boldsymbol{c}^\gamma$'s are genuine generalizations of the graph cycles and cocycles, and we use the same terminology to designate them. Notably, the orthogonality relations in Eq. (14) and (17) apply directly to the new sets $\{\boldsymbol{c}^\gamma\}$ and $\{\boldsymbol{c}^\alpha\}$, which opens to the possibility of interpreting them as geometrical objects on hypergraphs.

We conclude with a remark. The vectors $\boldsymbol{c}^\alpha$'s and the $\boldsymbol{c}^\gamma$'s are not the only bases of $\operatorname{Ker} \mathbb{S}$ and $\operatorname{Im} \mathbb{S}^\top$ (for instance the first $M$ columns of $\mathbb{S}$ span $\operatorname{Im} \mathbb{S}^\top$). The interest of the definitions of cycles and cocycles we put forward is that they allow for a physical decomposition of the chemical affinities and currents (see Sec. IV A), and that they can be used to build a geometrical representation of the forces and currents on hypergraphs, as we present now. We stress that, although chemical cycles, spanning Ker $\mathbb{S}$, are known to play an important role [8, 10], the basis $\{\boldsymbol{c}^\alpha\}$ that we introduce brings a new physical content, through a one-to-one correspondence between a chord and its associated cycle – which is very helpful to map precisely which reactions are affected by non-equilibrium drive. (See also Appendix D for a definition of oblique projectors based on the $\boldsymbol{c}^\alpha$'s and $\boldsymbol{c}^\gamma$'s that generalizes to arbitrary CRNs those defined in Ref. [17] for noninteracting CRNs.)

---

[3] In the reduced row echelon form, the left part need not always be an identity matrix. The pivots, i.e. the columns containing a leading one and zeroes in all the other entries are generally scattered in the matrix. In our case, since we placed $M$ independent columns first in $\mathbb{S}$ (by permutation), the reduced row echelon form presents an identity matrix as in Eq. (22). In fact, row reduction provides another way of permuting the columns of $\mathbb{S}$: if $\mathbb{S}$ is not already having its $M$ first columns independent, the row echelon form provides $M$ pivots, their position being that of $M$ independent columns [49], which can then be placed first. For consistency with the geometrical analysis introduced for noninteracting CRNs in Sec. II C, we place in Eq. (22) the 0's on the first lines instead of the last lines (as often done).

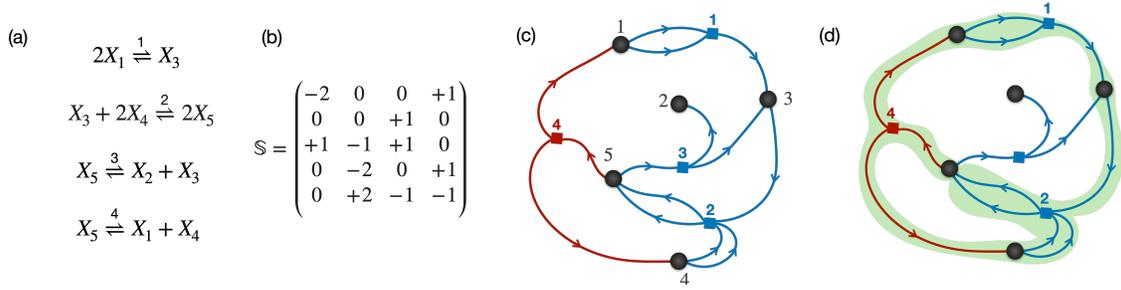

FIG. 6. (a) Example of an interacting CRN involving four reactions and five species $\{X_1, X_2, X_3, X_4, X_5\}$. (b) The corresponding stoichiometric matrix $\mathbb{S}$ is not longer an incidence matrix: its columns contain more than two entries and the values of the stoichiometric coefficients are in general different from ±1. As reported in the main text the rank $\mathbb{S} = 3$ and the matrix has two left nullvectors and one cycles, respectively $\boldsymbol{\ell}_1 = (1, 2, 1, 2, 0)$, $\boldsymbol{\ell}_2 = (0, 0, 1, 1, 1)$ and $\boldsymbol{c} = (1/2, 1/2, 0, 1)^\top$. (c) The hypergraphical representation of the reaction scheme in terms of nodes (species) and hyperedges (reactions). Following the convention described in the main text the independent reactions (cochords in blue) are numbered first and the dependent reaction (chord in red) associated to the cycle $\boldsymbol{c}$ is numbered last. (d) The reactions (1, 2 and 4) involved in the cycle $\boldsymbol{c}$ are highlighted in green. Upon performing each reaction $\rho$ a (fractional) number of time $c_\rho$, each species $X_i$ is consumed and produced the same amount.

### B. Geometry of hypergraphs

#### 1. Cycles

The geometrical aspect of cycles and cocycles is rooted in the orthogonality relations Eqs. (14) and (17). Those are the expression of the one-to-one correspondence between cocycles and independent reactions on the one side, and cycles and dependent reactions, on the other side. In the previous section we pointed out that for interacting CRNs the entries of the cycles vectors $\boldsymbol{c}^\alpha$ may be fractional. Contrary to the case of noninteracting CRNs, cycles are decorated with weights given by the entries of the $\mathbb{T}$ matrix in Eq. (22). Intuitively, these physical weights express the (fractional) number of times that each reaction must be performed along a cycle in order to leave the state of the system invariant. For illustrative purposes, we report in Fig. 6 the example of an interacting CRN with five chemical species and four reactions. In this case the stoichiometric matrix has rank $M = 3$ and exhibits $N - M = 2$ conservation laws and a number of cycles $R - M = 1$. From Eq. (22) one obtains $\mathbb{T} = (-1/2, -1/2, 0)^\top$ and $\boldsymbol{c} = (1/2, 1/2, 0, 1)^\top$ (we drop the index $\alpha = 1$ for simplicity). One sees that identifying the cycle graphically is not straightforward on the hypergraph (Fig. 6c). Nevertheless, cycles are still an important feature of the dynamics: we will see their relevance for nonequilibrium steady states and the practical consequences of the duality between cycles and cocycles (i.e. that they are described using the same matrix $\mathbb{T}$).

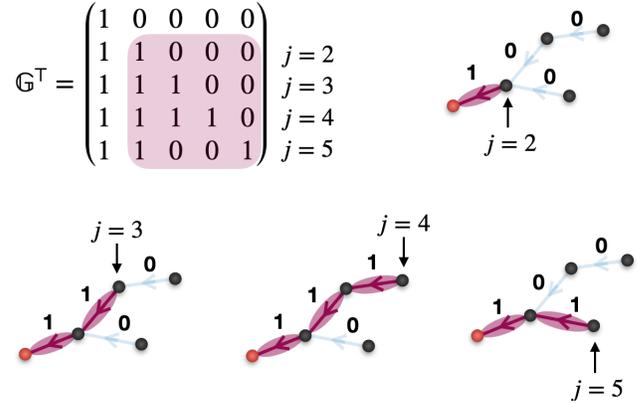

FIG. 7. One reads along the line of the bottom-right $(N-1) \times (N-1)$ block of the matrix $\mathbb{G}^\top$, defined by the transpose of Eq. (19), the integration paths along the spanning tree which are used in Eq. (20). Those correspond to the unique escape routes connecting any node $j \neq$ root to the root. Here we report $\mathbb{G}^\top$ for the example in Fig. 3 where the corresponding block is highlighted in purple. Notice that every node $j \neq$ root is the source $s(\gamma)$ of exactly one edge in the spanning tree (see Fig. 5a). Hence, for a fixed $j \neq$ root, the sum $\sum_\gamma \mathbb{G}^\top_{js(\gamma)}$ runs over all the edges which connect the node $j$ to the root. For the example of Fig. 5a, we depict graphically the escape routes for the various nodes $2 \leq j \leq N = 5$.

#### 2. Conservative forces can be integrated

We now ask the question of what is the geometrical meaning of the weights in the matrix $\mathbb{T}$, underlying both cycles and cocycles. First let us recall the integration matrix $\mathbb{G}^\top$ previously introduced for noninteracting CRNs. It was explicitly constructed by fixing a spanning tree $T_\mathcal{G}$ and one root [see Eq. (19) and Appendix A] such that the $(N-1) \times (N-1)$ bottom-right block contains the set of paths on the spanning tree along which we integrate the conservative force $\boldsymbol{F}$ to define the potential $V$ [Eq. (20)]. From a purely graphical viewpoint, each path can be interpreted as the unique 'escape route' in $T_\mathcal{G}$ along which a unit 'charge' placed on a given node is expelled through the root leaving no trace along the way (see

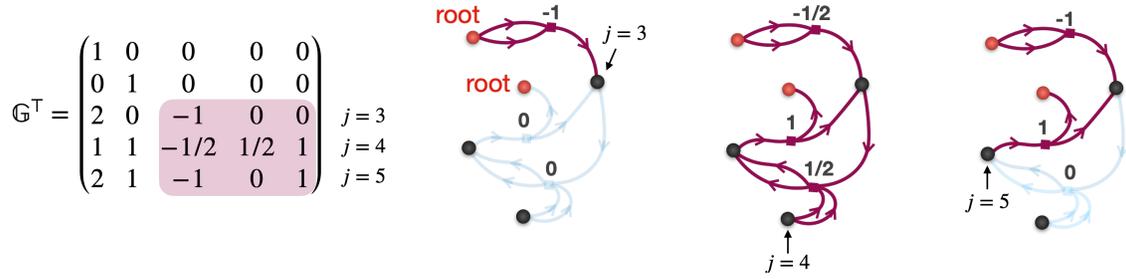

FIG. 8. The escape routes are read as lines of the bottom-right ($M \times M$) block in the $\mathbb{G}^\top$ matrix. For interacting CRNs, there is no simple graphical procedure to fill the entries of $\mathbb{G}^\top$. Nevertheless, an algebraic algorithm to identify the entries of $\mathbb{G}^\top$ is described in Appendix A. Here we report the $\mathbb{G}^\top$ for the example of Fig. 6. The numbering of the nodes follows the convention discussed in the main text: the two roots are labelled as node 1 and 2 and the rest of the nodes are numbered last (see Fig. 6). In this case the rank $\mathbb{S} = M = 3$ hence we highlight in purple the bottom-right ($3 \times 3$) block. We also represent the escape routes graphically together with the entries of the corresponding line. By construction, the escape routes are constrained to live on the independent reactions (cochords); for this reason we have removed from the hypergraph reaction 4 which is the reaction associated to the cycle $c$ (the red chord in Fig. 6).

Fig. 7). Thus, following this geometrical view, we may re-express matrix $\mathbb{G}^\top$ as

$$\mathbb{G}^\top = \begin{pmatrix} 1 & 0 & \cdots & 0 \\ \vdots & & & \\ 1 & & \text{escape routes} & \end{pmatrix}, \qquad (25)$$

where the escape routes constitute the bottom-right $(N-1) \times (N-1)$ submatrix of $\mathbb{G}^\top$. For graphs, the escape routes involve a succession of adjacent edges, irrespective of the connectivity of each node, which is reflected in the entries of $\mathbb{G}^\top$ being 0 or 1. This is no longer the case in hypergraphs due to the presence of branching in the hyperedges. Notice that in Eq. (25) the root is naturally associated to the conservation law $\ell_0$ which appears as the first column of $\mathbb{G}^\top$. Thus for the case of interacting CRNs it is natural to generalize the structure in Eq. (25) by picking a root for each of the (now possibly multiple) conservation laws of $\mathbb{S}$. Doing so, one obtain a set of roots, each one associated to a given conservation law, and we may ask what are the corresponding escape routes on the hypergraph, i.e. the 'hyper-paths' along which a charge placed on any node is expelled through the roots leaving no trace. In the absence of a spanning tree we lack a graphical procedure to find such escape routes; nevertheless in Appendix A we show that, given a suited set of roots[4], the escape routes can be obtained algebraically and are uniquely defined. By construction, they involve the $M$ independent reactions (the cochords) in analogy to the escape routes defined from the spanning tree in simple graphs.

---

[4] In the case of a simple noninteracting CRN, the choice of the root is fully arbitrary since all the $N$ species in the system are equally constrained by the left nullvector $\ell_0 \in \text{Ker} \, \mathbb{S}^\top$. It is no longer the case for an interacting CRNs for which the conservation laws are shaped by the interactions and will typically involve subsets of species. Then, given a conservation law one chooses a root among the subset of species that are constrained by that conservation law. The procedure is repeated for every conservation law. One species cannot be picked twice as a root.

Accordingly, we introduce a generalized matrix $\mathbb{G}^\top$

$$\mathbb{G}^\top = \begin{pmatrix} \text{csv} & \begin{matrix} 0 & \cdots & 0 \\ 0 & \cdots & 0 \end{matrix} \\ \text{laws} & \text{escape routes} \end{pmatrix} \begin{matrix} \updownarrow N-M \\ \updownarrow M \end{matrix} \qquad (26)$$

$$\underbrace{\phantom{XXXXXXXXXXX}}_{\mathbb{G}^\top_M}$$

where the conservation laws (spanning $\text{Ker} \, \mathbb{S}^\top$) make the first $N - M$ columns, the top-right block is padded with 0's and the $M \times M$ bottom-right square matrix contain the escape routes from each node $\notin$ {roots}. In Fig. 8 we represent such escape routes for the example given in Fig. 6. Compared to the case of simple graphs, each escape route is now a 'multi-path', i.e. a combination of the hyperedges, and is decorated with weights which tell how many times each cochord reaction is applied for the unit charge to vanish through the roots. As such, they constitute the generalization to hypergraphs of the simple escape routes identified from the spanning tree in graphs. Most notably, the matrix in Eq. (26) realizes the row reduction of Eq. (22); see Appendix A for a proof. It corresponds to a geometrically-informed choice for the matrix of the row-reduction.

Like for the graph, we can use the matrix $\mathbb{G}^\top$ to invert the relation $\boldsymbol{F} = -\mathbb{S}^\top \boldsymbol{V} = \sum_\gamma F^c_\gamma \boldsymbol{c}^\gamma$, for any conservative force $\boldsymbol{F} \in \text{Im} \, \mathbb{S}^\top$. Physically it means that the matrix $\mathbb{G}^\top$ allows one to integrate any conservative force $\boldsymbol{F}$ on the hypergraph in order to find the corresponding potential landscape $\boldsymbol{V}$. In particular, denoting $\mathbb{G}^\top_M$ the $N \times M$ right-block obtained by excluding the conservation laws from $\mathbb{G}^\top$ [see Eq. (26)] one finds:

$$\boldsymbol{V} = \mathbb{G}^\top_M \begin{pmatrix} F^c_\gamma \end{pmatrix} . \qquad (27)$$

Notice that first $N - M$ rows of $\mathbb{G}^\top_M$ are padded with 0's, which corresponds to fix the potential $V_{\text{root}} = 0 \; \forall$ root.

What we discussed so far holds generally for any conservative force defined on the network. In Sec. IV we will make contact between the integration procedure, that we just detailed, and the dynamics and thermodynamics of mass-action systems: if $\boldsymbol{F}$ is the chemical affinity vector whose components are defined in Eq. (6), and the system is closed, the integration procedure in Eq. (27) yields the chemical potential of thermodynamics $\boldsymbol{V} = \boldsymbol{\mu}$. Similarly, if $F_\gamma^c = \log\left(k_\gamma^+/k_\gamma^-\right)$, the integration in Eq. (27) leads to the standard chemical potential of thermodynamics $\boldsymbol{V} = \boldsymbol{\mu}^\ominus$. (Notice that such potentials are defined up to linear combinations of the conservation laws which can always be added to $\boldsymbol{V}$ while leaving $\boldsymbol{F}$ unchanged.)

### 3. Cocycles

For simple graphs $\mathcal{G}$, we have identified each cocycle $\boldsymbol{c}^\gamma$ with a binary splitting of $\mathcal{G}$ into a source and a target island. This is no longer true for the hypergraph which is not necessarily split in two disconnected islands when a cocycle is removed (see for instance Fig. 9). We thus ask what is the geometrical interpretation of cocycles for the hypergraph.

We have seen how the matrix $\mathbb{G}_M^\top$ transforms a set of conservative forces defined on the cochords into a set of potentials defined on the nodes (with zero potential on the chosen roots). Also, this matrix directly relates to the family of cocycles since, from Eq. (22), we have $-(\mathbb{S}^\top \mathbb{G}_M^\top)_\gamma = \boldsymbol{c}^\gamma$. We thus define the characteristic potential $\boldsymbol{v}^\gamma = (\mathbb{G}_M^\top)_\gamma$, $\forall \gamma$, such that $\boldsymbol{c}^\gamma = -\mathbb{S}^\top \boldsymbol{v}^\gamma$: each potential $\boldsymbol{v}^\gamma$, differentiated with $\mathbb{S}^\top$, generates the cocycle $\boldsymbol{c}^\gamma$. By analogy with graphs, we define the source island of $\boldsymbol{c}^\gamma$ from the set of nodes $i$ such that $v_i^\gamma \neq 0$. Then the cocycle $\boldsymbol{c}^\gamma$ is the boundary of the island $\boldsymbol{v}^\gamma$, i.e. the set of reactions that connect this island to the rest of the nodes. This is useful in metabolic reconstruction (see Sec. V B) because identifying the internal and external reactions involved in a cocycle yields exact relations between their corresponding currents. Also, because the entries of $\boldsymbol{v}^\gamma$ are no longer simply 0's or 1's, its associated island on the hypergraph now has a geography. Namely, each node $i$ is given an altitude $v_i^\gamma$ which quantifies the impact of $i$ onto the outward flux along the cocycle: as shown in Appendix A, $\boldsymbol{v}^\gamma$ is the potential landscape that ensures the cochord $\gamma$ to present a unit current.

### 4. Coarse-graining of the dynamics based on cocycles

Finally, let us connect the geometrical pictures of the islands, identified by the columns of $\mathbb{G}_M^\top$, with the dynamics. Applying the $\mathbb{G}$ matrix to Eq. (5) one finds that each island concentration $z_\gamma(t) = (\mathbb{G}_M \boldsymbol{x}(t))_\gamma$ evolves by the corresponding cocycle flow, such that:

$$\frac{d}{dt} z_\gamma(t) = \boldsymbol{J} \cdot \boldsymbol{c}^\gamma \quad \forall \gamma . \tag{28}$$

Equation (28) can be seen as an integrated continuity equation, where $z_\gamma(t)$ is the sum of the concentrations of the nodes of the island, weighted by the components of $\boldsymbol{v}^\gamma$ (which can be seen as the elevation map of the island) and $\boldsymbol{c}^\gamma \cdot \boldsymbol{J}$ is the total current flowing across its boundary. Islands thus constitute a geometrical coarse-graining of the $N$ species into $M$ independent and macroscopic degrees of freedom, which describe the relaxation of the system to its steady state.

Assume now that one is able to cancel the current $\boldsymbol{c}^\gamma \cdot \boldsymbol{J}$, i.e. effectively 'remove' the cocycle $\boldsymbol{c}^\gamma$ from the hypergraph. Then from Eq. (28) one sees that the (weighted) concentration of the island is conserved $z_\gamma = $ const. Accordingly, the vector $\boldsymbol{v}^\gamma$ can be seen as a new conservation law which emerges when removing $\boldsymbol{c}^\gamma$. While this is expected for the graph, for which the removal of a cocycle always generates a new disconnected component, it is nontrivial for the hypergraph. In fact, at the level of the full hypergraph (see Fig. 9), source and target island are still connected after the removal of the cocycle and no new component necessarily arises. The physical interpretation is the following: exchange between the source and the target islands is still possible after the removal of the cocycle and $\boldsymbol{v}^\gamma$ is an emergent conservation laws but not necessarily a mass-conservation law (i.e. its entries can be negative), in contrast to the case of the graph.

We end this section by putting forward a possible application of this formalism to control, in the chemical setting. In chemistry, molecular inhibitors are often employed to delay, slow or prevent chemical reactions. Like in an inverse catalysis, the inhibitor acts by suppressing the reaction rate constants $k_\rho^\pm \to 0$ of a target reaction $\rho$, thus introducing a slow timescale at the kinetic level. In complex CRNs it is not clear *a priori* what is the effect of suppressing a reaction on (*i*) the macroscopic relaxation timescale and (*ii*) the steady-state concentrations reached in the long-time limit. It depends on a number of factors including the initial condition, the distribution of reaction rate constants and the topology of the network. Nevertheless, some insights follow directly from our algebraic approach. In particular, we have identified the cocycles as the relaxation modes of the dynamics, whose removal leads to the emergence of new conservation laws (left zero modes of $\mathbb{S}$). As such, we expect them to have a strong impact on the timescales of relaxation. In Fig. 9 we show the finite time relaxation of the island concentrations $z_\gamma$'s [Eq. (28)] for the example in Fig. 6 when a cocycle is suppressed, compared to the case when a non-cocycle reaction is suppressed. As anticipated, the behavior strongly differs. The removal of a reaction that is not a cocycle affects minimally the finite-time dynamics, leaving unchanged the characteristic relaxing time and the equilibrium steady state. On the contrary, upon decreasing the reaction rates of a cocycle, the dynamics develops a *plateau* which corresponds to a new timescale controlled by the inhibitor. In the limit of complete suppression of the cocycle the system relaxes to a new equilibrium state, which is a sign of the emergence of a new conservation law.

## IV. THE PHYSICS OF CHEMICAL CURRENTS AND AFFINITIES

In this section, we make contact between the geometrical framework developed so far and the physical and thermody-namic description of CRNs. First, we use cycles and cocycles to represent currents $J$ and affinities $A$ for arbitrarily CRNs. Those are the vectors defined in Eqs. (4)-(6) and that control the dynamics and thermodynamics of the system. In particular, we put forward a decomposition for $A$ and $J$ respectively into conservative and non-conservative forces and into transient and steady-state currents. Then, building on such decompositions, we show how (*i*) for closed systems the (conservative) chemical affinity relates to the chemical potentials of thermodynamics; (*ii*) the potential condition for the affinity breaks down for chemostatted systems; (*iii*) the *a priori* deterministic notion of chemical affinity relates to the entropy production rate of stochastic thermodynamics.

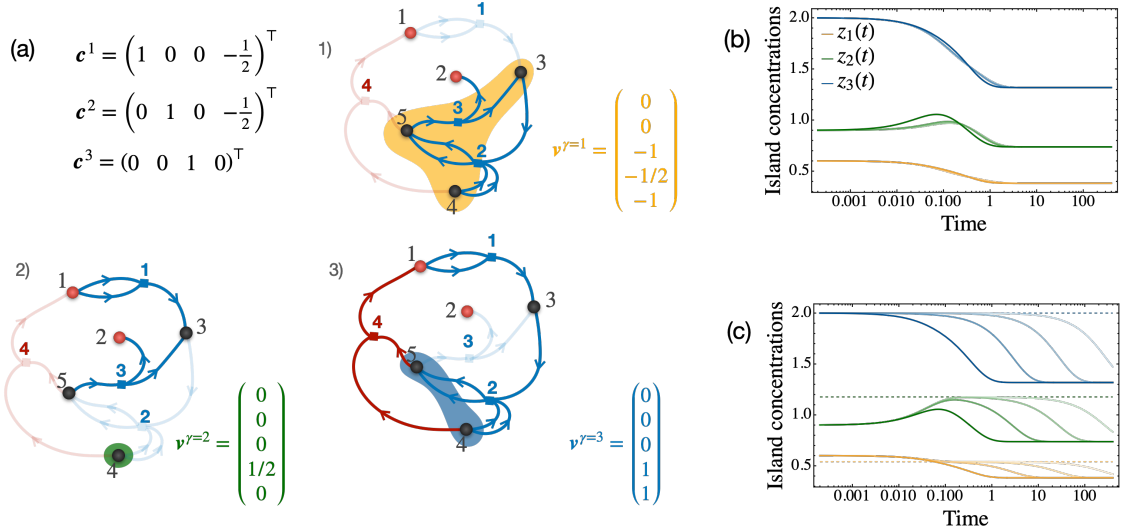

FIG. 9. **(a)** We report the cocycles $c^\gamma$'s, with $1 \leq \gamma \leq 3$, for the example of interacting CRN introduced in Fig. 6. Removing a cocycle results in the emergence of a new conservation law $v^\gamma$ which can be read from the columns of $\mathbb{G}_M^\top$. We interpret the nodes belonging to the new conservation law as the source island of the corresponding cocycle, where each node has a certain weight (altitude). For instance, by removing the cocycle $c^1$, one identifies from $v^{\gamma=1} = (\mathbb{G}_M^\top)_{\gamma=1}$ a source island (orange) containing nodes 3, 4 and 5 with different (negative) weights. The conservation laws are reported for every cocycle, $c^1$ to $c^3$. We stress that, in contrast to the case of simple graphs, the source islands are generally not disconnected from the roots, as in the case of the orange and blue islands. Let us focus on cocycle $c^3$, which consists of the sole reaction 3. We compare the dynamical relaxation of the island concentrations $z_\gamma(t) = v^\gamma \cdot x(t)$ upon suppressing (to various degrees) reaction 2 **(b)** and reaction 3 **(c)**. Eq. (5) is solved numerically with initial condition $x_i(0) = 1 \, \forall i$. The rates are chosen in accordance to Wegscheider criterion (37) with $k_\rho^\pm \sim O(1) \, \forall \rho$. In **(c)** the rates of reaction 3 are suppressed by a factor $\varepsilon = 10^{-1}, 10^{-2}, 10^{-3}$ (different shadows) such that $k_{\rho=3}^\pm \mapsto \varepsilon k_{\rho=3}^\pm$ and $k_{\rho=3}^+/k_{\rho=3}^- = $ const. The dashed lines correspond to the equilibrium steady-state value obtained for $\varepsilon = 0$, i.e. full suppression of reaction 3. As anticipated, in this case $z_3$ becomes a constant of motion.

### A. Geometrical decomposition of currents and affinities

Generalizing the graph approach of Ref. [17], we introduce a decomposition of the affinity vector $A \in \mathbf{R}^R$ in terms of cocycles and chords as:

$$A = \sum_\gamma A_\gamma^c c^\gamma + \sum_\alpha A_\alpha^e e^\alpha. \qquad (29)$$

Notice that this decomposition is not orthogonal since $c^\gamma \cdot e^\alpha \neq 0$. Nevertheless, it bears a clear physical interpretation. We recognize in the first term of Eq. (29) the conservative part of the affinity in accordance to Eq. (18), and now prove that the second term contains the non-conservative part. In continuous space, a test for conservativeness is Stokes' theorem, which states that the circulation – the line integral along any closed path – of a conservative force field vanishes. In our setting, any closed path in the space of reactions can be expressed as a linear combination of the basis of cycles, as it belongs to Ker $\mathbb{S}$ [5]. Thus to compute the circulation of $A$ it is sufficient to compute its scalar product with the cycles $c^\alpha$'s. Using the orthogonality conditions Eqs. (14) and (17), one gets:

$$c^\alpha \cdot A = A_\alpha^e \quad \forall \alpha. \qquad (30)$$

Thus, the coefficient $A_\alpha^e$ in Eq. (29) results from integrating the affinity vector along cycle $c^\alpha$, and quantifies the deviation from Stokes' theorem. Using Eq. (6) in Eq. (30) one sees that the coefficients $\{A_\alpha^e\}$ do not depend on the system concentrations [11]: they constitute a set of parameters that are intrinsic to the dynamics and quantify the nonequilibrium drive. Indeed, we can express the conservative condition for $A$ as the requirement for all coefficients $A_\alpha^e$ to vanish:

$$c^\alpha \cdot A = A_\alpha^e = 0 \quad \forall \alpha, \qquad (31)$$

which for graphs is Kirchhoff Voltage Law (KVL) [11] and we will keep the same name for generic CRNs. Whenever Eq. (31) is fulfilled, the full affinity vector reduces to the conservative part [Eq. (18)] and can be integrated using the pro-cedure described in Secs. II C and III B 2. Importantly, having a conservative affinity is equivalent to having stochastic reversibility of the underlying dynamics at the level of populations (proofs are given in Appendix B). Hence, although the condition to have a conservative affinity seems to only pertain to the deterministic level, it applies also to the stochastic one, relating to the notion of detailed balance in stochastic population dynamics (see also the discussion is Sec. IV D). Such characterizations of reversibility are analogous to that of a Langevin equation of the form $\partial_t \mathbf{x}(t) = \mathbf{F}(\mathbf{x}(t)) + \eta(t)$ (with $\mathbf{F}(\mathbf{x})$ the force and $\eta(t)$ a centered Gaussian white noise). Indeed, there the process $\mathbf{x}(t)$ is stochastically reversible if and only if the force derives from a potential, if and only if its circulation along any loop is zero. All in all, Eq. (29) can be viewed as a generalization of the Helmholtz–Hodge decomposition of the affinities on a hypergraph.

---

[5] For simplicity consider the case of a closed path on a graph. It may pass through bridges, but since any bridge must be crossed an even number of times in opposite directions along such path, the bridges do not contribute.

Let us now introduce the complementary decomposition for the vector of currents $\boldsymbol{J} \in \mathbf{R}^R$ in terms of cochords and cycles:

$$\boldsymbol{J} = \sum_\gamma J_\gamma^e \boldsymbol{e}^\gamma + \sum_\alpha J_\alpha^c \boldsymbol{c}^\alpha. \tag{32}$$

Once again, one can identify an analog of it for vector calculus in continuous space. In the same way as a closed surface splits $\mathbf{R}^3$ into an inner and an outer region, every cocycle $\boldsymbol{c}^\gamma$ splits the network (either a graph or a hypergraph) into a source and target islands (see Figs. 4 and 9). By construction, any flux between the source and the target must flow through the cocycle itself. If we take the scalar product between Eq. (32) and a cocycle $\boldsymbol{c}^\gamma$ one gets:

$$\boldsymbol{c}^\gamma \cdot \boldsymbol{J} = J_\gamma^e \quad \forall \gamma. \tag{33}$$

Thus, the coefficient $J_\gamma^e$ in Eq. (32) represents the total current flowing from the source to the target along the corresponding cocycle $\boldsymbol{c}^\gamma$. Eq. (33) corresponds to a surface integral of the current, i.e. a 'flux' across the 'boundary' of the source island. Using the definition of $\boldsymbol{v}^\gamma$, one has

$$\boldsymbol{c}^\gamma \cdot \boldsymbol{J} = (-\mathbb{S}^\top \boldsymbol{v}^\gamma) \cdot \boldsymbol{J} = -\boldsymbol{v}^\gamma \cdot (\mathbb{S} \boldsymbol{J}), \tag{34}$$

where on the r.h.s. we recognize the divergence $\mathbb{S} \boldsymbol{J}$ entering in Eq. (5). Hence, Eq. (34) can be seen as a divergence theorem for CRNs: the total current outward from the source island equates the volume integral (weighted by $\boldsymbol{v}^\gamma$) over the source island of the divergence of the current. Finally, the coefficient $J_\alpha^e = \boldsymbol{e}^\alpha \cdot \boldsymbol{J}$ is the local current flowing along chord $\alpha$.

Assuming that in the long-time limit the dynamics reaches a stationary state, $\boldsymbol{x}^* = \lim_{t \to \infty} \boldsymbol{x}(t)$, then the stationary current vector $\boldsymbol{J}^* = J(\boldsymbol{x}^*)$ belongs to the kernel of $\mathbb{S}$ [see Eq. (5)]:

$$\mathbb{S} \boldsymbol{J}^* = 0. \tag{35}$$

For graphs, Eq. (35) corresponds to the Kirchhoff Current Law (KCL) [11] and we keep the same name for generic CRNs. It ensures the balance between all currents entering and exiting at each node. Using Eq. (32), it expresses that $\boldsymbol{J}^*$ reduces to a linear combination of cycles: $\boldsymbol{J}^* = \sum_\alpha J_\alpha^{c,*} \boldsymbol{c}^\alpha$. It follows that the currents $J_\gamma^e$ flowing from the source to the target of each cocycle are transient and they vanish at steady state.

Notice that the cocycle/chord and cochord/cycle bases used in the decompositions of Eq. (29) and (32) are different than the bases of Eq. (16). We stress that the advantage of these two decompositions lies in their direct physical interpretation. Indeed the conservative condition is expressed in the vanishing of the cycle affinities at all times, Eq. (31) (this requirement is sometimes called thermodynamic feasibility). Also, the condition of stationarity is expressed in the vanishing (in the long-time limit) of the transient currents, Eq. (33). Table I in the Outlook summarizes these results.

We conclude the section by recalling the definition of entropy production rate $\sigma$ in terms of currents and affinities [50]:

$$\sigma = \boldsymbol{J} \cdot \boldsymbol{A} = \sum_\gamma J_\gamma^e A_\gamma^c + \sum_\alpha J_\alpha^c A_\alpha^e, \tag{36}$$

where in the second equality used Eqs. (29) and (32). Notice that the first contribution vanishes at steady state while the second contribution vanishes for reversible dynamics.

### B. Conservative affinities and thermodynamic potential(s)

For a closed system, i.e. in the absence of couplings with external reservoirs [51], we expect the concentrations $\boldsymbol{x}(t)$ governed by Eq. (5) to relax to an equilibrium state $\boldsymbol{x}^{\text{eq}} = \lim_{t \to \infty} \boldsymbol{x}(t)$, fixed by the initial conditions. In this scenario, the internal currents are driven by nonequilibrium initial conditions and are expected to vanish at steady state, $\boldsymbol{J}(\boldsymbol{x}^{\text{eq}}) = 0$. This is guaranteed by a choice of reaction rates in accordance with the Wegscheider criterion [52], which states that the product of the forward rates along any cycle $\boldsymbol{c}^\alpha$ is equal to that of the backward rates:

$$\prod_{\rho=1}^R \left(\frac{k_\rho^+}{k_\rho^-}\right)^{c_\rho^\alpha} = 1, \quad \forall \alpha. \tag{37}$$

Equation (37) is a necessary and sufficient condition for the dynamics in Eq. (5) to relax to an equilibrium steady state with $\boldsymbol{J}(\boldsymbol{x}^{\text{eq}}) = 0$ ($\Leftrightarrow \boldsymbol{A}(\boldsymbol{x}^{\text{eq}}) = 0$). As proved in Appendix B, it is equivalent to having stochastic reversibility for the underlying population dynamics, or equivalently to the existence of a vector $\boldsymbol{\mu}^\ominus$ such that

$$\frac{k_\rho^+}{k_\rho^-} = \exp\left[-(\mathbb{S}^\top \boldsymbol{\mu}^\ominus)_\rho\right]. \tag{38}$$

One recognizes $\boldsymbol{\mu}^\ominus$ to play the role of the (dimensionless) standard chemical potential of equilibrium thermodynamics [51]. In fact, if one pictures the chemical reaction $\rho$ as a transition between molecular conformations in the (standard) landscape of possible chemical combinations of atoms, then the r.h.s. of Eq. (38) is the ratio of Kramers transition rates in such landscape.

For noninteracting CRNs, Eq. (38) is equivalent to a local detailed balance condition [7, 10]:

$$\frac{k_\rho^+}{k_\rho^-} = \exp\left(\mu_{s(\rho)}^\ominus - \mu_{t(\rho)}^\ominus\right), \tag{39}$$

as $\mathbb{S}$ is an incidence matrix. Notice that Eq. (39) is also the standard condition of detailed balance for Markov chains with respect to a configuration probability $\mathcal{P}_i \propto \exp(-\mu_i^\ominus)$. Reading from Eq. (38) that the forces $\log(k_\rho^+/k_\rho^-)$ derive from the potential $\mu^\ominus$, we can use the integration procedure of Sec. II C: fixing a root on the graph, we have

$$\exp(\mu_i^\ominus) = \prod_{\rho \in [\text{root} \to i]} \frac{k_\rho^+}{k_\rho^-}, \quad (40)$$

where the product is taken along any arbitrary path on the graph $\mathcal{G}$ from the root to node $i$. The result does not depend on the choice of path, thanks to Eq. (37). The choice of root changes $\mu^\ominus$ by a global constant without affecting Eq. (39).

For interacting CRNs, the generalization of this integration procedure is described in Sec. III B 2: Eq. (38) tells that the force of components $F_\rho = \log(k_\rho^+/k_\rho^-)$ derives from the potential $\mu^\ominus$; hence, one fixes a set of roots from the conservation laws[6], and $\mu_i^\ominus$ is obtained by a weighted summation of the $F_\rho$'s along the multi-path that connects the roots to node $i$ (see Fig. 8 for an example). Here, different choices of roots will lead to expressions of $\mu^\ominus$ that differ by a linear combination of conservation laws [which does not affect Eq. (38)].

The correspondence with equilibrium thermodynamics also includes the chemical potential $\mu_i(t)$ of species $i$ defined as $\mu_i(t) = \mu_i^\ominus + \log x_i(t)$. Using Eqs. (6) and (38), this yields an expression for the vector of affinities in a closed system (valid at all times):

$$\boldsymbol{A} = -\mathbb{S}^\top \boldsymbol{\mu}. \quad (41)$$

Picturing $\boldsymbol{A}(\boldsymbol{x})$ as the chemical force [see Eq. (7)], one sees that, in closed (equilibrium) systems, it derives from a potential which is precisely the chemical potential $\boldsymbol{\mu}$. We prove in Appendix B that the converse is true. Worthily, $\boldsymbol{\mu}$ can be reconstructed from $\boldsymbol{A}$ using the integration procedure described in Sec. III B 2.

Finally, via Eq. (41) one sees that, at the level of the decomposition Eq. (29), Wegscheider condition ensures $A_\alpha^e = 0$ at all times. Hence, for closed systems (i.e. reversible dynamics) only the cocycle contribution in Eq. (29) survive at finite time, and vanish in the long-time limit $A_\gamma^c(t \to \infty) = 0$; this describes the process of relaxation to equilibrium. Algebraically this implies from Eq. (41) that the equilibrium chemical potential $\boldsymbol{\mu}^{\text{eq}} = \lim_{t \to \infty} \boldsymbol{\mu}(t)$ is a left nullvector of the stoichiometric matrix, $\boldsymbol{\mu}^{\text{eq}} \cdot \mathbb{S} = 0$, i.e. it is a linear combination of conservation laws (see also Ref. [55] for insights on the role of conservation laws). In particular, for closed noninteracting CRNs, $\boldsymbol{\mu}^{\text{eq}}$ is proportional to the mass conservation law $\ell_0$ and equilibrium is reached when species all have the same chemical potential.

---

[6] For generic CRNs, we assume there is at least one mass-like conservation law. Interestingly, this is related to Gordan's theorem (see e.g. [29, 53]), which states that the two following conditions are mutually exclusive: (i) There exists a left nullvector $\boldsymbol{\ell}$ of $\mathbb{S}$, with all components positive ($\ell_i \geq 0$). Such vector $\boldsymbol{\ell}$ is a mass-like conservation law. (ii) There exists a vector $\boldsymbol{j}$ such that $\mathbb{S}\boldsymbol{j} > 0$. We exclude case (ii) because, from the rate equation (5), it implies that the set of reactions can create matter, which cannot occur for a closed system – this would contradict Lavoisier's principle [54]. (Notice that chemostatted autocatalytic reactions fall in case (ii) [29].)

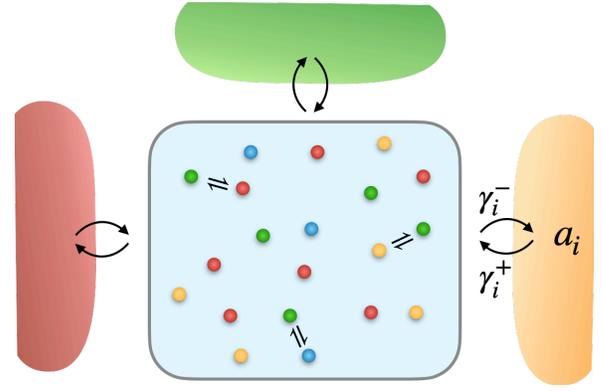

FIG. 10. Sketch of a system of chemical species put in contact with external chemostats. The chemostats are treated as infinite reservoirs of a single chemical species. For illustration, the chemostat $i$ (here in orange) exchanges orange particles with the system and it is characterized by a pair of reaction rates $\gamma_i^\pm$ and a driving parameter $a_i$.

### C. How chemostatting breaks conservative conditions

Suppose now that the reaction rate constants $k_\rho^\pm$ do not fulfill the Wegscheider condition in Eq. (37). This is often the case in phenomenological models of evolutionary games [56], gene regulatory networks [57] or theoretical ecology [58] where effective reactions are typically irreversible. As a result, the dynamics evolves towards a nonequilibrium steady state or a limit cycle or a more complex behavior. To make contact with thermodynamics while still being irreversible, in CRNs, the breakdown of the Wegscheider condition is usually prescribed through the coupling with different chemostats which drive the system out of equilibrium [59]. Each chemostat is depicted as a reservoir of a single chemical species which is put in contact with the system and exchanges molecules (Fig. 10). Conventionally, the corresponding reaction reads:

$$\emptyset \underset{\gamma_i^-}{\overset{\gamma_i^+}{\rightleftharpoons}} Y_i, \quad (42)$$

where $Y_i$ is the chemostatted species and $\gamma_i^\pm$ are the rate of particles exchange with the chemostat, which, contrarily to the bulk rates $\{k_\rho^\pm\}$, are not thermodynamically constrained by the Wegscheider condition. We parametrize the effect of chemostat $i$ via a driving parameter $a_i$ defined from

$$\frac{\gamma_i^-}{\gamma_i^+} = \exp(\mu_i^\ominus - a_i). \quad (43)$$

Following Refs. [8, 10, 60], we label the set of chemostatted species with $Y$ and the remaining species with $X$, such that $X \cup Y$ forms the set of all chemical species. Accordingly, we group internal and external reactions into an $N \times (R + |Y|)$ extended stoichiometric matrix $\mathbb{S}_{\text{res}}$, namely:

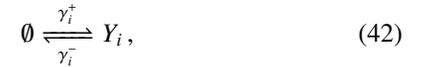

$$\mathbb{S}_{\text{res}} = (\mathbb{S}_Y \mid \mathbb{S}) \quad \text{with} \quad \mathbb{S}_Y = \begin{pmatrix} \mathbb{1}_{|Y|} \\ \mathbb{0} \end{pmatrix}, \quad (44)$$

where the first $|Y|$ columns of $\mathbb{S}_{\text{res}}$ represent the chemostatting reactions [Eq. (42)]. The evolution of the corresponding concentrations $\boldsymbol{y}(t)$ and $\boldsymbol{x}(t)$ reads $\frac{d}{dt}(\boldsymbol{y}(t), \boldsymbol{x}(t)) =$ $\mathbb{S}_{\text{res}}\boldsymbol{J}(\boldsymbol{x}(t),\boldsymbol{y}(t))$, the first $|Y|$ component of $\boldsymbol{J}$ being the external currents of the chemostatting reactions:

$$J_i = \gamma_i^+ - \gamma_i^- y_i = \gamma_i^+\left[1 - \exp\left(\mu_i - a_i\right)\right] \quad (\text{for } 1 \le i \le |Y|) \quad (45)$$

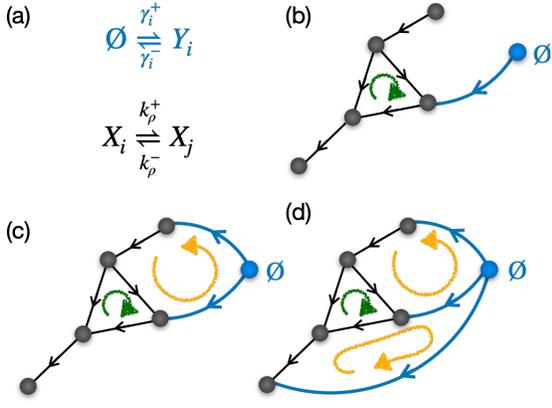

FIG. 11. Example of the graphical representation of chemostatting for unimolecular reactions. **(a)** Species labeled $Y_i$ are subjected to chemostatting reactions (in blue) with rates $\gamma_i^\pm$, describing the contact with reservoirs. **(b)** Chemostatting one species corresponds to the simple addition of one edge in the graph which connects the node $\emptyset$ to the corresponding chemostatted species/node. In this case, the only cycle present in the graph is the internal one (in green). **(c)** Chemostatting a second species results in the emergence of a new cycle, depicted in yellow. **(d)** The same holds true when chemostatting a third species, and so on. Such emergent cycles (in yellow) are not constrained by the Wegscheider condition on the rates and they can drive the system out of equilibrium.

This means that, for open systems, the full affinity vector $\boldsymbol{A}$ takes the following form

$$\boldsymbol{A} = -\mathbb{S}_{\text{res}}^\top \boldsymbol{\mu} + \boldsymbol{a}, \quad (46)$$

where $\boldsymbol{a} \in \mathbf{R}^R$ is the vector of driving affinities, $a_\rho = a_i\,\delta_{\rho,i}$ (for $1 \le i \le |Y|$). Eq. (46) is valid arbitrarily far from equilibrium and expresses how the chemostatting in general leads to irreversibility, since when $\boldsymbol{a} \notin \text{Im}\,\mathbb{S}^\top$ the potential condition Eq. (31) breaks down.

In practice, in order to drive nonequilibrium, one needs at least two chemostats held at different values of the driving parameter $a$, which is why often in the literature one encounters '$\Delta\mu$' as the effective driving force [61, 62]. *Proof.* For a single chemostat, the driving affinity vector takes the form $\boldsymbol{a} = (a, 0\ldots 0)^\top$. By construction, $\boldsymbol{\ell}_0\cdot\mathbb{S}_{\text{res}} = (1, 0\ldots 0)^\top$ which constitutes a generating vector for the space where $\boldsymbol{a}$ lives. It follows that $\boldsymbol{a} = a\,\boldsymbol{\ell}_0 \cdot \mathbb{S}_{\text{res}}$ and the affinity becomes:

$$\boldsymbol{A} = -\mathbb{S}_{\text{res}}^\top\left[\boldsymbol{\mu}(t) - a\,\boldsymbol{\ell}_0\right]. \quad (47)$$

The total affinity vector $\boldsymbol{A} \in \text{Im}\,\mathbb{S}_{\text{res}}^\top$ is therefore still conservative, the dynamics still reversible, and the equilibrium state is given by $\boldsymbol{\mu}^{\text{eq}} = a\,\boldsymbol{\ell}_0$. □

For noninteracting CRNs, the emergence of nonequilibrium behavior can be understood graphically. To do so, we augment the original graph by an extra node $\emptyset$ and, for each chemostatting reaction in Eq. (42), we introduce a new edge linking the chemostatted species $Y_i$ to the node $\emptyset$ (see Fig. 11). By construction, the incidence matrix of the augmented graph is endowed with an extra line:

$$\left(\begin{array}{c|c} -1 \cdots -1 & 0 \ldots 0 \\ \hline \mathbb{S}_{\text{res}} \end{array}\right) \quad (48)$$

which bookkeeps the exchange of particles between the chemostatting node $\emptyset$ and the system. As shown in Fig. 11d, adding more than one chemostatting edges results in the appearance of new cycles in the graph. Those 'emergent' cycles are associated to cycle affinities $A_\alpha^e$ that do not necessarily satisfy the Wegscheider condition, and thus play the role of driving the system out of equilibrium. Again, one sees that at least two chemostats are needed for an emergent cycle to appear. Following the procedure outlined in Sec. II A one finds the corresponding basis for cocycles and cycles. Those now include both the internal and the emergent cycles, which are graphically identified. One verifies that the number of emergent cycles is given by[7]:

$$\#\,\text{emergent cycles} = \dim\text{Ker}\,\mathbb{S}_{\text{res}} - \dim\text{Ker}\,\mathbb{S} \quad (49)$$

in accordance with the algebraic result of Ref. [10] (for generic interacting CRNs).

Other chemostatting procedures [8, 10, 11] consist in fixing the concentrations of the chemostatted species (treated as external parameters), but then $\mathbb{S}$ is no longer an incidence matrix and one loses the graph-theory approach. In contrast, the present setting preserves the geometrical insights introduced in Sec. II A-II C when going from closed to open systems.

### D. Connection with the entropy production rate

The thermodynamic equilibrium conditions discussed in Sec. IV B are equivalent to requiring the reversibility of the underlying stochastic process (as described in Appendix B). Considering the dynamics of the vector of the population $\boldsymbol{n} = (n_1, ..., n_N)$ of the species, it means that every transition associated to a reaction $\rho$ in Eq. (2) verifies the detailed balance condition

$$\frac{W_\rho^+(\boldsymbol{n})}{W_\rho^-(\boldsymbol{n}+\mathbb{S}_\rho)} = \frac{\mathcal{P}^{\text{eq}}(\boldsymbol{n}+\mathbb{S}_\rho)}{\mathcal{P}^{\text{eq}}(\boldsymbol{n})} \quad \forall \boldsymbol{n}. \quad (50)$$

Here $W_\rho^\pm(\boldsymbol{n}) = W(\{n_i \mapsto n_i \pm \mathbb{S}_{i\rho}\})$ denotes the transition rate at population level, and the equilibrium distribution $\mathcal{P}^{\text{eq}}(\boldsymbol{n})$ is a product-form Poisson-like law (constrained by the conservation laws) of parameters $\boldsymbol{x}^{\text{eq}}$. We refer to Appendix B for an explicit expression of $\mathcal{P}^{\text{eq}}(\boldsymbol{n})$ and to Appendix C for the transition rates. The vector $\boldsymbol{x}^{\text{eq}}$ represents the average value of the species concentrations in the long-time limit and depends on the initial condition.

---

[7] By construction, any right nullvector of Eq. (48) is also a right nullvector of $\mathbb{S}_{\text{res}}$. Furthermore, the extra line in the matrix of Eq. (48) is not linearly independent from the rows of $\mathbb{S}_{\text{res}}$ and it can be obtained by taking the negative sum of all the rows of $\mathbb{S}_{\text{res}}$, which is full-row rank as recalled previously. It follows that the kernel of the two matrices must be the same

Introducing the quasipotential $\Phi(\boldsymbol{n}) = -\frac{1}{\Omega} \log \mathcal{P}^{\text{eq}}(\boldsymbol{n})$ associated to the equilibrium law, where $\Omega$ is the system's volume, Eq. (50) rewrites

$$\frac{W_\rho^+(\boldsymbol{n})}{W_\rho^-(\boldsymbol{n} + \mathbb{S}_\rho)} = \exp\left\{-\Omega\left[\Phi(\boldsymbol{n} + \mathbb{S}_\rho) - \Phi(\boldsymbol{n})\right]\right\}. \quad (51)$$

At large volume $\Omega \gg 1$ with fixed $\boldsymbol{x} = \boldsymbol{n}/\Omega$, one finds that $\Phi(\boldsymbol{n}) \to \phi(\boldsymbol{x})$ with

$$\phi(\boldsymbol{x}) = \sum_i \left(x_i \log x_i - x_i - x_i \log x_i^{\text{eq}} - x_i\right). \quad (52)$$

Here one recognizes that $\phi(\boldsymbol{x})$ is the free energy density, that is, the difference between the energy density and the entropy density. Notice that $\phi(\boldsymbol{x})$ is minimum (and cancels) at $\boldsymbol{x} = \boldsymbol{x}^{\text{eq}}$. Defining (for any $\boldsymbol{n}$ and $\boldsymbol{x} = \boldsymbol{n}/\Omega$) $x_i = \exp(\mu_i - \mu_i^{\ominus})$, one can expand Eq. (51) for $\Omega \gg 1$, which yields the expression of the entropy production $\Sigma_\rho$ of reaction $\rho$:

$$\Sigma_\rho(\boldsymbol{x}) = \log \frac{W_\rho^+(\boldsymbol{n})}{W_\rho^-(\boldsymbol{n} + \mathbb{S}_\rho)} = -(\mathbb{S}^\top \boldsymbol{\nabla}\phi(\boldsymbol{x}))_\rho \quad (53)$$

$$(\boldsymbol{\nabla}\phi(\boldsymbol{x}))_i = \log x_i - \mu_i^{\text{eq}} + \mu_i^{\ominus} = \mu_i - \mu_i^{\text{eq}}. \quad (54)$$

We kept the usual denomination of entropy production for this ratio of rates at population level, but we remark that it plays the same role as the affinity at the deterministic level of reaction rate constants [see Eq. (6)]. We observe that the constant $\mu_i^{\text{eq}}$ in Eq. (54) plays no role in Eq. (53) since $\boldsymbol{\mu}^{\text{eq}} \in \text{Ker } \mathbb{S}^\top$ [8]. We can thus write for the vector $\boldsymbol{\Sigma}$ of the $\Sigma_\rho$'s:

$$\boldsymbol{\Sigma}(\boldsymbol{x}) = -\mathbb{S}^\top \boldsymbol{\mu}(\boldsymbol{x}). \quad (55)$$

Here $\boldsymbol{\Sigma}(\boldsymbol{x})$ comes from the stochastic dynamics [the l.h.s. of Eq. (50)] while the associated chemical potential $\boldsymbol{\mu}(\boldsymbol{x})$ comes from the equilibrium distribution [the r.h.s. of Eq. (50)], and is thus of thermodynamic nature. We stress that Eq. (55) can be read for any occupation state of the system. If evaluated for $\boldsymbol{x}(t)$ solution of the rate equation (5), we see that $\boldsymbol{\Sigma}(\boldsymbol{x}(t))$ becomes equal to the affinity $\boldsymbol{A}(\boldsymbol{x}(t))$ and goes to zero at large time, as expected in equilibrium.

Last, let us consider the case of an open system in contact with reservoirs. Following the chemostatting procedure of Sec. IV C, one defines the vector of the entropy production rates of individual reactions as in Eq. (53), including now chemostatting reactions. Accordingly, one finds at large $\Omega$

$$\boldsymbol{\Sigma}(\boldsymbol{x}) = -\mathbb{S}_{\text{res}}^\top \boldsymbol{\mu}(\boldsymbol{x}) + \boldsymbol{a}, \quad (56)$$

where the computation is done directly from the expression of the ratio of transition rates. Interestingly, Eq. (56) expresses the condition of local detailed balance [63, 64] (see Ref. [65] for a review and Refs. [55, 66] for the case of CRNs), with $\boldsymbol{a}$ playing the role of a chemical drive. As before for Eq. (55), this equation holds for any occupation $\boldsymbol{n}$ (at large $\Omega$) through $\boldsymbol{x} = \boldsymbol{n}/\Omega$, and not just for the deterministic concentration $\boldsymbol{x}(t)$; it makes the link between stochastic aspects (on the l.h.s.) and thermodynamic quantities (the r.h.s. being expressed as a function of chemical potential and drive). If evaluated for the deterministic $\boldsymbol{x}(t)$, Eq. (56) takes exactly the form of the affinity $\boldsymbol{A}$ [see Eq. (46)]. We refer the reader to Refs. [66–68] for relations to the second law.

## V. CLOSE-TO-EQUILIBRIUM REGIME

### A. Linear responses for CRNs

We now apply our framework to study the response of interacting networks to small perturbations out of equilibrium stationary states. As discussed in Sec. IV B, the Wegscheider condition for closed CRNs ensures the existence of an equilibrium steady state $\boldsymbol{x}^{\text{eq}}$, fixed by the initial conditions, and characterized by the vanishing of all the currents and affinities, $J_\rho^{\text{eq}} = A_\rho^{\text{eq}} = 0 \;\forall \rho$. Close to this equilibrium state, we can linearize Eq. (7) and re-express it in matrix form:

$$\boldsymbol{J} \simeq \mathbb{A}\boldsymbol{A}, \quad (57)$$

where $\mathbb{A}$ is the $R \times R$ diagonal matrix of linear susceptibility defined by the diagonal entries $(\mathbb{A})_{\rho\rho} = \lambda_\rho^+(\boldsymbol{x}^{\text{eq}}) = \lambda_\rho^-(\boldsymbol{x}^{\text{eq}}) = \lambda_\rho^{\text{eq}}$. Despite the familiar form of a linear phenomenological relation [50], Eq. (57) is little informative on the system's response. It describes the local response of each current $J_\rho$ to a small perturbation of the corresponding affinity $A_\rho$ without taking into account the cross-couplings between chemical reactions. In this sense, the Onsager reciprocal relations [69] are trivially satisfied being $\mathbb{A}$ a diagonal matrix. Furthermore, Eq. (57) is blind to the underlying network topology: we know that only $M$ out of the $R$ reactions (and the corresponding currents) in the network are linearly independent due to cycles. A natural question is thus how cross-couplings among reactions emerge in this context and how they relate to the network structure. The decompositions introduced in Sec. IV A will provide a natural framework to address these aspects.

Eq. (57) is valid whenever the affinity is small, but one may further assume that the system is (i) closed, with affinity $\boldsymbol{A} \ll 1$ remaining small and conservative while relaxing to zero; (ii) open, with an external source (e.g. a chemostat) providing a constant non-conservative contribution to the total affinity $\boldsymbol{A} \ll 1$. In the first case, the system exhibits a transient relaxation towards $\boldsymbol{x}^{\text{eq}}$. In the second case, the system reaches a nonequilibrium steady state $\boldsymbol{x}^*$, close to $\boldsymbol{x}^{\text{eq}}$, with positive entropy production. We shall treat these two cases separately before revealing the connections between them, in the spirit of the Einstein relation between diffusivity and mobility.

#### 1. Transient response

We have already seen how the finite-time relaxation is fully captured by the $M$ cocycle currents $J_\gamma^e$ in Eq. (28) (and this even outside the linear-response regime, when a steady state is reached). By substituting the decomposition Eq. (32) in Eq. (5), one directly sees that the currents $J_\alpha^c$ do not contribute to the time evolution of $x(t)$, since $c^\alpha \in \text{Ker}\,\mathbb{S}$. Accordingly, we can plug Eq. (57) into the definition of $J_\gamma^e$ and get

$$J_\gamma^e = c^\gamma \cdot J = c^\gamma \cdot \mathbb{A} A = \sum_{\gamma'} \underbrace{c^{\gamma\top}\mathbb{A} c^{\gamma'}}_{(\mathbb{L}_Q)_{\gamma\gamma'}} A_{\gamma'}^c, \quad (58)$$

where we used the decomposition in Eq. (29) together with KVL, Eq. (31). Equation (58) describes the linear relation between transient currents $J_\gamma^e$ and conservative affinities $A_\gamma^c$. They vanish together in the long-time limit, as $x(t) \to x^{\text{eq}}$. Accordingly, we identify the matrix $\mathbb{L}_Q$ in Eq. (58) as an $M \times M$ relaxation matrix. It is symmetric and positive-defined in accordance with Onsager reciprocal relations.

---

[8] In fact, if one evaluates Eq. (51) starting from the l.h.s. instead of the r.hs. of Eq. (50), one arrives directly at Eq. (55).

We introduce the distances from equilibrium for the concentration $x(t)$ and the chemical potential $\mu(t)$ as:

$$\delta x(t) = x(t) - x^{\text{eq}} \quad (59)$$
$$\delta \mu(t) = \mu(t) - \mu^{\text{eq}}, \quad (60)$$

so that $\delta x$, $\delta \mu \xrightarrow{t\to\infty} 0$. Then, from Eq. (41), the affinity vector becomes

$$A = -\mathbb{S}^\top \mu = -\mathbb{S}^\top \delta\mu = -\mathbb{S}^\top (\mathbb{X}^{\text{eq}})^{-1} \delta x \quad (61)$$

where in the second equality we have introduced a diagonal matrix $\mathbb{X}^{\text{eq}}$ whose entries are given by $(\mathbb{X}^{\text{eq}})_i = x_i^{\text{eq}}$. Also, by applying the matrix $\mathbb{G}$ to $\delta x(t)$ one gets:

$$\mathbb{G}\,\delta x(t) = \begin{pmatrix} \mathbf{0} \\ \hline \delta z(t) \end{pmatrix} \begin{matrix} \updownarrow N-M \\ \updownarrow M \end{matrix}, \quad (62)$$

where $\delta z(t)$ is the vector containing the distance to equilibrium for the $z_\gamma$ variables, $\delta z_\gamma = z_\gamma(t) - z_\gamma^{\text{eq}}$. The first $N-M$ zeroes in Eq. (62) correspond to the conservation laws $\ell$ (by definition, $\ell \cdot \delta x(t) = 0$). The relation in Eq. (62) can be inverted using the structure of the row reduction [see Eq. (A16) in Appendix A], and one gets:

$$\delta x(t) = \mathbb{S}_M \,\delta z(t), \quad (63)$$

where we recall that $\mathbb{S}_M$ is the matrix consisting of the $M$ first columns of $\mathbb{S}$. We replace Eq. (63) into Eq. (61) to obtain an expression for the affinity as a function of the reduced set of variables $\delta z_\gamma$. In particular, for the $M$ cocycle affinities $A_\gamma^c$ we find:

$$A_\gamma^c = -\sum_{\gamma'} \Big(\underbrace{\mathbb{S}_M^\top (\mathbb{X}^{\text{eq}})^{-1} \mathbb{S}_M}_{\mathbb{H}_Q}\Big)_{\gamma,\gamma'} \delta z_{\gamma'}, \quad \forall \gamma. \quad (64)$$

Notice that the $M \times M$ matrix $\mathbb{H}_Q$ defined in this relation is symmetric and positive-defined, in accordance to the conservative nature of $A_\gamma^c$. Finally, combining Eqs. (28), (58) and (64) we obtain the linear evolution of $\delta z$:

$$\frac{d}{dt}\delta z(t) = -\mathbb{B}\,\delta z(t), \quad (65)$$

where $\mathbb{B} = \mathbb{L}_Q \mathbb{H}_Q$ is the stability matrix whose spectrum controls the relaxation to the equilibrium state and is strictly positive, $\text{Sp}\,\mathbb{B} = \text{Sp}\,\mathbb{L}_Q \mathbb{H}_Q > 0$. As a consequence, the system relaxes monotonically to the equilibrium steady state, which in the theory of dynamical systems is called a stable node. Interestingly, the matrices $\mathbb{L}_Q$ and $\mathbb{H}_Q$, which appear naturally in our deterministic framework, bear a physical meaning in the underlying stochastic dynamics. It is known that Gaussian temporal fluctuations around equilibrium are well described by the (linearized) chemical Langevin equation [70]. In Appendix C, we show that the Onsager matrix $\mathbb{L}_Q$ appears to be the covariance matrix of the Gaussian noise entering the Langevin description, where, in the large but finite $\Omega$ asymptotics, $\delta z(t)$ becomes a stochastic process. The matrix $\mathbb{H}_Q$ appears as the Hessian matrix associated to the quadratic quasipotential $\varphi(\delta z) = \frac{1}{2}\delta z^\top \mathbb{H}_Q\, \delta z$ from which the conservative force $-\mathbb{L}_Q \nabla \Phi$ of the Langevin equation is obtained [see Eq. (C12)] and that describes the equilibrium Gaussian distribution $\propto \exp[-\Omega \varphi(\delta z)]$ of the deviation $\delta z$ around its average value $\mathbf{0}$.

### 2. Steady-state response

For an open system, relaxation to equilibrium is impeded by the continuous supply of external currents, as described in Sec. IV C. Then, the overall affinity is non-conservative and takes the explicit form given by Eq. (46) with the non-vanishing circulations $A_\alpha^e$ determined by the chemostatting parameters $a$:

$$A_\alpha^e = c^\alpha \cdot A = c^\alpha \cdot a \neq 0 \quad \forall \alpha. \quad (66)$$

Consider for simplicity the case of a time-independent chemostatting, $a \ll 1$, so that the system reaches a nonequilibrium steady state $x^*$ linearly close to the equilibrium state, $\delta x^* = x^* - x^{\text{eq}} \ll 1$. In the linear regime, we can replace Eq. (57) into Eq. (66) so that

$$A_\alpha^e = c^\alpha \cdot \mathbb{A}^{-1} J \stackrel{t\to\infty}{=} \sum_{\alpha'} \underbrace{c^{\alpha\top}\mathbb{A}^{-1} c^{\alpha'}}_{(\mathbb{L}_P)_{\alpha\alpha'}} J_{\alpha'}^{c,*}, \quad (67)$$

where in the last equality we used the current decomposition Eq. (32) under the steady-state condition, i.e. when only the cyclic currents survive $J_\alpha^{c,*} = \lim_{t\to\infty} J_\alpha^c(t)$. The matrix $\mathbb{L}_P$ in Eq. (67) describes the linear relation between the non-conservative affinities maintaining the system out of equilibrium and the non-zero currents characterizing the steady state. As such, it corresponds to the Onsager matrix of the steady-state response and one verifies that it is symmetric and positive-defined. Notice that the response in Eq. (67) corresponds to the one initially studied by Schnakenberg [11]. In his analysis, Schnakenberg emphasizes the thermodynamic significance of cycles. Indeed, we see that the steady-state response is fully determined by a number of currents and affinities given by the number of cycles in the underlying topology.

Notably, the dimensions of the Onsager matrices controlling transient response $\mathbb{L}_Q$ and steady-state response $\mathbb{L}_P$ are not the same: they are fixed by the number of cocycles and cycles, respectively. In both cases, the off-diagonal contributions to the response emerge once we restrict the analysis to the subset of physically relevant currents and affinities. Formally, this is done by projecting the total currents and affinities on the subspaces defined by the cycles and the cocycles (see Appendix D). A natural question is how the two Onsager matrices $\mathbb{L}_P$ and $\mathbb{L}_Q$ are related, given that (*i*) they describe respectively relaxation to equilibrium and response to a small drive (that one thus expects to be related fluctuation-dissipation); (*ii*) they live in (complementary) spaces of different dimensions. We address this question in next paragraph.

### 3. Hidden fluctuation-dissipation symmetries

Following the same convention for the ordering of the reactions as in Sec. II B, we subdivide the diagonal matrix $\mathbb{A}$ as:

$$\mathbb{A} = \begin{pmatrix} \mathbb{A}_M & 0 \\ 0 & \mathbb{A}_{R-M} \end{pmatrix}, \tag{68}$$

where the upper diagonal block $\mathbb{A}_M$ corresponds to the $M$ cochord reactions and the lower diagonal block $\mathbb{A}_{R-M}$ to the $R-M$ chord reactions. Doing so, from Eqs. (58) and (67), the Onsager matrices $\mathbb{L}_Q$ and $\mathbb{L}_P$ explicitly read

$$\mathbb{L}_Q = \mathbb{A}_M + \mathbb{T}\mathbb{A}_{R-M}\mathbb{T}^\top \tag{69}$$

$$\mathbb{L}_P = \mathbb{A}_{R-M}^{-1} + \mathbb{T}^\top \mathbb{A}_M^{-1} \mathbb{T}, \tag{70}$$

with no apparent connections between them for generic $\mathbb{A}$.

In order to unveil such connection, we perform the following diagonal transformation for the variables:

$$\hat{\boldsymbol{J}} = \mathbb{A}^{-1/2}\boldsymbol{J}, \quad \hat{\boldsymbol{A}} = \mathbb{A}^{+1/2}\boldsymbol{A}. \tag{71}$$

One sees from Eq. (57) that such change of variable corresponds to a rescaling of the linear-regime current/affinity relation, such that $\hat{J} = \hat{A}$. Moreover, it preserves the orthogonality structure between the potential condition $\boldsymbol{A} \in \operatorname{Im} \mathbb{S}^\top$ and the stationary condition $\boldsymbol{J} \in \operatorname{Ker} \mathbb{S}$ discussed in Sec. IV A: for the new variables, these conditions become

$$\hat{A} \in \operatorname{Im}(\mathbb{S}\mathbb{A}^{1/2})^\top, \quad \text{for conservative affinities}, \tag{72}$$

$$\hat{J} \in \operatorname{Ker} \mathbb{S}\mathbb{A}^{1/2}, \quad \text{for stationary currents}, \tag{73}$$

involving complementary orthogonal subspaces,

$$\operatorname{Im}(\mathbb{S}\mathbb{A}^{1/2})^\top \perp \operatorname{Ker}\mathbb{S}\mathbb{A}^{1/2}. \tag{74}$$

The matrix $\mathbb{A}$ being invertible, one easily verifies that $\{\mathbb{A}^{1/2}\boldsymbol{c}^\gamma\}$ constitutes a basis for the subspace in Eq. (72) while $\{\mathbb{A}^{-1/2}\boldsymbol{c}^\alpha\}$ forms a basis for the subspace in Eq. (73). Accordingly, we can introduce rescaled cocycles and cycles defined as $\hat{\boldsymbol{c}}^\gamma = \mathbb{A}^{1/2}\boldsymbol{c}^\gamma \mathbb{A}_\gamma^{-1/2}$ and $\hat{\boldsymbol{c}}^\alpha = \mathbb{A}^{-1/2}\boldsymbol{c}^\alpha \mathbb{A}_\alpha^{1/2}$.

The new cycles and cocycles still satisfy the orthogonality relations Eqs. (14) and (17) and constitute a basis in $\mathbf{R}^R$, namely:

$$\{\hat{\boldsymbol{c}}^\gamma, \hat{\boldsymbol{c}}^\alpha\} = \begin{pmatrix} \mathbb{1}_M & -\hat{\mathbb{T}} \\ \hat{\mathbb{T}}^\top & \mathbb{1}_{R-M} \end{pmatrix}, \tag{75}$$

where $\hat{\mathbb{T}} = \mathbb{A}_M^{-1/2} \mathbb{T} \mathbb{A}_{R-M}^{1/2}$. As a consequence, the decompositions for the affinity Eq. (29) and the current Eq. (32) readily generalize to the new representation, with macroscopic components defined as: $\hat{J}_\gamma^e = \hat{c}^\gamma \cdot \hat{\boldsymbol{J}}$, $\hat{J}_\alpha^c = e^\alpha \cdot \hat{\boldsymbol{J}}$, $\hat{A}_\gamma^c = e^\gamma \cdot \hat{\boldsymbol{A}}$, $\hat{A}_\alpha^e = \hat{c}^\alpha \cdot \hat{\boldsymbol{A}}$. Finally, in the linear regime, the Onsager matrices $\hat{\mathbb{L}}_Q$ and $\hat{\mathbb{L}}_P$ such that $\hat{J}_\gamma^e = \sum_{\gamma'}(\hat{\mathbb{L}}_Q)_{\gamma\gamma'}\hat{A}_{\gamma'}^c$, and $\hat{A}_\alpha^e = \sum_{\alpha'}(\hat{\mathbb{L}}_P)_{\alpha,\alpha'}\hat{J}_{\alpha'}^{c,*}$ are obtained following the same procedure as before and read:

$$\hat{\mathbb{L}}_Q = \mathbb{1}_M + \hat{\mathbb{T}}\hat{\mathbb{T}}^\top = \mathbb{A}_M^{-1/2} \mathbb{L}_Q \mathbb{A}_M^{-1/2} \tag{76}$$

$$\hat{\mathbb{L}}_P = \mathbb{1}_{R-M} + \hat{\mathbb{T}}^\top \hat{\mathbb{T}} = \mathbb{A}_{R-M}^{1/2} \mathbb{L}_P \mathbb{A}_{R-M}^{1/2}. \tag{77}$$

Interestingly, the two matrices $\hat{\mathbb{T}}\hat{\mathbb{T}}^\top$ and $\hat{\mathbb{T}}^\top\hat{\mathbb{T}}$ share the same non-zero eigenvalues, meaning that the Onsager matrices $\hat{\mathbb{L}}_Q$ and $\hat{\mathbb{L}}_P$ also have the same spectrum up to the multiplicity of eigenvalue $\lambda = 1$.

*Proof.* Let us consider an eigenvector $\boldsymbol{w}$ and the corresponding eigenvalue $\lambda \neq 0$ of the matrix $\hat{\mathbb{T}}^\top \hat{\mathbb{T}}$ so that

$$\exists \boldsymbol{w} : \hat{\mathbb{T}}^\top \hat{\mathbb{T}} \boldsymbol{w} = \lambda \boldsymbol{w}. \tag{78}$$

By multiplying by $\hat{\mathbb{T}}$ on the left one gets $\hat{\mathbb{T}}\hat{\mathbb{T}}^\top \hat{\mathbb{T}}\boldsymbol{w} = \lambda \hat{\mathbb{T}}\boldsymbol{w}$. If $\hat{\mathbb{T}}\boldsymbol{w}$ is different from zero, then $\lambda$ is also an eigenvalue of the matrix $\hat{\mathbb{T}}\hat{\mathbb{T}}^\top$. *Ad absurdum* let us assume that $\hat{\mathbb{T}}\boldsymbol{w} = 0$. From Eq. (78) we see that this implies $\lambda \boldsymbol{w} = 0$ which is against the original assumption ($\lambda \neq 0$). As a consequence, for any non-zero eigenvalues $\lambda$,

$$\lambda \in \operatorname{Sp}\hat{\mathbb{T}}^\top\hat{\mathbb{T}} \Leftrightarrow \lambda \in \operatorname{Sp}\hat{\mathbb{T}}\hat{\mathbb{T}}^\top \quad \square \tag{79}$$

The diagonal transformation in Eq. (71) reveals a hidden symmetry in the spectrum of the Onsager matrices of complex CRNs. It links the transient relaxation of the system produced by a spontaneous (or imposed) fluctuation to the stationary response of the system to a drive:

$$\operatorname*{Sp}_{\neq 1} \hat{\mathbb{L}}_P = \operatorname*{Sp}_{\neq 1} \hat{\mathbb{L}}_Q, \tag{80}$$

which generalizes to network topologies the 1D Einstein relation $\mu_E = D$ between the mobility $\mu_E$ and the diffusion constant $D$ (setting $k_B T = 1$). For CRNs, the Onsager matrices $\mathbb{L}_P$ and $\mathbb{L}_Q$ plays the respectively role of a matricial mobility and diffusivity (see Appendix C). We stress that this symmetry is nontrivial: the two matrices have different dimensions due to the existence of conservation laws and cycles.

### B. Linear-regime thermodynamically feasible reconstruction of metabolic networks

Typically, in metabolomics some species are injected in the cell by external reactions $\emptyset \rightleftharpoons Y$ of the kind introduced in Sec. IV C. Such chemostatting reactions naturally give rise to emergent cycles whose driving affinities are generally nonzero [10]. In contrast, there is no drive associated with internal cycles, which have vanishing affinity according to KVL [Eq. (31)]. This grants thermodynamic feasibility, that is, the existence of a potential vector $\boldsymbol{\mu}$ such that $A_\rho = -\mathbb{S}_\rho^\top \boldsymbol{\mu}$ and $J_\rho A_\rho \geq 0$ for any internal reaction[9]. At the same time,

---

[9] Mass-action kinetics implies through Eq. (7) that a positive (negative) force $A_\rho$ produces a positive (negative) current $J_\rho$. Accordingly, a feasible inter-

mass balance is ensured by the continuity equation, Eq. (5), which at stationarity reduces to KCL [Eq. (35)].

In metabolic reconstruction, a subset of external currents are treated as known (fixed) parameters and the problem consists in predicting a thermodynamically feasible value (or range of values) for the remaining set. We label $J_Y$ the known external currents and refer to the remaining set of reactions as internal, with (unknown) current state $J$. Then, KCL can be expressed as:

$$\mathbb{S} J = -\mathbb{S}_Y J_Y , \qquad (81)$$

where we have separated the known external reactions in $\mathbb{S}_Y$ from the remaining reactions in $\mathbb{S}$. The presence of internal cycles for $\mathbb{S}$ makes that the linear problem in Eq. (81) is generally under-determined and the solution space multi-dimensional. It is the case in standard cellular networks, for which additional constraints such as upper/lower current bounds, or the optimization of objective functions, are typically imposed to reduce the solution space. In practice, solving Eq. (81) while properly taking into account the thermodynamic constraints has proven computationally hard due to the nonlinearity of KVL. On another hand, if all the reactions in the network are independent (i.e. there are no internal cycles), the solution to Eq. (81) is unique and fully determined by the set of external currents $J_Y$. In this case, $\mathbb{S}$ is full column rank ($M = R$) and $J = \mathbb{G}_M \mathbb{S}_Y J_Y$ with $\mathbb{G}_M$ defined in Eq. (26) [where $\mathbb{G}$ associated with the stoichiometric matrix $\mathbb{S}$ of the internal reactions].

### 1. Insights from geometry

For an arbitrary network $\mathbb{S}$ of internal reactions, we can use the framework introduced in Sec. III A to identify independent and dependent reactions, the cochords $\{\gamma\}$ and chords $\{\alpha\}$. Such decomposition is not unique, however it allows one to identify a limited number of degrees of freedom affected by the nonlinearity of the KVL constraint.

Consider a cocycle $c^\gamma$ and its associated island $v^\gamma = (\mathbb{G}_M^\top)_\gamma$ [see Sec. III B 3]. At steady state, its population $v^\gamma \cdot x$ is constant, so that $0 = v^\gamma \cdot \partial_t x = v^\gamma \cdot (\mathbb{S} J + \mathbb{S}_Y J_Y)$. Using that $c_\gamma = -\mathbb{S}^\top v_\gamma$, one finds:

$$c^\gamma \cdot J = (\mathbb{G}_M \mathbb{S}_Y J_Y)_\gamma . \qquad (82)$$

Physically, Eq. (82) establishes a balance between the flux of internal currents (l.h.s) and that of the external currents (r.h.s) at the boundary of the island $v_\gamma$. Noticing that $c^\gamma \cdot J$ is nothing but the component $J^e_\gamma$ in the current decomposition Eq. (32), we see that $J^e_\gamma$ is fully determined by the external currents $J_Y$, depends linearly on them and is not affected by the thermodynamic constraint of KVL. In particular, this is true for 'bridge' reactions, i.e. reactions that do not enter

---

nal reaction with nonvanishing affinity contributes positively to the entropy production rate, in accordance with second law of thermodynamics, see Eq. (36). When mass-action kinetics is not assumed, thermodynamic feasibility is in general encoded in the condition that $A_\rho$ and $J_\rho$ have the same sign for internal reactions [43, 46].

any internal cycles, for which $c^\gamma = e^\gamma$. The identification of bridges is independent from the choice of chords and cochords, and one may wonder what is the biological advantage of having bridges in a metabolic pathway. Those are internal reactions whose current $J_\gamma = (\mathbb{G}_M \mathbb{S}_Y J_Y)_\gamma$ is fully determined by the external environment of the cell (characterized by the uptake and secretion rates $J_Y$).

More generally, the full set of components of the decomposition Eq. (32) of $J$ on cochords/cycles takes the form

$$\begin{pmatrix} J^e_\gamma \\ J^c_\alpha \end{pmatrix} = \begin{pmatrix} \mathbb{G}_M \mathbb{S}_Y J_Y \\ \hline \text{function}(J_Y) \end{pmatrix} . \qquad (83)$$

This choice of basis thus tells that $M$ linear combinations of the internal currents, the $J^e_\gamma$'s, are *independent of the cycle currents $J^c_\alpha$'s* and *fully determined by an explicit linear function* of the external currents. We surmise that such strong constraints between the components of $J$ is of interest in metabolic reconstruction. As a consequence, all the difficulty of imposing KVL is condensed into the determination of the $J^c_\alpha$'s as, in general, a non-linear function of the external currents [the last $R - M$ lines on the r.h.s. of Eq. (83)]. Clearly, imposing $J^c_\alpha = e_\alpha \cdot J = 0$, i.e. effectively removing chords from the network, provides a feasible reference state $J_0$

$$J_0 = \begin{pmatrix} \mathbb{G}_M \mathbb{S}_Y J_Y \\ 0 \end{pmatrix} \qquad (84)$$

which only lives on the chosen $M$ independent (cochord) reactions. Such solution is simple in the sense that the internal current state is a linear function of the external currents, and the currents associated to every cycle are zero (so that KVL is trivially verified). As detailed below, this is what is practically done in state-of-the-art algorithms of metabolic reconstruction, which *de facto* lack a procedure to fully explore the role of cycles in metabolic pathways.

We now study some examples to understand how the unknown function in Eq. (83) can be determined in a linear regime assumption close to equilibrium, before presenting a generic algorithmic procedure valid in this regime. To keep things simple, we consider here non-interacting networks, but the method fully applies to interacting ones (since, as will become apparent, it relies on the algebraic tools we have presented).

### 2. Noninteracting network, one internal cycle

As the simplest possible case consider the network

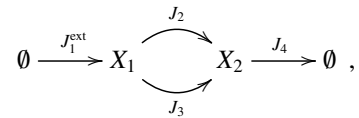

where for the sake of ease we momentarily loosen the assumption that there are no multiple reactions with the same stoichiometry. We assume that $J_1^{\text{ext}}$ is known, and that all other currents have to be reconstructed. Notice that $J_4$ is also an external current, which we could also fix, but then we should keep in mind that by mass conservation $J_4^{\text{ext}} = J_1^{\text{ext}}$. In general, not any external current can be independently fixed, so some care is needed in choosing the boundary data.

We now want to impose dynamic and thermodynamic constraints. KCL at the nodes of the network clearly implies

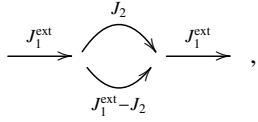

where current $J_2$ now acts as a free parameter. As far as KCL is involved, this parameter could take any real value. We will now implement thermodynamic feasibility to reduce the span of $J_2$. Given the constitutive equation Eq. (6), and identifying the internal cycle

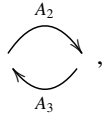

KVL yields

$$\lambda_2^+ \lambda_3^- = \lambda_2^- \lambda_3^+ , \tag{85}$$

where the $\lambda_\rho^\pm$'s are evaluated at the steady state $\lambda_\rho^\pm = \lambda_\rho^\pm(x^*)$. Let us now rewrite this in terms of the external current $J_1^{\text{ext}}$, of the internal current $J_2$ (that we take as free parameter), and of the velocities $\lambda_2^-$ and $\lambda_3^-$ (chosen arbitrarily). After a little work we obtain

$$J_2 = \frac{1}{1 + \lambda_3^- / \lambda_2^-} J_1^{\text{ext}}. \tag{86}$$

In reconstruction problems the actual values of $\lambda_2^-$ and $\lambda_3^-$ are usually not known. However, by the fact that they are positive, this latter equation implies

$$0 \leq J_2 \leq J_1^{\text{ext}}. \tag{87}$$

This restricts the range of feasible values for $J_2$ and, most importantly, prescribes the current directionality. For positive $J_1^{\text{ext}}$ both currents $J_2$ and $J_3$ have to flow left-to-right, which makes physical sense: one would not expect a river that bifurcates around an island to have upward flows along one of its branches!

However, reconstruction procedures that just implement KCL may fail to impose this constraint, thus producing thermodynamically infeasible behaviors. For instance, if $J_1^{\text{ext}} = 0$, one could have a perpetuum mobile in the absence of forces:

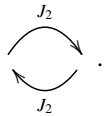

The same solution as Eq. (86) can be found by the linear-regime approach. In view of Eq. (57), KVL prescribes

$$0 = A_2 - A_3 \approx \frac{J_2}{\lambda_2^{\text{eq}}} - \frac{J_1^{\text{ext}} - J_2}{\lambda_3^{\text{eq}}}, \tag{88}$$

leading to

$$J_2 = \frac{1}{1 + \lambda_3^{\text{eq}} / \lambda_2^{\text{eq}}} J_1^{\text{ext}}. \tag{89}$$

This, in fact, is almost identical to Eq. (86), but for the fact that it is written in terms of equilibrium values of the velocities. For the sake of our analysis, notice that, provided that the velocities are just some positive quantities, Eqs. (86) and (89) impose the exact same constraint. Linearization here has no actual consequence.

### 3. Noninteracting network, two cycles

Let us now consider

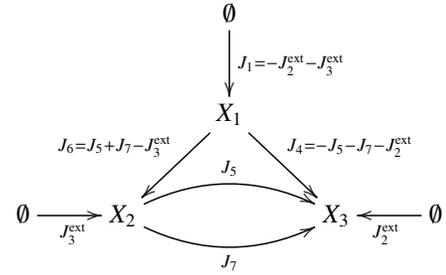

where we already implemented KCL in terms of the external currents $J_2^{\text{ext}}, J_3^{\text{ext}}$ and of the internal currents $J_5, J_7$.

KVL on the two internal cycles instead prescribes:

$$\lambda_5^+ \lambda_7^- = \lambda_5^- \lambda_7^+ \tag{90}$$
$$\lambda_4^+ \lambda_5^- \lambda_6^- = \lambda_4^- \lambda_5^+ \lambda_6^+ . \tag{91}$$

Selecting $\lambda_4^-, \lambda_5^-, \lambda_6^-$ as free parameters, after some work from the first we find the linear equation $J_7 = J_5 \lambda_7^- / \lambda_5^-$, and letting $\beta = 1 + \lambda_7^- / \lambda_5^-$ from the second we obtain the quadratic equation

$$\frac{\beta}{\lambda_5^- \lambda_6^-} J_5^2 + \left( \frac{1}{\lambda_5^-} + \frac{\beta}{\lambda_4^-} + \frac{\beta}{\lambda_6^-} - \frac{J_3^{\text{ext}}}{\lambda_5^- \lambda_6^-} \right) J_5$$
$$+ \frac{J_2^{\text{ext}}}{\lambda_4^-} - \frac{J_3^{\text{ext}}}{\lambda_6^-} = 0. \tag{92}$$

Once again, notice that when there is no external current $J_2^{\text{ext}} = J_3^{\text{ext}} = 0$ we get $J_5 = J_7 = 0$, no perpetuum mobile. Otherwise, for given values of $J_2^{\text{ext}}, J_3^{\text{ext}}$ one can use this equation to explore the possible values of the internal currents in terms of arbitrarily chosen parameters $\lambda_4^-, \lambda_5^-, \lambda_6^-$.

This quadratic problem is already becoming complicated (and it is easy to foresee that for more complicated topologies this will give rise to higher-order polynomial systems). Given that we are interested in some bulk characterization of the landscape, a full solution may be an overshooting. There-

fore, like in the previous example, let us proceed by linearization of KVL:

$$0 = A_5 - A_7 \approx \frac{J_5}{\lambda_5^{eq}} - \frac{J_7}{\lambda_7^{eq}} \tag{93}$$

$$0 = A_4 - A_5 - A_6 \approx \frac{J_4}{\lambda_4^{eq}} - \frac{J_5}{\lambda_5^{eq}} - \frac{J_6}{\lambda_6^{eq}}. \tag{94}$$

The first easily gives $J_7 = J_5 \lambda_7^{eq}/\lambda_5^{eq}$, and letting $\beta^{eq} = 1 + \lambda_7^{eq}/\lambda_5^{eq}$ the second yields

$$\left(\frac{1}{\lambda_5^{eq}} + \frac{\beta^{eq}}{\lambda_4^{eq}} + \frac{\beta^{eq}}{\lambda_6^{eq}}\right) J_5 + \frac{J_2^{ext}}{\lambda_4^{eq}} - \frac{J_3^{ext}}{\lambda_6^{eq}} = 0. \tag{95}$$

But now this is a simple linear equation that, given the external currents, provides a reconstruction for any given choice of positive real $\lambda$'s. Notice that this equation can also be obtained from Eq. (92) by disregarding terms of order $J^2$. Thus this reconstruction, being the limiting case of a feasible reconstruction, is also thermodynamically feasible.

### 4. Considerations and problem setting

The key takeaways of these examples are the following. Both ways, direct imposition of Kirchhoff's laws and linearization grant thermodynamic feasibility of reconstruction. The first is consistent with the basic tenets of reaction rate theory but increasingly complicated with topology. Known algorithms that deal with it rely on the identification and removal of infeasible cycle currents from solutions of Eq. (81). This is done by sampling and postprocessing the solution space of Eq. (81) using linear optimization procedures [46, 71, 72]. In practice, such algorithms are bound to hit the boundary of the space of solutions, where internal cycles are effectively removed, leading to 'trivial' solution of the type of Eq. (84). In the first example, there are two of such trivial solutions, respectively $J_2 = 0$ or $J_1^{ext} - J_2 = 0$.

Our stance is that, given that metabolic reconstruction is a very under-determined problem and that one is more interested in spanning a space of viable solutions rather than into specific solutions, we can resort to linearization to obtain a broad bulk of feasible reconstructions. Given the stoichiometric matrix of a metabolic network, the first step is to split it in terms of internal reactions $\mathbb{S}$, that we want to reconstruct, and external reactions $\mathbb{S}_Y$, for which there exist data or that we want to control. Notice that from Eq. (81) any left-null vector $\ell$ of $\mathbb{S}$ which is not a left-null vector of $\mathbb{S}_Y$ imposes a constraint (interdependence) among the parameters $J_Y$. Once a good choice of independent external reactions is made, we fix some input currents $J_Y$ for those, and want to produce a reconstructed current state $J$ that satisfies both KCL and KVL. We base our solution on the linear-regime assumption.

### 5. Linear-regime reconstruction algorithm

1) Input the metabolite stoichiometric matrix and split it as $(\mathbb{S}_Y, \mathbb{S})$ in terms of external reactions $\mathbb{S}_Y$ and internal reactions $\mathbb{S}$. In the following we refer to $R$ as the number of internal reactions and $M$ as the rank of $\mathbb{S}$;

2) Input the external currents $J_Y$;

3) For all left nullvectors $\ell$ of $\mathbb{S}$ check that $\ell \cdot \mathbb{S}_Y J_Y = 0$, else revise the input currents or reduce the number of external reactions and go back to 1);

4) Row-reduce the matrix $\mathbb{S}$ to obtain the matrix $\mathbb{GS}$;

5) Reorder reactions in such way that $\mathbb{GS}$ takes the form Eq. (22); Reorder $\mathbb{S}$ accordingly;

6) Input $R$ real positive parameters $\lambda_\rho^{eq}$;

7) Let $\mathbb{A} = \text{diag}\{\lambda_\rho^{eq}\}_\rho$; Let $\text{diag}(\mathbb{A}_M, \mathbb{A}_{R-M}) = \mathbb{A}$; Let $\mathbb{L}_P$ as per Eq. (70);

8) Let

$$J = \begin{pmatrix} (\mathbb{1}_M - \mathbb{T}\mathbb{L}_P^{-1}\mathbb{T}^\top \mathbb{A}_M^{-1}) \mathbb{G}_M \mathbb{S}_Y J_Y \\ \mathbb{L}_P^{-1}\mathbb{T}^\top \mathbb{A}_M^{-1} \mathbb{G}_M \mathbb{S}_Y J_Y \end{pmatrix}. \tag{96}$$

In Appendix E we derive Eq. (96) and prove that it satisfies both KCL and KVL. Namely, every current state $J$ obtained from Eq. (96) by fixing $J_Y$ and the $\lambda$'s is thermodynamically feasible. Most importantly, the *full space* of feasible solution (within the linear regime) is explored by varying the $\lambda$'s in Eq. (96).

The rationale behind the algorithm can be understood by re-expressing the solution as follows:

$$J = \begin{pmatrix} \mathbb{1}_M & -\mathbb{T} \\ \mathbb{0} & \mathbb{1}_{R-M} \end{pmatrix} \begin{pmatrix} J_\gamma^e \\ J_\alpha^c \end{pmatrix}, \text{ with } \begin{pmatrix} J_\gamma^e \\ J_\alpha^c \end{pmatrix} = \begin{pmatrix} \mathbb{G}_M \mathbb{S}_Y J_Y \\ \mathbb{L}_P^{-1}\mathbb{T}^\top \mathbb{A}_M^{-1} \mathbb{G}_M \mathbb{S}_Y J_Y \end{pmatrix} \tag{97}$$

where we made contact with the current decomposition of Eq. (32), by noticing that cochords $\{e^\gamma\}$ and cycles $\{c^\alpha\}$ form respectively the first $M$ and last $R - M$ columns of the left matrix in Eq. (97). Then Eq. (97) corresponds to a perturbative solution around the loop-less reference point Eq. (84) where one introduces back the chords (and the internal cycles) and assume the validity of the linear regime approximation for the full set of currents. In other words, the term $\mathbb{L}_P^{-1}\mathbb{T}^\top \mathbb{A}_M^{-1} \mathbb{G}_M \mathbb{S}_Y J_Y$ in Eq. (97) is a linear-regime expression of the unknown function$(J_Y)$ in Eq. (84).

The reconstruction problem in our approach is thus particularly simple: one just needs focusing on the chord currents $J_\alpha^c$ ($< 10\%$ of the total reactions in realistic networks [71]) for which we provide a linear-response solution in terms of a susceptibility matrix $\mathbb{L}_P^{-1}\mathbb{T}^\top \mathbb{A}_M^{-1}$. The latter contains ratio of $\lambda$'s and has well-defined asymptotics as the $\lambda_\rho$'s are sent to 0 or $+\infty$ (see Appendix E and Fig. 12 for an example), which makes it possible to meaningfully explore all the thermodynamically feasible values for $J_\alpha^c$, irrespectively of the particular choice of dependent reactions associated with Eq. (84).

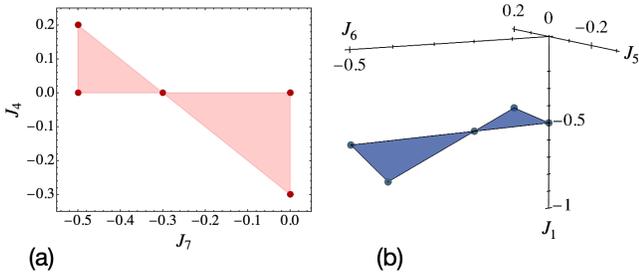

FIG. 12. Metabolic reconstruction based on Eq. (96) for the example introduced in Sec. V B 3. We fix the values of the external currents, respectively $J_2^{\text{ext}} = 0.3$ and $J_3^{\text{ext}} = 0.2$ and choose $J_4$ and $J_7$ to be the chord-currents $J_\alpha^c$'s associated to the two internal cycles. The details of the computations are provided in Appendix F. **(a)** The full range of feasible values for the chords currents is shown in red and obtained by exploring the asymptotics of the susceptibility matrix $\mathbb{L}_P^{-1} \mathbb{T}^\top \mathbb{A}_M^{-1}$ for $0 \le \lambda$'s $\le \infty$. The five red points are the values of $(J_4, J_7)$ obtained by sending every $\lambda$ to either 0 or $\infty$. They correspond to the boundary solutions of the type of Eq. (84) where cycles are effectively removed from the network (Notice that the multiplicity of such solutions is given by the number of spanning trees, which in this case is precisely 5). The monotonicity of the susceptibility matrix (see Appendix E) allows us to interpolate between such solutions, and explore the full space of feasible currents as a function of the $\lambda$'s $\ge 0$. Likewise in Eq. (87), the positivity of the $\lambda$'s constrains the feasible values of the cycle currents. Then any point within the red region is thermodynamically feasible and one would need some biological knowledge to further reduce the span of possible values. **(b)** Once identified the solution space for the chords, one uses Eq. (96) to obtain the corresponding solution space for the remaining cochords, in this case $J_1$, $J_5$ and $J_6$. Notice that $J_1$ is independent from the chords currents and always equal to $J_1 = -J_3^{\text{ext}} - J_2^{\text{ext}}$. It is an example of bridge current, which is not affected by the presence of internal cycles in the network.

In our approach, note that we stay close to the equilibrium case (where the stationary solution of the rate equation is unique, once the conserved quantities are fixed): the description of transitions between multiple attractors (including possibly limit cycles) is beyond the reach of our perturbative approach, and remains unaddressed in metabolic reconstruction.

Our algorithm is different from simply linearizing the currents in classical frameworks such as ll-COBRA or CycleFreeFlux [46, 71, 72] because there many cycle currents are simply set to zero. The free parameters in our approach are the coefficients $\lambda_\rho \ge 0$'s for each reaction $\rho$ involved in internal cycles. In the Outlook, we comment on their possible biochemical interpretation.

## VI. OUTLOOK

Interacting mass-action CRNs present a host of behaviors coming from the multiplicity of fixed points [73, 74] and ranging from non-linear oscillations [75] to chaos [76]. In our work, we focused on the stationary state and on relaxation properties, but the decomposition of currents and affinities that we identified could be useful tools to study such non-stationary phenomena. Also, naturally, the geometrical tools that we identified could help to study the role played by deficiency [23, 77, 78] in dissipation and noise [57, 79]. The symmetries in linear response could be compared to the recent approach of Ref. [80]. The separation of timescales that we identified in Sec. III B 4 through the evolution of the population of islands associated to cocycles bears strong ties with the control of chemical kinetics through catalysts and inhibitors, following for instance the recent methods presented in Ref. [16]. Also, autocatalysis plays an essential role in biochemical processes and it is only recently that the classifications of CRNs leading to this type of self-replication have been identified [29]. Such classification could be investigated in the light of our geometrical tools. Another geometrical approach was recently proposed in Ref. [81] where a notion of Hessian geometry of CRNs is constructed; it would be worth to identify the link between such approach and the notions of cycles and cocycles we have put forward. In the same way, geometrical decompositions were identified in Markov processes [82] and in field theories [33], which may be related to the ones we define. And of course much could be gained from going beyond the well-stirred limit by considering extended systems where spatial inhomogeneities play a role.

On the graph-theoretical side, a known duality exists between vertices and faces for planar graphs, which, in our language exchanges the roles of cycles and cocycles. For non-planar graphs, the concept of matroid [83] allows one to treat abstract independence sets based on circuits and to generalize dualities. For noninteracting CRNs, such duality thus implies a mapping between stationary currents (supported by cycles) and transient ones (supported by cocycles). It would be interesting to investigate the consequences of such a mapping. Our definition of cycles and cocycles of the hypergraph associated to a generic CRN opens natural questions: can such duality be extended to a class of hypergraphs, and what could we learn from it? Also, cycles were recently shown to control several aspects of fluctuations and large deviations in the graph associated to Markov jump processes [84–86]. Such results could be extended to dynamics on hypergraphs using the tools we have put forward.

Regarding the reconstruction algorithm of Sec. V B, the main open questions are about how to further constrain solutions with empirical data or reasonable target functions, and whether the linear-regime assumption is consistent with physiological conditions. About this latter, further analysis is needed to characterize the difference between the linear-regime landscape and the algebraic variety of solutions of the nonlinear KVL. About the former, the main virtue of our proposal is that the coefficients $\lambda_\rho^{\text{eq}}$'s are independent one of each other and can take any real value, while previous reconstruction efforts had to deal with non-convex spaces of parameters where optimization algorithms could get stuck into subspaces or at boundaries. Let us argue that these parameters also make biochemical sense, by going back to their linear-response meaning. By the fluctuation-dissipation paradigm, the coefficients $\lambda_\rho^{\text{eq}}$'s quantify the spontaneous activity of a system at equilibrium, that is, in the absence of external currents. In theory, one would have to realize the sole reaction $\rho$ *in vitro* and measure its activity. In practice, given that a single reaction's activity can be associated to the expression of the enzyme that catalyzes it, we propose that $\lambda_\rho^{\mathrm{eq}}$ could be roughly proportional to the abundance of the corresponding enzyme, for which there could be available data.

| Algebraic definitions | Geometrical interpretation | Physical significance |
|---|---|---|
| Cycles $$(c^\alpha) = \begin{pmatrix} -\mathbb{T} \\ \mathbb{1}_{R-M} \end{pmatrix}$$ See Sec. III A | Maps dissipative currents on the hypergraph See Sec. IV A | Basis for steady-state currents (KCL): $\mathbb{S} J = 0 \iff J = \sum_\alpha J_\alpha^c c^\alpha$ [Eq. (35)] Onsager response to external drive: $(\mathbb{L}_P)_{\alpha,\alpha'} = c^{\alpha\top} \mathbb{A}^{-1} c^{\alpha'}$ [Eq. (67)] |
| Cocycles $$(c^\gamma) = \begin{pmatrix} \mathbb{1}_M \\ \mathbb{T}^\top \end{pmatrix}$$ See Sec. III A | Maps transient currents on the hypergraph Boundary of island $v^\gamma$ See Sec. III B 3 and IV A | Basis for conservative forces (KVL): $c^\alpha \cdot A = 0 \iff A = \sum_\gamma A_\gamma^c c^\gamma$ [Eq. (31)] Onsager relaxation to equilibrium: $(\mathbb{L}_Q)_{\gamma\gamma'} = c^{\gamma\top} \mathbb{A} c^{\gamma'}$ [Eq. (58)] |
| Relation between the matrices $\mathbb{T}$ and $\mathbb{S}$: $-\mathbb{G}_M \mathbb{S} = (\mathbb{1}_M \ \mathbb{T})$ See Sec. III A and Eq. (22) | $\mathbb{G}_M$ identifies islands whose boundary are the cocycles $c^\gamma$'s $\mathbb{G}_M^\top$ is an integrator on the hypergraph: $V = \sum_\gamma (\mathbb{G}_M^\top)_\gamma A_\gamma^c$ See Sec. III B 2 and Eq. (27) | Coarse-grained evolution of islands population: $\frac{d}{dt}(\mathbb{G}_M x)_\gamma = J \cdot c^\gamma$ [Eq. (28)] Constraints in metabolic reconstruction: $J^e_\gamma = \mathbb{G}_M \mathbb{S}_Y J_Y$ [Eq. (82)] |

TABLE I. Main concepts and results for interacting CRNs.

In this paper, we treated mass-action CRNs, but, most of the results being of topological nature, they apply to more generic reaction kinetic laws (such as effective enzymatic models) with the only requirement that there exist conjugate currents and forces such that $J > 0$ if and only if $A > 0$, and $J = 0$ if and only if $A = 0$. What is special about mass-action kinetics is that the cycle affinities $A_\alpha^e$ do not depend explicitly on the populations, and therefore are constants of motion.

In this respect, an interesting direction to explore is that of reaction networks where species are not chemical, but biological. There, no notion of thermodynamic feasibility imposes that the affinities of internal cycles have to be zero, but the decompositions of affinities and currents, together with the geometrical an physical interpretation that we put forward (Table I), still apply. In such context, migration from regional pools of species can also play the role of chemostatting. Ecological and evolutionary models are known to present a variety of phenomena such as strong space-time fluctuations [87], chaos [88] or sensitivity to noise [89]. Systems modeled by (generalized) Lotka–Voltera equations [87, 88, 90] are particularly amenable to the tools we propose, as, at the population level, they can be put in correspondence with CRNs [91]. The geometrical concepts we have identified can help to study such problems.

### ACKNOWLEDGMENTS

SDC and VL acknowledge support by the ANR-18-CE30-0028-01 Grant LABS, the EverEvol CNRS MITI project and an IXXI project. The research was supported by the National Research Fund Luxembourg (project CORE Thermo-Comp C17/MS/11696700) and by the European Research Council, project NanoThermo (ERC-2015-CoG Agreement No. 681456). The authors apply a Creative Commons Attribution 4.0 International (CC BY 4.0) license to any version of this manuscript. SDC is grateful to Emanuele Penocchio and Gianmaria Falasco for helpful discussions on chemical reaction networks. MP thanks Daniele De Martino, Alexander Skupin and Susan Ghaderi for useful insights and discussions on metabolic reconstruction issues. We warmly thank Muhittin Mungan for a critical read of a first version of this manuscript, Delphine Ropers for discussions and references on metabolic reconstruction and Pierre Recho for stimulating discussions that inspired this work.

## Appendix A: Integration and differentiation on the networks of chemical reactions

### 1. Noninteracting CRNs: integration on spanning trees

Consider a set of unimolecular reactions, as in Sec. II, and assume that the corresponding graph $\mathcal{G}$ is connected. The stoichiometric matrix satisfies $(\mathbb{S}^\top V)_\rho = V_{t(\rho)} - V_{s(\rho)}$ so that $\mathbb{S}^\top$ can be seen as a gradient operator, which transforms a potential $V$ defined on every species/node $i$ into a (chemical) force field between the source $s(\rho)$ and target $t(\rho)$ of every reaction/edge $\rho$. In this Appendix, we build an explicit 'integrator': that is, if a conservative force $A$ belongs to $\mathrm{Im}\,\mathbb{S}^\top$ we want to build a potential $V$ such that $A = -\mathbb{S}^\top V$. This will be done by defining an integrator matrix $\mathbb{G}^\top$ from the entries of $\mathbb{S}$. Then, we present how these two matrices are related.

We remark in advance that $V$ is not unique: if $V$ and $V'$ yield the same $A$, we have that $\mathbb{S}^\top(V' - V) = 0$ so that $V' - V \in \mathrm{Ker}\,\mathbb{S}^\top$, meaning that the two potentials are equal up to a global constant, since $\mathrm{Ker}\,\mathbb{S}^\top$ is spanned by $\ell_0 = (1, \cdots, 1)$. This is similar to what happens in the continuum when integrating a function: a primitive is defined « up to a constant ».

Since the labeling and the orientation of reactions is arbitrary, we can redefine them to our best convenience. To do so, we first select arbitrarily one of the nodes, that will play the role of the 'root' of the graph. Then, we fix a spanning tree $T_\mathcal{G}$ (see Fig. 4), that is, a set of $M$ independent reactions. This allows one to fix the orientations: the edges in $T_\mathcal{G}$ are directed toward the root while the other edges (the chords) are oriented arbitrarily. We then turn to labeling, starting by the nodes. The root is node 1, and we label the other nodes incrementally from the root along $T_\mathcal{G}$ as follows (see Fig. 3): at every branching point of $T_\mathcal{G}$, we pick one of the branches and continue the numbering of species incrementally, until we reach a 'leaf' (i.e. a node of the graph without further edge). We then come back to the last branching point and continue the procedure until every remaining node is exhausted. Second, we label the edges. From node 2, a single edge points towards the root, which we label as edge 1. Recursively, the edge pointing out of node $\gamma + 1$ (for $1 \leq \gamma < M$) is labeled as edge $\gamma$. This exhausts the $M = N - 1$ cochords, labeled from 1 to $M$. The remaining $R - M$ chords (equal, in number, to the number of cycles) are labeled arbitrarily from $M + 1$ to $R$. Doing so, for the first $M$ columns of $\mathbb{S}$ (indexed by $1 \leq \gamma \leq M$) we have $s(\gamma) = \gamma + 1$ and $t(\gamma) \leq \gamma$. This implies that the stoichiometric matrix takes the form

$$\mathbb{S} = \underbrace{\begin{pmatrix} 1 & 1 & & & \\ -1 & & 1 & & \\ & -1 & & & \\ & & & \ddots & \\ & & (0) & & -1 \end{pmatrix}}_{\mathbb{S}_M} \middle| \mathbb{S}_{\text{dep}} \quad (A1)$$

where: (i) on the $N \times M$ block $\mathbb{S}_M$, the matrix $-\mathbb{1}_M$ lies on the lower diagonal and, on every column $\gamma$, there is a single entry 1 on line $t(\gamma) \leq \gamma$; and (ii) the last $R - M$ columns correspond to the chords, which are the dependent reactions. This means there exists a $M \times (R - M)$ matrix $\mathbb{T}$ such that $\mathbb{S}_{\text{dep}} = \mathbb{S}_M \mathbb{T}$, encoding the fact that every column of $\mathbb{S}_{\text{dep}}$ can be expressed as a linear combination of the $M$ independent columns of $\mathbb{S}_M$. In fact, this encodes a graphical property: every chord is part of a cycle (and every cycle has exactly one chord, see Fig. 4), and the algebraic dependency we just explain encodes that the chord reaction can be obtained by applying all the cochord reactions of the cycle (with the adequate orientation). Notice last that the first line of $\mathbb{S}$ contains only positive entries, since by our convention the root only has entering edges.

We define an $N \times N$ matrix $\mathbb{G}$ from Eq. (19) and we recall that $\mathcal{U}(j)$ is the set of nodes (including $j$) that are upstream of $j$ on $T_\mathcal{G}$. From a potential $V$ defined on the nodes (and imposed to verify $V_{\text{root}} = 0$), we define a set of forces

$$A_\gamma = V_{s(\gamma)} - V_{t(\gamma)} = -(\mathbb{S}^\top V)_\gamma \quad (A2)$$

for every cochord. Because the line $i$ of $\mathbb{G}^\top$ contains 1 for every node located in between the root and node $i$ (i.e. for every node $j$ such that $i \in \mathcal{U}(j)$) we have, by telescoping sum

$$V_i = V_i - V_{\text{root}} = \sum_\gamma A_\gamma \delta_{i \in \mathcal{U}(s(\gamma))} = \sum_\gamma (\mathbb{G}^\top)_{i,s(\gamma)} A_\gamma. \quad (A3)$$

Notice that, for every node $j \neq \text{root}$ there is exactly one cochord $\gamma$ on the spanning tree such that $j = s(\gamma)$. This allows one to express the sum in Eq. (A3) as a integration along the unique path of cochords $\gamma$ connecting node $i$ to the root along the spanning tree. Equations (A2) and (A3) express a one-to-one relation between a set of $M$ forces $A_\gamma$ defined on the cochords, and a set of $N = M + 1$ potentials $V_i$ (including $V_{\text{root}} = 0$) defined on the nodes. These two equations thus encode the differentiation and the integration of a conservative force on a graph: indeed if $A \in \text{Im } \mathbb{S}^\top$ is a 'gradient', the potential $V$ defined through the 'integral' (A3) of $M$ components of $A$ generates the full vector $A$ through $A = -\mathbb{S}^\top V$. As remarked above, such potential $V$ is unique up to a constant, and the condition $V_{\text{root}} = 0$ fixes $V$ uniquely (independently of the choice of the spanning tree).

We now identify an algebraic relation between $\mathbb{G}$ and $\mathbb{S}$. To do so, one defines for every node $j \neq \text{root}$ (i.e. $2 \leq j \leq N$) a unit 'charge' as the potential $\mathcal{V}^j$, a vector of entries $\mathcal{V}^j_i = \delta_{ij}$. The corresponding forces $\mathcal{A}^j$ defined on the cochords from Eq. (A2) have entries

$$\mathcal{A}^j_\gamma = \begin{cases} -1, & \text{if } t(\gamma) = j \\ 1, & \text{if } s(\gamma) = j \\ 0, & \text{otherwise,} \end{cases} \quad \text{i.e.,} \quad \mathcal{A}^j_\gamma = -(\mathbb{S}^\top_M{}')_{\gamma j} \quad (A4)$$

where $\mathbb{S}^\top_M{}'$ is the $M \times M$ matrix constituted of the $M$ last columns of $\mathbb{S}^\top_M$ [the transpose of the matrix defined in Eq. (A1)]. By direct application of Eq. (A3), we see that the unit potential $\mathcal{V}^j$ is obtained from the force $\mathcal{A}^j$ as

$$\mathcal{V}^j_i = \delta_{ij} = \sum_\gamma (\mathbb{G}^\top)_{i,s(\gamma)} \mathcal{A}^j_\gamma. \quad (A5)$$

We now interpret this relation algebraically. Since $s(\gamma) = \gamma + 1$, we define $\mathbb{G}_M$ as the $M \times M$ submatrix of $\mathbb{G}$ deprived from its first line and column (i.e. $\mathbb{G}_M$ is the black submatrix of $\mathbb{G}$ in Fig. 5b). Then the identity (A5) yields, from Eq. (A4)

$$-\mathbb{G}^\top_M \mathbb{S}^\top_M{}' = \mathbb{1}_M. \quad (A6)$$

See Refs. [31, 32] for related results in incidence matrix inversion and Ref. [16] for applications in chemistry.

Before going on, let us take a closer look at this relation. Because $\mathbb{G}^\top_M$ is lower triangular with only 1's on the diagonal, it is invertible and its inverse is given by $-\mathbb{S}^\top_M{}'$. We thus read Eq. (A6) as follows: the subset of $M$ independent reactions between $M$ independent species are described by the lines of matrix $\mathbb{S}^\top_M{}'$, which constitutes an invertible 'core' of the full (and transposed) stoichiometric matrix $\mathbb{S}^\top$. What we have done so far to arrive at Eq. (A6) is to realize that $\mathbb{S}^\top_M{}'$ defines a set of forces on the cochords whose integration along the path from root to any node $j \neq \text{root}$ gives the 'unit charge' potential $\mathcal{V}^j$ defined above – which is quite natural from the graph perspective. This provides an electrostatic picture of the incidence matrix of $\mathcal{G}$. To proceed, one now remarks that Eq. (A6) implies

$$-\mathbb{S}_M^{\top\prime} \mathbb{G}_M^{\top} = \mathbb{1}_M, \qquad (A7)$$

which is algebraically trivial but not obvious from the underlying graph-theoretical viewpoint. Yet, completing the matrices, it implies that

$$-\mathbb{S}_M^{\top}\mathbb{G}^{\top} = -\left((\mathbb{S}_M^{\top})_1 \;\middle|\; \mathbb{S}_M^{\top\prime}\right)\begin{pmatrix}1 & 0 & \cdots & 0 \\ \vdots & & & \\ 1 & & \mathbb{G}_M^{\top} & \end{pmatrix} = \begin{pmatrix} 0 \\ \vdots \\ 0 \end{pmatrix}\mathbb{1}_M \quad (A8)$$

where $(\mathbb{S}_M^{\top})_1$ is the first column of $\mathbb{S}_M^{\top}$ and where we used the fact that $\ell_0^{\top}$ is a right nullvector of $\mathbb{S}^{\top}$ (and thus of $\mathbb{S}_M^{\top}$). Transposing this relation and using the structure of $\mathbb{S}$ given by Eq. (A1) with $\mathbb{S}_{\text{dep}} = \mathbb{S}_M \mathbb{T}$ we obtain

$$-\mathbb{G}\mathbb{S} = \begin{pmatrix} \mathbf{0} & \mathbf{0} \\ \mathbb{1}_M & \mathbb{T} \end{pmatrix}, \qquad (A9)$$

where the first line is full of zeroes. This is Eq. (21) announced in the main text. Physically, it encodes the fact that $\mathbb{S}^{\top}$, seen as a gradient operator, can be inverted on the cochords by the matrix $-\mathbb{G}^{\top}$; and that, if cochord forces are conservative, $A_\gamma = -(\mathbb{S}^{\top}V)_\gamma = -(\mathbb{S}_M^{\top}V)_\gamma$, then the corresponding chord forces write as

$$A_\alpha = -(\mathbb{S}^{\top}V)_\alpha = -(\mathbb{T}^{\top}\mathbb{S}_M^{\top}V)_\alpha = \sum_\gamma \mathbb{T}_{\gamma\alpha} A_\gamma \qquad (A10)$$

i.e., they are expressed as a linear combination of the $A_\gamma$'s. Mathematically, Eq. (A9) encodes the row-reduction of $\mathbb{S}$ in an echelon form (see e.g. Ref. [49]). Such identity is at the basis of our geometrical analysis of complex CRNs in Sec. III, and of the algebraic analysis presented in the next paragraph.

### 2. Interacting CRNs: integration on multi-paths

To generalize the construction presented in the previous paragraph, we now follow a complementary path. The matrix $\mathbb{S}^{\top}$ can still be understood as a (weighted) discrete gradient: every line $\rho$ of $\mathbb{S}^{\top}$ tells for a given reaction $\rho$ how the products (resp. reactants) contribute positively (resp. negatively) to the affinity $A_\rho$. Our aim is to explain how to 'invert' that gradient and to define an integration that allows one to reconstitute explicitly a potential $V$ such that $A = -\mathbb{S}^{\top}V$ if $A \in \text{Im}\,\mathbb{S}^{\top}$ is a conservative affinity. For complex CRNs and their corresponding hypergraph (see Sec. III and Fig. 1), there exists no such notion as a spanning tree, and the topological construction of the previous paragraph (which consists in integrating from a chosen root to a node along the spanning tree) cannot be generalized.

Here we follow a mirror procedure, starting from algebra to build a geometrical picture. Denoting by $M$ the rank of $\mathbb{S}$, we re-order the reactions and species so that the first $M$ reactions are independent, and the last $M$ species are independent.

Namely, the first $M$ columns of $\mathbb{S}$ are linearly independent, and the same holds for its last $M$ lines. The row-reduction of $\mathbb{S}$ in echelon form (see e.g. Ref. [49]) ensures that there exists an invertible $N \times N$ matrix $\mathbb{G}$ such that

$$-\mathbb{G}\mathbb{S} = \begin{pmatrix} \mathbb{0} & 0 \\ \mathbb{1}_M & \mathbb{T} \end{pmatrix} \begin{matrix} \updownarrow N-M \\ \updownarrow M \end{matrix} \qquad (A11)$$
$$\underbrace{\phantom{xxxxxx}}_{R-M}$$

(notice that the ordering conventions we are using makes that the lines of 0's are placed first compared to the canonical row reduction). The matrix $\mathbb{G}$ is not unique and its entries can be found by the Gauss–Jordan elimination procedure through elementary line and columns operations [49]. This ensures that the entries of $\mathbb{G}$ can be taken as rational when, as in our case, $\mathbb{S}$ has integer entries.

To learn about geometry, it is convenient not to rely on Gauss–Jordan elimination and instead to build the matrix $\mathbb{G}$ in an explicit manner. We first fix a basis of $\text{Ker}\,\mathbb{S}^{\top}$ as $N - M$ column vectors representing conservation laws (whose entries are taken as rational). Then, postulating the following form:

$$\mathbb{G}^{\top} = \left(\begin{array}{c|c} \text{csv} & \begin{matrix} 0 \cdots\cdots 0 \\ 0 \cdots\cdots 0 \end{matrix} \\ \text{laws} & \begin{matrix} \text{escape routes} \\ (\mathbb{G}_M^{\top\prime}) \end{matrix} \end{array}\right) \begin{matrix} \updownarrow N-M \\ \updownarrow M \end{matrix} \qquad (A12)$$
$$\underbrace{\phantom{xxx}}_{\mathbb{G}_M^{\top}}$$

we want to show that there exists a $M \times M$ matrix $\mathbb{G}_M^{\top\prime}$ which ensures that the following key relation is satisfied:

$$-\mathbb{S}^{\top}\mathbb{G}^{\top} = \begin{pmatrix} \mathbb{0} & \mathbb{1}_M \\ 0 & \mathbb{T}^{\top} \end{pmatrix}. \qquad (A13)$$

To do so, we notice that the $N - M$ columns of conservation laws in Eq. (A12) ensure the $N - M$ first columns of 0's in Eq. (A13). Then, by our ordering conventions, the first $M$ columns of $\mathbb{S}$ are independent and form an $N \times M$ matrix $\mathbb{S}_M$, so that we can organize $\mathbb{S}^{\top}$ as

$$\mathbb{S}^{\top} = \left(\begin{array}{c} \mathbb{S}_M^{\top} \\ \hline \mathbb{S}_{\text{dep}}^{\top} \end{array}\right) \begin{matrix} \updownarrow M \\ \updownarrow R-M \end{matrix} . \qquad (A14)$$

There, $\mathbb{S}_{\text{dep}}$ are the $R - M$ last columns of $\mathbb{S}$, which depend on the first $M$ ones; this means that they can be expressed as a linear combination of those, i.e. that there exists a $M \times (R - M)$ matrix $\mathbb{T}$ such that $\mathbb{S}_{\text{dep}} = \mathbb{S}_M \mathbb{T}$. With this property, we see that the proof of Eq. (A13) reduces to showing that

$$-\mathbb{S}_M^{\top\prime}\mathbb{G}_M^{\top\prime} = \mathbb{1}_M \qquad (A15)$$

where $\mathbb{S}_M^{\top\prime}$ is the $M \times M$ matrix consisting of the last $M$ columns of $\mathbb{S}_M^{\top}$, and $\mathbb{G}_M^{\top\prime}$ is the $M \times M$ matrix consisting of the last $M$ lines of $\mathbb{G}_M^\top$ [defined in Eq. (A12)]. Physically, $\mathbb{S}_M^{\top\prime}$ represents a 'core' set of $M$ independent reactions between $M$ independent species. Crucially, it is an invertible matrix, since the last $M$ species are independent[10]. This implies that one can *define* $\mathbb{G}_M^{\top\prime}$ as the inverse of $\mathbb{S}_M^{\top\prime}$, which ensures the relation (A15) to be satisfied. Since $\mathbb{S}_M^{\top\prime}$ has integer entries, we obtain that $\mathbb{G}$ has rational entries (as is also the case when defining $\mathbb{G}$ through Gauss–Jordan elimination).

As we just described, this shows that the form of $\mathbb{G}^\top$ given in Eq. (A12) allows for the row-reduction of $\mathbb{S}$ as in Eq. (A11), with the 'escape routes' in Eq. (A12) being precisely given by the $M \times M$ matrix $\mathbb{G}_M^{\top\prime}$ defined from Eq. (A15). Before showing that the elements of that matrix play the geometrical role of escape routes indeed, it remains to prove that the row-reducing matrix $\mathbb{G}$ defined in Eq. (A12) is invertible. This is done by exhibiting its inverse: one checks with Eqs. (A14) and (A15) that

$$\mathbb{G}^{-1} = \begin{pmatrix} \mathbb{1}_{N-M} & \mathbb{S}_M \\ 0 & \end{pmatrix} \quad (A16)$$

is the inverse of $\mathbb{G}$, provided the conservation laws in Eq. (A12) are organized (as columns) as:

$$\begin{pmatrix} \text{csv} \\ \text{laws} \end{pmatrix} = \begin{pmatrix} \mathbb{1}_{N-M} \\ -\mathbb{U}^\top \end{pmatrix}. \quad (A17)$$

Up to now, the specific choice of basis for the conservation laws was left undetermined, and this form fixes it. Its existence is shown *ad absurdum*.

*Proof*: Consider an arbitrary choice of basis for the $N-M$ conservation laws, and split it as follows:

$$\begin{pmatrix} \text{csv} \\ \text{laws} \end{pmatrix} = \begin{pmatrix} \mathbb{C}_1 \\ \mathbb{C}_2 \end{pmatrix} \begin{matrix} \updownarrow N-M \\ \updownarrow M \end{matrix}. \quad (A18)$$

Correspondingly, split the $N$ lines of the matrix $\mathbb{S}$ as

$$\mathbb{S} = \begin{pmatrix} \mathbb{S}_1 \\ \mathbb{S}_2 \end{pmatrix} \begin{matrix} \updownarrow N-M \\ \updownarrow M \end{matrix}, \quad (A19)$$

where by hypothesis the $M$ lines of $\mathbb{S}_2$ are independent while the $N-M$ lines of $\mathbb{S}_1$ depend on those of $\mathbb{S}_2$, meaning that

---

[10] The proof goes as follows: if $\mathbb{S}_M^{\top\prime}$ is not invertible, there exists a vector $z \neq 0$ such that $\mathbb{S}_M^{\top\prime} z = 0$. Then, defining $\check{z} \neq 0$ as $N-M$ lines of 0 followed by the $M$ components of $z$, one has $\mathbb{S}_M^\top \check{z} = 0$ and also $\mathbb{S}_{\text{dep}}^\top \check{z} = \mathbb{T}^\top \mathbb{S}_M^\top \check{z} = 0$. From Eq. (A14), this implies $\mathbb{S}^\top \check{z} = 0$, but this is absurd, since this represents a linear dependency between the last $M$ columns of $\mathbb{S}^\top$ (i.e. between the last $M$ species) – which are independent.

there exists a $(N-M) \times M$ matrix $\mathbb{U}$ such that

$$\mathbb{S}_1 = \mathbb{U}\mathbb{S}_2. \quad (A20)$$

This identity and the decompositions above imply from the definition of conservation laws (they span $\text{Ker}\,\mathbb{S}^\top$) that

$$\mathbb{S}_2^\top \mathbb{U}^\top \mathbb{C}_1 + \mathbb{S}_2^\top \mathbb{C}_2 = 0. \quad (A21)$$

Let us now show that $\mathbb{C}_1$ is invertible. *Ad absurdum*, if this is not the case, there exists a vector $x \neq 0$ such that $\mathbb{C}_1 x = 0$. From Eq. (A21) this implies $\mathbb{S}_2^\top \mathbb{C}_2 x = 0$, and since the $M$ columns of $\mathbb{S}_2^\top$ are independent, we have also $\mathbb{C}_2 x = 0$. From Eq. (A18), we then read

$$\begin{pmatrix} \text{csv} \\ \text{laws} \end{pmatrix} x = 0 \quad (A22)$$

which is absurd, since the column vectors of the matrix of conservation laws are independent. Hence $\mathbb{C}_1$ is an invertible matrix. Multiplying Eq. (A18) by $\mathbb{C}_1^{-1}$ on the right, we see that the conservation laws can be organized as in Eq. (A17), as announced. [Notice that matrix $\mathbb{U}$ in Eqs. (A17) and (A20) are the same, as seen from Eq. (A21).] □

To summarize, the stoichiometric matrix can be row-reduced in echelon form as in Eq. (A11), with an invertible matrix $\mathbb{G}$ taking the form Eq. (A12) and whose inverse takes the explicit form Eq. (A16) provided the columns of conservation laws in Eq. (A12) are organized as in Eq. (A17). We now depict how these algebraic results can be translated in geometrical terms.

The interpretation of Fig. 5 of the matrix $\mathbb{G}$ for graphs can be generalized to hypergraphs, without relying on the notion of spanning tree. To do so, one defines an 'escape' protocol as follows. The $N-M$ dependent species are labeled as 'roots'. For each of the $M$ independent species, we place a unit charge in its corresponding node $i$, and we ask how many times each of the $M$ independent reactions have to be applied (possibly a fractional and/or negative number of times) in order to expel completely the charge from $i$ through the set of roots. How each reaction acts on the charges is governed by the stoichiometry of Eq. (2). Because the matrix $\mathbb{S}_M^{\top\prime}$ precisely represents the action of the independent reactions on the independent species, we see that Eq. (A15), rewritten as

$$-\mathbb{G}_M^{\top\prime} \mathbb{S}_M^{\top\prime} = \mathbb{1}_M, \quad (A23)$$

tells that the *line* entries of $\mathbb{G}_M^{\top\prime}$ precisely solve the escape problem. Indeed, since $\sum_k (\mathbb{G}_M^{\top\prime})_{ik} (\mathbb{S}_M^{\top\prime})_{kj} = -\delta_{ij}$ we see that applying each independent reaction $k$ (with $1 \leq k \leq M$) a number $(\mathbb{G}_M^{\top\prime})_{ik}$ of times will expel a unit charge from the node $i$, while leaving empty the rest of the nodes $\notin \{\text{root}\}$ (see Fig. 8). We thus see how the algebraically trivial passage from Eq. (A15) to Eq. (A23) allows one to build a geometrical interpretation of the lines of the row-reducing matrix $\mathbb{G}^\top$. In Sec. III B 3, we explain how this leads to a generalization of the notion of cocycle from graphs to hypergraphs.

There also exists a geometrical interpretation of Eq. (A15). Here instead, one places a unit force on a cochord (i.e. independent reaction) $\gamma$ and one asks which charge have to be set on the independent nodes (which are not roots) so as to produce this force. The solution is now given by the *columns* of $\mathbb{G}_M^\top{}'$. Indeed, we read from $\sum_k (\mathbb{S}_M^\top{}')_{\gamma k}(\mathbb{G}_M^\top{}')_{k\gamma'} = -\delta_{\gamma\gamma'}$ that column $\gamma$ of $\mathbb{G}_M^\top{}'$ gives the set of charges on the set of independent species $k$ that generates a unit force on the cochord $\gamma$ (and 0 on the other cochords $\gamma' \neq \gamma$). Notice that on the chords, the generated force is not necessarily equal to 0 (it is in fact given by the entries of $\mathbb{T}$ which is used to define cocycles, see Sec. III B 3). For every $\gamma$, this set of charges can be seen as the 'elevation map' of an 'island' associated to $c^\gamma$. Such elevation map is a potential landscape that generates a force given by the entries of cocycle $c^\gamma$ (see Fig. 9 for an example).

As a last remark, we explain how, following an argument similar to that leading to the form of Eq. (A17) for the conservation laws, one can find a basis of the chemical cycles (that is, of Ker $\mathbb{S}$) such that

$$\begin{pmatrix} \text{cycles} \end{pmatrix} = \begin{pmatrix} -\mathbb{T} \\ \mathbb{1}_{R-M} \end{pmatrix}, \tag{A24}$$

where $\mathbb{T}$ is the matrix that expresses the dependency $\mathbb{S}_{\text{dep}} = \mathbb{S}_M \mathbb{T}$ of the dependent reactions of $\mathbb{S}_{\text{dep}}$ as a function of the independent ones of $\mathbb{S}_M$ (see the decomposition in Eq. (A14)). *Proof*: We start from a basis of Ker $\mathbb{S}$ composed of $R - M$ column vectors written as

$$\begin{pmatrix} \mathfrak{C}_2 \\ \mathfrak{C}_1 \end{pmatrix} \updownarrow {}^M_{R-M}. \tag{A25}$$

The definition of cycles tells that $\mathbb{S}_M(\mathbb{T}\mathfrak{C}_1 + \mathfrak{C}_2) = 0$. The same *ad absurdum* argument as above tells that $\mathfrak{C}_1$ is invertible so that we can multiply the basis (A25) on the right by $\mathfrak{C}_1^{-1}$, and still keep a matrix whose columns are a basis for the cycles. It takes the announced form in Eq. (A24) since the relation above implies $\mathbb{T}\mathfrak{C}_1 + \mathfrak{C}_2 = 0$, as the columns of $\mathbb{S}_M$ are independent □

This proof is the algebraic counterpart, for generic CRNs, of the property that the same matrix $\mathbb{T}$ controls the dependency between reactions and the organization of cycles – a property we also obtained in Sec. II B from graph theory for unimolecular reactions.

## Appendix B: Reversibility and the Wegscheider–Kolmogorov condition

Consider the set of $R$ complex reactions (2) between $N$ species $X$, described by the $N \times R$ stoichiometric matrix $\mathbb{S}$. These reactions describe, at the microscopic level, a stochastic population process on the numbers $n$ of each chemical species $X$; and at the macroscopic level, a continuity Equation (5) for the evolution of the concentrations $x(t)$, which involves the current $J$ of Eq. (4), expressed in terms of the affinities defined in Eq. (6). In this Appendix, we present in a unified manner the equivalence between the so-called Wegscheider condition [52] [equivalent to Kolmogorov's criterion [92] in the language of Markov chains (see e.g. [93])] and varied notions of reversibility, both at the microscopic population level and the macroscopic concentrations level.

We stress that the rate constants $k_\rho^\pm$ of the reactions (2) are *macroscopic* in the sense that they enter in the deterministic description Eq. (5) of the real-valued concentrations $x$, and do not depend on the system's size. They differ from the microscopic rates of the individual reactions which, at the molecular level, scale with the system's volume $\Omega$ as

$$\kappa_\rho^\pm = \Omega \frac{k_\rho^\pm}{\Omega^{\nu^{\pm\rho}}} \qquad \forall \rho, \tag{B1}$$

where $\Omega^{\nu^{\pm\rho}} = \Omega^{\sum_i \nu_i^{\pm\rho}}$. This describes that, at fixed number of molecules, reactions involving (the collision of) several species are rarer as $\Omega$ gets larger (see e.g. §7.5.3 in [94]). Notice that when we discuss the population dynamics stochastic process, the vector $x$ denotes the rational-valued vector $n/\Omega$ representing the discrete concentrations of species. See also Appendix C for a discussion on the large-$\Omega$ asymptotics.

We recall (see Sec. III) that we can fix a basis of $R - M$ cycles $c^\alpha$ which span the right nullspace Ker $\mathbb{S}$ of $\mathbb{S}$, of dimension $R - M$. The following properties are equivalent:

I. Wegscheider's condition:

$$\forall \alpha, \quad \prod_\rho \left( \frac{k_\rho^+}{k_\rho^-} \right)^{c_\rho^\alpha} = 1, \tag{B2}$$

i.e. the product of macroscopic transition rates of every cycle is the same in both directions along the cycle.

I'. Kolmogorov's condition:

$$\forall \alpha, \quad \prod_\rho \left( \frac{\kappa_\rho^+}{\kappa_\rho^-} \right)^{c_\rho^\alpha} = 1, \tag{B3}$$

i.e. the product of microscopic transition rates of every cycle is the same in both directions along the cycle.

II. Existence of the standard chemical potential $\boldsymbol{\mu}^\ominus$:

$$\exists \boldsymbol{\mu}^\ominus: \quad \forall \rho, \; \frac{k_\rho^+}{k_\rho^-} = \exp\left[-\left(\mathbb{S}^\top \boldsymbol{\mu}^\ominus\right)_\rho\right]. \tag{B4}$$

For noninteracting CRNs, this is detailed balance, see Eq. (39).

III. Existence of concentrations canceling affinities:
$$\exists\, x^{\text{eq}}: \quad \forall \rho,\, A_\rho(x^{\text{eq}}) = 0\,. \tag{B5}$$

IV. Existence of concentrations canceling currents:
$$\exists\, x^{\text{eq}}: \quad \forall \rho,\, J_\rho(x^{\text{eq}}) = 0\,. \tag{B6}$$

V. Reversible constrained product Poisson law at the population level:
$$\exists\, x^{\text{eq}}: \quad |\mathcal{P}^{\text{eq}}\rangle \propto \sum_{\boldsymbol{n}} \frac{(\Omega x^{\text{eq}})^{\boldsymbol{n}}}{\boldsymbol{n}!}\, \delta(\boldsymbol{\ell}(\boldsymbol{n}) - \mathcal{L})\,|\boldsymbol{n}\rangle \tag{B7}$$

is an equilibrium steady state of the microscopic dynamics of occupation numbers. We use the Doi–Peliti ket notation $|\cdot\rangle$ for occupation states (see the proof, Sec. B). The components of $\boldsymbol{\ell}(\boldsymbol{n})$ are the conserved quantities, their values being the components of $\mathcal{L}$ (fixed by initial condition). Vector notations are used ($\boldsymbol{n}! = \prod_{i=1}^{N} n_i!$, etc.).

VI. Microscopic reversibility: the stochastic dynamics of occupation numbers $\boldsymbol{n}$ verifies detailed balance.

VII. Gradient condition on affinities:
$$\forall t,\, \boldsymbol{A}(x(t)) \in \operatorname{Im} \mathbb{S}^\top\,. \tag{B8}$$

*Remark*: in the proofs, we will often make use of the following writings of the affinity of a reaction $\rho$, that comes from Eq. (6):
$$A_\rho(x) = \log \frac{k_\rho^+}{k_\rho^-} - \overbrace{(\mathbb{S}^\top \log x)_\rho}^{\in \operatorname{Im} \mathbb{S}^\top}\,. \tag{B9}$$

#### 1. Proof of I⇔I'

For every cycle $c^\alpha$, we have
$$\prod_\rho \left(\frac{\kappa_\rho^+}{\kappa_\rho^-}\right)^{c_\rho^\alpha} = \prod_\rho \left(\frac{k_\rho^+}{k_\rho^-}\right)^{c_\rho^\alpha} \Omega^{(\boldsymbol{\nu}^{-\rho} - \boldsymbol{\nu}^{+\rho}) c_\rho^\alpha} = \prod_\rho \left(\frac{k_\rho^+}{k_\rho^-}\right)^{c_\rho^\alpha} \tag{B10}$$

since $\prod_\rho \Omega^{(\boldsymbol{\nu}^{-\rho} - \boldsymbol{\nu}^{+\rho}) c_\rho^\alpha} = \Omega^{\sum_{i\rho} \mathbb{S}_{i\rho} c_\rho^\alpha} = \Omega^{\sum_i (\mathbb{S} c^\alpha)_i}$ and $\mathbb{S} c^\alpha = 0$ by definition. The conditions (B2) and (B3) are thus the same $\square$

#### 2. Proof of I⇔II

If II holds, then for any cycle $c^\alpha$, since $\mathbb{S} c^\alpha = 0$, one has
$$\prod_\rho \left(\frac{k_\rho^+}{k_\rho^-}\right)^{c_\rho^\alpha} = \exp\left(-c^\alpha \cdot \mathbb{S}^\top \boldsymbol{\mu}^\ominus\right) = \exp\left(-\boldsymbol{\mu}^\ominus \cdot \mathbb{S} c^\alpha\right) = 1 \tag{B11}$$

which yields I. Conversely, if I holds, for any cycle $c \in \operatorname{Ker} \mathbb{S}$ we have
$$\boldsymbol{A} \cdot c = \underbrace{\log \prod_\rho \left(\frac{k_\rho^+}{k_\rho^-}\right)^{c_\rho}}_{=0 \text{ from Eq. (B2)}} + \underbrace{\log x^{-\mathbb{S}c}}_{=0} = 0\,, \tag{B12}$$

which implies that $\boldsymbol{A} \in (\operatorname{Ker} \mathbb{S})^\perp = \operatorname{Im} \mathbb{S}^\top$; combining then with Eq. (B9) we obtain $\log \frac{k^+}{k^-} \in \operatorname{Im} \mathbb{S}^\top$, which is precisely Eq. (B4) $\square$

In practice, if the rates verify Wegscheider's condition (B2), identifying a standard corresponding chemical potential $\boldsymbol{\mu}^\ominus$ can be done using the hypergraph integration procedure described in Sec. III B 2.

#### 3. Proof of II⇔III

If II holds, then
$$\forall \rho,\, A_\rho(x^{\text{eq}}) = 0$$
$$\Leftrightarrow \forall \rho,\, (\mathbb{S}^\top \log x)_\rho = \log \frac{k_\rho^+}{k_\rho^-} \stackrel{\text{(B4)}}{=} -(\mathbb{S}^\top \boldsymbol{\mu}^\ominus)_\rho \tag{B13}$$
$$\Leftrightarrow \boldsymbol{\mu}^\ominus - \log x^{\text{eq}} \in \operatorname{Ker} \mathbb{S}^\top \tag{B14}$$

but $\operatorname{Ker} \mathbb{S}^\top$ is never an empty set, so that we can find $x^{\text{eq}}$ canceling all affinities, which is III. Conversely, if III holds, Eq. (B9) implies that $\log \frac{k^+}{k^-} \in \operatorname{Im} \mathbb{S}^\top$, which is precisely II $\square$

#### 4. Proof of III⇔IV

It is obvious from the expression (4) of the currents as function of the affinities. Notice that interestingly, this means that for complex CRNs, stochastic reversibility is equivalent to the existence of a fixed point with zero macroscopic current for the deterministic dynamics of Eq. (5).

#### 5. Proof of III⇔V

We use Fock space notations for occupation vectors $|\boldsymbol{n}\rangle$ and the Doi–Peliti operators [95, 96] to represent the reactions at the microscopic level of occupation numbers (see [97–99] for reviews). We attach an annihilation operator $a_i$ and a creation one $a_i^\dagger$ to every species $i$. For a single species, they act as $a|n\rangle = n|n-1\rangle$, $a^\dagger|n\rangle = |n+1\rangle$, while for several species they only act on their attached species. The number operator $\hat{n}_i = a_i^\dagger a_i$ is diagonal and $\hat{n}_i|\boldsymbol{n}\rangle = n_i|\boldsymbol{n}\rangle$. The action of the creation/annihilation operators on (arbitrarily normalized) unconstrained Poisson laws is well known and easily checked:
$$a_i \sum_{\boldsymbol{n}} \frac{x^{\boldsymbol{n}}}{\boldsymbol{n}!} |\boldsymbol{n}\rangle = x_i \sum_{\boldsymbol{n}} \frac{x^{\boldsymbol{n}}}{\boldsymbol{n}!} |\boldsymbol{n}\rangle \tag{B15}$$
$$a_i^\dagger \sum_{\boldsymbol{n}} \frac{x^{\boldsymbol{n}}}{\boldsymbol{n}!} |\boldsymbol{n}\rangle = \frac{\hat{n}_i}{x_i} \sum_{\boldsymbol{n}} \frac{x^{\boldsymbol{n}}}{\boldsymbol{n}!} |\boldsymbol{n}\rangle\,. \tag{B16}$$

When constraints are present inside the Poisson law as on the r.h.s. of Eq. (B7), similar replacement rules $a_i \mapsto x_i$ and $a_i^\dagger \mapsto \frac{\hat{n}_i}{x_i}$ hold as in Eqs. (B15)-(B16), provided the operators on the l.h.s. leave the conserved quantities $\boldsymbol{\ell}(\boldsymbol{n})$ unchanged.

In the Doi–Peliti approach, the Markov dynamics in the population space is represented as a linear operator $\mathbb{W}$ acting on the probability vector $|P(t)\rangle = \sum_{\boldsymbol{n}} P(\boldsymbol{n}, t)|\boldsymbol{n}\rangle$. For the

$R$ reactions of the form (2), we decompose $\mathbb{W} = \sum_\rho \mathbb{W}_\rho$ with

$$\mathbb{W}_\rho = \kappa_\rho^+\left[(a^\dagger)^{\nu^{-\rho}} a^{\nu^{+\rho}} - \hat{n}^{\nu^{+\rho}}\right] + \kappa_\rho^-\left[(a^\dagger)^{\nu^{+\rho}} a^{\nu^{-\rho}} - \hat{n}^{\nu^{-\rho}}\right] \quad (B17)$$

where the microscopic rates $\kappa_\rho^\pm$ are defined in Eq. (B1). Every reaction respects the conservation laws of $\mathbb{S}$ so that one can apply the replacement rules mentioned above to compute the action of $\mathbb{W}_\rho$ on the vector $|\mathcal{P}^{\text{eq}}\rangle$ defined in Eq. (B7). It's a matter of simple algebra, using the definition of affinity of Eqs. (4)-(6), to find

$$\mathbb{W}_\rho|\mathcal{P}^{\text{eq}}\rangle \qquad (B18)$$
$$= -\Omega\left[k_\rho^-\left(1-e^{A_\rho}\right)\left(\frac{\hat{n}}{\Omega}\right)^{\nu^{-\rho}} + k_\rho^+\left(1-e^{-A_\rho}\right)\left(\frac{\hat{n}}{\Omega}\right)^{\nu^{+\rho}}\right]|\mathcal{P}^{\text{eq}}\rangle$$

where $A_\rho = A_\rho(x^{\text{eq}})$. Let us now come to the proof of the equivalence III⇔V. If III holds, then we have the existence of a vector of concentrations $x^{\text{eq}}$ which cancels every affinity, see Eq. (B5). From the identity (B18), we find that $\mathbb{W}_\rho|\mathcal{P}^{\text{eq}}\rangle = 0$, where $|\mathcal{P}^{\text{eq}}\rangle$ defined in Eq. (B7) is evaluated on the $x^{\text{eq}}$ we just found (whose components are thus promoted from being average concentrations to being a parameters of a constrained product Poisson law). This proves that $|\mathcal{P}^{\text{eq}}\rangle$ is a steady state of $\mathbb{W}$. To check explicitly that it verifies detailed balance, one introduces a diagonal operator $\hat{\mathcal{P}}^{\text{eq}}$ whose components along the diagonal are those of the vector $|\mathcal{P}^{\text{eq}}\rangle$ in Eq. (B7). Detailed balance is then equivalent to checking that $\mathbb{W}\hat{\mathcal{P}}^{\text{eq}} = \hat{\mathcal{P}}^{\text{eq}}\mathbb{W}^\top$. Using the identities $a_i\hat{\mathcal{P}}^{\text{eq}} = \Omega x_i^{\text{eq}}\hat{\mathcal{P}}^{\text{eq}}(a_i^\dagger)^\top$ and $a_i^\dagger\hat{\mathcal{P}}^{\text{eq}} = (\Omega x_i^{\text{eq}})^{-1}\hat{\mathcal{P}}^{\text{eq}}a_i^\top$ [11], one finds

$$\mathbb{W}_\rho\hat{\mathcal{P}}^{\text{eq}} - \hat{\mathcal{P}}^{\text{eq}}\mathbb{W}_\rho^\top \qquad (B19)$$
$$= \hat{\mathcal{P}}^{\text{eq}}\left\{\kappa_\rho^-\left(e^{A_\rho}-1\right)(a^{\dagger\nu^{+\rho}}a^{\nu^{-\rho}})^\top + \kappa_\rho^+\left(e^{-A_\rho}-1\right)(a^{\dagger\nu^{-\rho}}a^{\nu^{+\rho}})^\top\right\}$$

where $A_\rho$ denotes $A_\rho(x^{\text{eq}})$. If III holds, then from Eq. (B5) we obtain $\mathbb{W}_\rho\hat{\mathcal{P}}^{\text{eq}} = \hat{\mathcal{P}}^{\text{eq}}\mathbb{W}_\rho^\top$ ($\forall\rho$); hence, summing over $\rho$, detailed balance indeed holds. Conversely, if V holds, there exists a vector $x^{\text{eq}}$ such that $\mathbb{W}\hat{\mathcal{P}}^{\text{eq}} = \hat{\mathcal{P}}^{\text{eq}}\mathbb{W}^\top$ and Eq. (B19) yields

$$\sum_\rho\left\{\kappa_\rho^-\left(e^{A_\rho(x^{\text{eq}})}-1\right)a^{\dagger\nu^{+\rho}}a^{\nu^{-\rho}}\right.$$
$$\left. + \kappa_\rho^+\left(e^{-A_\rho(x^{\text{eq}})}-1\right)a^{\dagger\nu^{-\rho}}a^{\nu^{+\rho}}\right\} = 0, \qquad (B20)$$

since $\hat{\mathcal{P}}^{\text{eq}}$ is an invertible operator. Consider now a given reaction $\rho$, applied in the direction where it transforms $|n\rangle$ into $|n-\nu^{+\rho}+\nu^{-\rho}\rangle$. Since by hypothesis reaction $\rho$ is the only one performing that transformation, we obtain by taking the scalar product of Eq. (B20) between $\langle n-\nu^{+\rho}+\nu^{-\rho}|$ and $|n\rangle$ that $e^{-A_\rho(x^{\text{eq}})}-1 = 0$. We thus see that necessarily $A_\rho(x^{\text{eq}}) = 0$ ($\forall\rho$) which is precisely III □

### 6. Proof of V⇔VI

Obviously, V implies VI. Conversely, if the microscopic occupation-number dynamics verifies detailed balance, let us show that Kolmogorov's condition (B3) is verified, which will prove that I' and hence V holds (as we already showed). Consider a basis of cycles $c^\alpha$ of the stoichiometric matrix $\mathbb{S}$. They can be taken to have (positive or negative) integer entries. Then, a given $c^\alpha$ corresponds to a succession of reactions, where each reaction $\rho$ is used $c_\rho^\alpha$ times. In fact, depending on the precise order in which these reactions are applied, the algebraic cycle $c^\alpha$ corresponds to many possible cycles at the population level. We now consider a given $c^\alpha$ and choose arbitrary such an ordering. It leaves any configuration of the occupations invariant (and the same is true if the cycle is applied in reverse order). Detailed balance at the level of occupations implies that the product of transition rates of the cycle and its reverse are the same, at the level of population rates. Using the Doi–Peliti formalism, we express such product of rates, starting from configuration $n$, as follows:

$$\overleftarrow{\prod_\rho}(\kappa_\rho^+)^{c_\rho^\alpha}\langle n|(a^\dagger)^{c_\rho^\alpha\nu^{-\rho}}a^{c_\rho^\alpha\nu^{+\rho}}|n\rangle. \qquad (B21)$$

---

[11] These identities are verified by direct computation. We stress that in the canonical scalar product, the transpose operator does not allow one to switch between $a$ and $a^\dagger$. In fact from $\langle n|(a^\dagger)^\top = \langle n+1|$ and $\langle n|a^\top = \langle n-1|n$, one sees that $a^\top = \hat{n} a^\dagger$ and $(a^\dagger)^\top = a\frac{1}{\hat{n}}$, where $\hat{n}$ is the number operator $\hat{n}|n\rangle = n|n\rangle$.

The rate for reaction $\rho$ is raised to the power $c_\rho^\alpha$, and the arrow on the product sign indicates that the operators of the first reaction (in the considered ordering) are placed right of the ones of the next reaction, down to the last reaction involved. For the reversed reaction, the product of rates is

$$\overrightarrow{\prod_\rho}(\kappa_\rho^-)^{c_\rho^\alpha}\langle n|(a^\dagger)^{c_\rho^\alpha\nu^{+\rho}}a^{c_\rho^\alpha\nu^{-\rho}}|n\rangle. \qquad (B22)$$

Noticing now the identity $\langle n|(a^\dagger)^{c_\rho^\alpha\nu^{-\rho}}a^{c_\rho^\alpha\nu^{+\rho}}|n\rangle = \langle n|(a^\dagger)^{c_\rho^\alpha\nu^{+\rho}}a^{c_\rho^\alpha\nu^{-\rho}}|n\rangle$ [12], the equality of Eqs. (B21) and (B22) yields that for all $\alpha$, $\prod_\rho(\kappa_\rho^+)^{c_\rho^\alpha} = \prod_\rho(\kappa_\rho^-)^{c_\rho^\alpha}$ which is the announced Eq. (B3) □

### 7. Proof of I⇔VII

Every implication in I⇒VII⇒II is immediate, using the identity (B9). Then we already showed that II⇒I; so that finally we have both I⇒VII and VII⇒I □

Notice that, interestingly, in the implication VII⇒II inferred from Eq. (B9), we deduce a property valid independently of $x(t)$ (namely, $\log \frac{k^+}{k^-} \in \text{Im}\,\mathbb{S}^\top$) from a property depending on $x(t)$ (namely, $A(x(t)) \in \text{Im}\,\mathbb{S}^\top$).

## Appendix C: Effective Fokker–Planck and Langevin dynamics close to an equilibrium point

Consider an arbitrary function $f(n)$ of the population state, i.e. the number of particles $n$ for each chemical species $X$.

---

[12] This identity is shown by taking the transpose of $(a^\dagger)^{c_\rho^\alpha\nu^{-\rho}}a^{c_\rho^\alpha\nu^{+\rho}}$ and performing the similarity transformations $(\hat{n}!)^{-1}a_i\hat{n}! = a_i\hat{n}_i$ and $(\hat{n}!)^{-1}a_i^\dagger\hat{n}! = (\hat{n}_i)^{-1}a_i^\dagger$ for every species $i$ involved.

The master equation on the probability distribution $P(\boldsymbol{n},t)$ in the population space, for the chemical reactions (2), is equivalent to the following evolution equation for the average $\langle f \rangle = \sum_{\boldsymbol{n}} P(\boldsymbol{n},t) f(\boldsymbol{n})$:

$$\partial_t \langle f \rangle = \sum_{\boldsymbol{n},\rho} f(\boldsymbol{n}) \Big\{ W_\rho^+(\boldsymbol{n} - \mathbb{S}_\rho) P(\boldsymbol{n} - \mathbb{S}_\rho, t) - W_\rho^+(\boldsymbol{n}) P(\boldsymbol{n},t) \\ + W_\rho^-(\boldsymbol{n} + \mathbb{S}_\rho) P(\boldsymbol{n} + \mathbb{S}_\rho, t) - W_\rho^-(\boldsymbol{n}) P(\boldsymbol{n},t) \Big\} \quad \text{(C1)}$$

$$= \sum_{\boldsymbol{n},\rho} P(\boldsymbol{n},t) \Big\{ \big[ f(\boldsymbol{n} + \mathbb{S}_\rho) - f(\boldsymbol{n}) \big] W_\rho^+(\boldsymbol{n}) \\ + \big[ f(\boldsymbol{n} - \mathbb{S}_\rho) - f(\boldsymbol{n}) \big] W_\rho^-(\boldsymbol{n}) \Big\} \quad \text{(C2)}$$

where $\mathbb{S}_\rho$ designates the column vector of $\mathbb{S}$ describing reaction $\rho$, and $W_\rho^\pm(\boldsymbol{n}) = W(\{n_i \mapsto n_i \pm \mathbb{S}_{i\rho}\})$ are the transition rates at the species population level. In full generality, the transition rates are given by the product of the 'molecular' reaction rates $\kappa_\rho^\pm$ and the number of reactants $\boldsymbol{n}^{[\nu^{\pm\rho}]} = \boldsymbol{n}!/(\boldsymbol{n} - \nu^{\pm\rho})!$:

$$W_\rho^\pm(\boldsymbol{n}) = \kappa_\rho^\pm \, \boldsymbol{n}^{[\nu^{\pm\rho}]} = \Omega\, k_\rho^\pm \frac{\boldsymbol{n}^{[\nu^{\pm\rho}]}}{\Omega^{\nu^{\pm\rho}}}, \quad \text{(C3)}$$

where in the second equality we used Eq. (B1) to make the dependence of the molecular rates on system size explicit. Notably, the extensivity of the rates $\kappa_\rho^\pm$ depends on the stoichiometry of the corresponding reactions $\nu^{\pm\rho}$. This corresponds to the fact that collisions between particles, which are required for multiple-species reactions to occur, get rarer when $\Omega$ increases at fixed $\boldsymbol{n}$. Intuitively, this is even more so when the number of involved species is larger (see e.g. [12, 94, 100, 101]). The relevance of such scaling is for instance seen as follows: Using these rates in Eq. (C2) for $f(\boldsymbol{n}) = n_i$, one recovers the macroscopic rate Equation (5) as $\Omega \to \infty$, with $\boldsymbol{x} = \boldsymbol{n}/\Omega$ fixed (in the large-$\Omega$ limit where the average of concentrations product becomes the products of the average).

In the large-size asymptotics $\Omega \to \infty$, we expand Eq. (C2) for the rescaled function of the concentrations $\bar{f}(\boldsymbol{x}) = f(\Omega \boldsymbol{x})$ and for the probability density $\bar{P}(\boldsymbol{x},t)$, and define $\bar{W}_\rho^+(\boldsymbol{x}) = k_\rho^\pm \boldsymbol{x}^{\nu^{\pm\rho}}$, to get

$$\partial_t \langle \bar{f} \rangle = \int d^N \boldsymbol{x} \sum_\rho \bar{P}(\boldsymbol{x},t) \Big\{ \big[ \bar{W}_\rho^+(\boldsymbol{x}) - \bar{W}_\rho^-(\boldsymbol{x}) \big] \sum_i \mathbb{S}_{i\rho} \partial_i \bar{f}(\boldsymbol{x}) \\ + \frac{1}{2\Omega} \big[ \bar{W}_\rho^+(\boldsymbol{x}) + \bar{W}_\rho^-(\boldsymbol{x}) \big] \sum_{i,j} \mathbb{S}_{i\rho} \mathbb{S}_{j\rho} \partial_{ij} \bar{f}(\boldsymbol{x}) \Big\}, \quad \text{(C4)}$$

where we neglected terms of order $\Omega^{-2}$ and higher. We recognize the first square bracket to be the macroscopic current $J_\rho(\boldsymbol{x})$, see Eq. (4). The coefficient of $\partial_{ij}\bar{f}$ is proportional to the symmetric matrix $\mathbb{D}(\boldsymbol{x})$ of components

$$\mathbb{D}_{ij}(\boldsymbol{x}) = \sum_\rho \frac{1}{2} \mathbb{S}_{i\rho}(\bar{W}_\rho^+(\boldsymbol{x}) + \bar{W}_\rho^-(\boldsymbol{x}))(\mathbb{S}^\top)_{\rho j} \quad \text{(C5)}$$

so that, overall, Eq. (C4) becomes

$$\partial_t \langle \bar{f} \rangle = \Big\langle \sum_i (\mathbb{S}\boldsymbol{J}(\boldsymbol{x}))_i \partial_i \bar{f}(\boldsymbol{x}) + \frac{1}{\Omega} \sum_{ij} \mathbb{D}_{ij}(\boldsymbol{x}) \partial_{ij} \bar{f}(\boldsymbol{x}) \Big\rangle. \quad \text{(C6)}$$

Formally, the evolution equation (C6) for the average of $\bar{f}(\boldsymbol{x})$ is the same as that of a Fokker–Planck equation corresponding to the Langevin equation

$$\partial_t \boldsymbol{x}(t) = \mathbb{S}\boldsymbol{J}(\boldsymbol{x}(t)) + \boldsymbol{\eta}(\boldsymbol{x}(t),t) \quad \text{(C7)}$$

with $\boldsymbol{\eta}(\boldsymbol{x},t)$ a Gaussian white noise of zero average and covariance $\langle \eta_i(\boldsymbol{x},t) \eta_j(\boldsymbol{x},t') \rangle = \frac{1}{\Omega} \mathbb{D}_{ij}(\boldsymbol{x}) \delta(t'-t)$ (notice that the time discretization of such multiplicative noise has no importance in the small-noise regime $\Omega \to \infty$ we are considering). However, such a formal treatment has the problem in that it discards possible scaling with $\Omega$ of the probability density $\bar{P}(\boldsymbol{x},t)$ itself (and consistently of $\bar{f}(\boldsymbol{x})$), which would invalidate the large-$\Omega$ expansion and truncation. This was noticed within a large variety of contexts in the literature [70, 94, 102–104]. A regime where the above expansion is necessarily valid is that of $\boldsymbol{x}(t)$ close to a stationary point $\boldsymbol{x}^*$, i.e. $\boldsymbol{x}(t) = \boldsymbol{x}^* + \delta\boldsymbol{x}(t)$ with $\mathbb{S}\boldsymbol{J}(\boldsymbol{x}^*) = 0$ and $\delta\boldsymbol{x}(t) = O(\Omega^{-1/2})$. Then, the Langevin equation (C7) reduces to:

$$\partial_t \delta\boldsymbol{x}(t) = \mathbb{S}\boldsymbol{J}(\boldsymbol{x}^* + \delta\boldsymbol{x}(t)) + \boldsymbol{\eta}(t), \quad \text{(C8)}$$

where $\boldsymbol{J}(\boldsymbol{x}^* + \delta\boldsymbol{x}(t))$ is understood as truncated to first order in $\delta\boldsymbol{x}(t)$ (i.e. the Langevin equation is linear) and the centered Gaussian noise $\boldsymbol{\eta}(t)$ is now additive with correlations

$$\langle \eta_i(t) \eta_j(t') \rangle = \frac{1}{\Omega} \mathbb{D}_{ij}^* \delta(t'-t) \quad \text{(C9)}$$

where the matrix $\mathbb{D}^*$ is obtained from Eq. (C5) and reads

$$\mathbb{D}^* = \mathbb{D}(\boldsymbol{x}^*) = \mathbb{S}\mathbb{A}^*\mathbb{S}^\top. \quad \text{(C10)}$$

Here, $\mathbb{A}^*$ the $R \times R$ diagonal matrix with the entries of the vector $\frac{1}{2}(k_\rho^+ \boldsymbol{x}^{*\nu^{+\rho}} + k_\rho^- \boldsymbol{x}^{*\nu^{-\rho}})$.

In general, in irreversible dynamics, the drift of this Langevin equation is not simply related to the noise covariance matrix $\mathbb{D}^*$ of Eq. (C10). Focusing now on conservative affinities as in Sec. V A, the dynamics is reversible and there exists an equilibrium stationary point $\boldsymbol{x}^* = \boldsymbol{x}^{\text{eq}}$ (see Appendix B) that cancels the current and the affinity vectors. Then using Eqs. (57) and (61) and remarking that $\mathbb{A}^* = \mathbb{A}$, one has $\boldsymbol{J}(\boldsymbol{x}^{\text{eq}} + \delta\boldsymbol{x}) = \mathbb{A}\boldsymbol{A}(\boldsymbol{x}^{\text{eq}} + \delta\boldsymbol{x}) = -\mathbb{A}\mathbb{S}^\top(\mathbb{X}^{\text{eq}})^{-1}\delta\boldsymbol{x}$ and thus from Eq. (C8)

$$\partial_t \delta\boldsymbol{x}(t) = -\mathbb{D}\,(\mathbb{X}^{\text{eq}})^{-1} \delta\boldsymbol{x}(t) + \boldsymbol{\eta}(t) \quad \text{(C11)}$$

with $\mathbb{D} = \mathbb{D}^* = \mathbb{D}(\boldsymbol{x}^{\text{eq}})$. Hence, the symmetric matrix $\mathbb{D} = \mathbb{S}\mathbb{A}\mathbb{S}^\top$ read from Eq. (C10) plays at the same time the role of the noise amplitude and the prefactor of the potential gradient in the Langevin equation (C11), which is an incarnation of the Onsager reciprocity [50, 69].

We now connect the previous analysis to the core of the paper. We first note that the rank of $\mathbb{D}$ is $M$ and not $N$. This means that in general some directions of the noise present a zero amplitude. This corresponds to the fact that the degrees of freedom $\boldsymbol{x}(t)$ representing the instantaneous concentrations at time $t$ present one or several conservation law(s), both at the deterministic level the rate equation (2) and at the stochastic level. In Sec. V A, we identified $M$ independent degrees of freedom $\delta z(t)$, defined in Eq. (62), at the deterministic level. Using the same procedure at the stochastic level, we define a stochastic process $\delta z(t)$ from $\delta x(t)$ [that satisfies Eq. (C11)]. Now, the noise that governs the evolution of $\delta z(t)$ is non-singular. Indeed, multiplying Eq. (C11) by $\mathbb{G}$, using Eq. (63) and taking the last $M$ components, we find by direct computation:

$$\partial_t \delta z(t) = -\mathbb{L}_Q \mathbb{H}_Q \, \delta z(t) + \tilde{\eta}(t), \tag{C12}$$

where the (now non-singular) centered Gaussian white noise $\tilde{\eta}(t) \in \mathbf{R}^M$ has correlations $\langle \tilde{\eta}_i(t) \tilde{\eta}_j(t') \rangle = \frac{1}{\Omega}(\mathbb{L}_Q)_{ij} \delta(t' - t)$. In these expressions, $\mathbb{L}_Q$ and $\mathbb{H}_Q$ are the $M \times M$ matrices defined in Eq. (58) and (64) respectively. As expected, the deterministic drift of Eq. (C12) is the same as the one derived at the deterministic level [see Eq. (65)]. At the stochastic level, for $\delta z(t)$, the matrix $\mathbb{L}_Q$ plays at the same time the role of a relaxation response matrix close to an equilibrium point, and of a correlation matrix for the noise that describes the small Gaussian fluctuations close to that point. Accordingly, the Gaussian stationary probability density of the linearized Langevin equation (C12) is $\bar{P}(\delta z) \propto \exp[-\frac{\Omega}{2} \delta z^\top \mathbb{H}_Q \, \delta z]$, where we thus identify the matrix $\mathbb{H}_Q$ as the Hessian matrix of the equilibrium quasipotential. This concludes our illustration that the two matrices $\mathbb{L}_Q$ and $\mathbb{H}_Q$ that appeared in Sec. V A in the analysis of the deterministic relaxation close to an equilibrium point in fact also play a role at the Gaussian stochastic level.

### Appendix D: Cycles, cocycles and oblique projectors

In this Appendix we show how the decompositions Eqs. (29) and (32) of affinity and current can be reformulated in terms of complementary oblique projectors. We follow the line of [17] where the formalism was first introduced and discussed for graphs (corresponding to unimolecular reactions). From the families of cocycles and cycles introduced in the main text, we define two $R \times R$ matrices as:

$$\mathbb{Q}^\top = \left( \overbrace{\begin{array}{c|c} \mathbb{1}_M & 0 \\ \mathbb{T}^\top & 0 \end{array}}^{c^\gamma} \right), \quad \mathbb{P} = \left( \begin{array}{c|c} 0 & \overbrace{-\mathbb{T}}^{c^\alpha} \\ 0 & \mathbb{1}_{R-M} \end{array} \right) \tag{D1}$$

By construction, their images correspond to the spaces spanned by the $c^\gamma$'s and the $c^\alpha$'s, namely $\text{Im}\,\mathbb{Q}^\top = \text{Im}\,\mathbb{S}^\top$ and $\text{Im}\,\mathbb{P} = \text{Ker}\,\mathbb{S}$. Taking the transpose of Eq. (D1) one obtains two more matrices,

$$\mathbb{Q} = \left( \overbrace{\begin{array}{c|c} \mathbb{1}_M & \mathbb{T} \\ 0 & 0 \end{array}}^{e^\gamma} \right), \quad \mathbb{P}^\top = \left( \begin{array}{c|c} 0 & \overbrace{0}^{e^\alpha} \\ -\mathbb{T}^\top & \mathbb{1}_{R-M} \end{array} \right) \tag{D2}$$

whose images are now spanned by the $e^\gamma$'s and $e^\alpha$'s, i.e. $\text{Im}\,\mathbb{Q} = \text{Span}(e^\gamma)$ and $\text{Im}\,\mathbb{P}^\top = \text{Span}(e^\alpha)$.

We recall that a square matrix $\mathbb{A}$ is a projector if and only if it is idempotent $\mathbb{A}^2 = \mathbb{A}$. It can be directly checked that this property holds for $\mathbb{Q}^\top$ and $\mathbb{P}$, as well as for $\mathbb{Q}$ and $\mathbb{P}^\top$ making them oblique projectors with $\mathbb{Q}^\top \neq \mathbb{Q}$ and $\mathbb{P} \neq \mathbb{P}^\top$ as soon as $\mathbb{T}$ is present. In particular, they form pairs of complementary oblique projectors such that $\mathbb{P} + \mathbb{Q} = \mathbb{P}^\top + \mathbb{Q}^\top = \mathbb{1}_R$ and $\mathbb{Q}\mathbb{P} = \mathbb{Q}^\top \mathbb{P}^\top = 0$.

As a consequence, we may re-express the decompositions of Eqs. (29) and (32) in the main text as:

$$\boldsymbol{A} = \sum_\gamma A^c_\gamma \, \boldsymbol{c}^\gamma + \sum_\alpha A^e_\alpha \, \boldsymbol{e}^\alpha = \mathbb{Q}^\top \boldsymbol{A} + \mathbb{P}^\top \boldsymbol{A} \tag{D3}$$

$$\boldsymbol{J} = \sum_\gamma J^e_\gamma \, \boldsymbol{e}^\gamma + \sum_\alpha J^c_\alpha \, \boldsymbol{c}^\alpha = \mathbb{Q} \boldsymbol{J} + \mathbb{P} \boldsymbol{J}. \tag{D4}$$

These expressions are analogous the ones reported in [17] with the main difference being that here the operators are not derived from the spanning tree of a graph but from the family of $\{c^\gamma\}$ and $\{c^\alpha\}$ constructed in Sec. III A using the reduced row echelon form of $\mathbb{S}$. Thus, the construction we put forward in Sec. III and in this Appendix generalizes the oblique projector method of Ref. [17] for the decomposition of currents and affinities from unimolecular CRNs (and graphs) to arbitrary CRNs (and their associated hypergraphs).

We conclude the section by pointing out a connection between the oblique projectors and the Onsager matrices of linear response (see Sec. V A). First, one can always define new projectors using a change of basis. In particular, we may define $\hat{\mathbb{Q}} = \mathbb{A}^{-1/2} \mathbb{Q} \mathbb{A}^{1/2}$ and $\hat{\mathbb{P}} = \mathbb{A}^{-1/2} \mathbb{P} \mathbb{A}^{1/2}$ which are still complementary oblique projectors. Then, one finds:

$$\mathbb{Q}\mathbb{A}\mathbb{Q}^\top = \left( \begin{array}{c|c} \mathbb{L}_Q & 0 \\ \hline 0 & 0 \end{array} \right) \quad \hat{\mathbb{Q}}\hat{\mathbb{Q}}^\top = \left( \begin{array}{c|c} \hat{\mathbb{L}}_Q & 0 \\ \hline 0 & 0 \end{array} \right) \tag{D5}$$

$$\mathbb{P}^\top \mathbb{A}^{-1} \mathbb{P} = \left( \begin{array}{c|c} 0 & 0 \\ \hline 0 & \mathbb{L}_P \end{array} \right) \quad \hat{\mathbb{P}}^\top \hat{\mathbb{P}} = \left( \begin{array}{c|c} 0 & 0 \\ \hline 0 & \hat{\mathbb{L}}_P \end{array} \right) \tag{D6}$$

Thus, in both representations, the Onsager matrices appear as the invertible cores of the symmetric $R \times R$ matrices constructed from the oblique projectors. In Ref. [17], the matrices in Eqs. (D5)-(D6) were shown to govern the different contributions to the entropy production in linear response; we thus have shown in Sec. V A that these matrix also control the macroscopic relation between currents and affinities in the linear-response regime for generic CRNs.

### Appendix E: Proof of reconstruction feasibility

We want to show that Eq. (96) is consistent with stationary KCL,

$$\left( \mathbb{S}_Y \mid \mathbb{S} \right) \begin{pmatrix} \boldsymbol{J}_Y \\ \boldsymbol{J} \end{pmatrix} = 0, \tag{E1}$$

and with linear-regime KVL, obtained by plugging Eq. (57) into Eq. (31), namely

$$\mathbb{P}^\top \mathbb{A}^{-1} \boldsymbol{J} = 0, \tag{E2}$$

where we made use of the oblique projector $\mathbb{P}$ introduced in Appendix D. Find $\mathbb{T}$ by row reduction of $\mathbb{S}$, and consider

$$\mathbb{Q} = \begin{pmatrix} -\mathbb{G}_M \mathbb{S} \\ 0 \end{pmatrix} = \begin{pmatrix} \mathbb{1}_M & \mathbb{T} \\ 0 & 0 \end{pmatrix} \tag{E3}$$

obtained by adding or removing sufficient zero rows to make it a square $R \times R$ matrix. The matrix $\mathbb{G}_M$ is defined in Eq. (26) in the main text. As explained in Appendix D, $\mathbb{Q}$ and $\mathbb{P}$ are complementary oblique projectors so that $\mathbb{P} = \mathbb{1} - \mathbb{Q}$ and $\mathbb{SP} = \mathbb{0}$. Then solutions of the system Eqs. (E1)-(E2) can be found by exploiting the projector algebra. In particular expanding the identity

$$\bm{J} = \mathbb{P}\bm{J} + \mathbb{Q}\bm{J} \tag{E4}$$

we find

$$\mathbb{SQ}\bm{J} = -\mathbb{S}_Y \bm{J}_Y \tag{E5}$$

$$\mathbb{P}^\top \mathbb{\Lambda}^{-1} \mathbb{P}\bm{J} + \mathbb{P}^\top \mathbb{\Lambda}^{-1} \mathbb{Q}\bm{J} = 0, \tag{E6}$$

where $\bm{J}_Y$ are the external currents. By applying the matrix $\mathbb{G}_M$ to the first, in view of Eq. (E3), we find

$$\mathbb{Q}\bm{J} = \begin{pmatrix} \mathbb{G}_M \mathbb{S}_Y \bm{J}_Y \\ \mathbb{0} \end{pmatrix}, \tag{E7}$$

where we used $\mathbb{Q}^2 = \mathbb{Q}$. Plugging this latter into the second we find:

$$\mathbb{P}^\top \mathbb{\Lambda}^{-1} \mathbb{P} \bm{J} = -\mathbb{P}^\top \mathbb{\Lambda}^{-1} \begin{pmatrix} \mathbb{G}_M \mathbb{S}_Y \bm{J}_Y \\ \mathbb{0} \end{pmatrix}. \tag{E8}$$

A solution $\bm{J}^+$ that is consistent with the above equation can be found by the Moore–Penrose pseudoinverse

$$\bm{J}^+ = -(\mathbb{P}^\top \mathbb{\Lambda}^{-1} \mathbb{P})^+ \mathbb{P}^\top \mathbb{\Lambda}^{-1} \begin{pmatrix} \mathbb{G}_M \mathbb{S}_Y \bm{J}_Y \\ \mathbb{0} \end{pmatrix}. \tag{E9}$$

Projecting once again $\mathbb{P}\bm{J} = \mathbb{P}\bm{J}^+$, using Eq. (E4), we finally find

$$\begin{aligned}
\bm{J} &= \left[\mathbb{1}_R - \mathbb{P}(\mathbb{P}^\top \mathbb{\Lambda}^{-1} \mathbb{P})^+ \mathbb{P}^\top \mathbb{\Lambda}^{-1}\right] \begin{pmatrix} \mathbb{G}_M \mathbb{S}_Y \bm{J}_Y \\ \mathbb{0} \end{pmatrix} \\
&= \begin{pmatrix} (\mathbb{1}_M - \mathbb{T}\mathbb{L}_P^{-1}\mathbb{T}^\top \mathbb{\Lambda}_M^{-1}) \mathbb{G}_M \mathbb{S}_Y \bm{J}_Y \\ \mathbb{L}_P^{-1}\mathbb{T}^\top \mathbb{\Lambda}_M^{-1} \mathbb{G}_M \mathbb{S}_Y \bm{J}_Y \end{pmatrix},
\end{aligned} \tag{E10}$$

where in the last expression we made explicit the projector-based solution in terms of known matrices. The fact that the above system is full-rank grants, *a posteriori*, that this solution is unique and correct.

Finally, let us comment on the structure of the matrix $\mathbb{L}_P^{-1}\mathbb{T}^\top \mathbb{\Lambda}_M^{-1}$ which, within our approach, controls the response of the network to the external current $\bm{J}_Y$ through the chords. Using Eq. (70), one sees that the matrix depends on the $\lambda$'s via ratios of the type $\lambda_\alpha/\lambda_\gamma$ $\forall$ cycle $\alpha$ and for any reaction $\gamma$ that belongs to it. Its asymptotics for the set of ratios read:

$$\mathbb{L}_P^{-1}\mathbb{T}^\top \mathbb{\Lambda}_M^{-1} \sim \mathbb{\Lambda}_{R-M}\mathbb{T}^\top \mathbb{\Lambda}_M^{-1} \qquad \text{for } \lambda_\alpha/\lambda_\gamma \ll 1 \tag{E11}$$

$$\mathbb{L}_P^{-1}\mathbb{T}^\top \mathbb{\Lambda}_M^{-1} \sim \left(\mathbb{T}^\top \mathbb{\Lambda}_M^{-1}\mathbb{T}\right)^{-1} \mathbb{T}^\top \mathbb{\Lambda}_M^{-1} \quad \text{for } \lambda_\alpha/\lambda_\gamma \gg 1 \tag{E12}$$

As expected, in the limit $\lambda_\alpha/\lambda_\rho \to 0$ the solution $\bm{J}$ falls back to Eq. (84), and the network behaves effectively as a tree-like networks where internal cycles have been knocked out. Remarkably, the opposite limit in Eq. (E12) does not depend on the $\lambda_\alpha$'s but only on the $\mathbb{\Lambda}_M$ (excluding bridges). This asymptotics is reached monotonously and does not diverge with $\mathbb{\Lambda}_M$, making it possible to explore the full set of feasible $\bm{J}$ as parametrized by the $\lambda_\gamma$'s. The monotonicity is inferred from the identity

$$\frac{\partial}{\partial \lambda_\rho}\left(\mathbb{L}_P^{-1}\mathbb{T}^\top \mathbb{\Lambda}_M^{-1}\right) = \left[\mathbb{1}_{R-M} + \check{\mathbb{T}}^\top \mathbb{T}\right]^{-1} \frac{\partial \check{\mathbb{T}}^\top}{\partial \lambda_\rho}\left[\mathbb{1}_M + \mathbb{T}\check{\mathbb{T}}^\top\right]^{-1} \tag{E13}$$

where we denoted $\check{\mathbb{T}}^\top = \mathbb{\Lambda}_{R-M}\mathbb{T}^\top \mathbb{\Lambda}_M^{-1}$ (notice that the matrices between the square brackets on the r.h.s. of Eq. (E13) have a positive spectrum).

**Appendix F: Computational details for the results of Fig. 12**

We detail in this appendix how the linear-regime metabolic reconstruction algorithm proposed in Sec. V B 5 applies in practice to the example presented in Sec. V B 3, leading to the results displayed in Fig. 12. We first split the reactions into external chemostatting reactions 2 and 3, and $R = 5$ internal reactions 1, 4, 5, 6, and 7. The resulting stoichiometric matrix of internal reactions has 2 cycles. We pick reactions 4 and 7 to be their corresponding chords, meaning that reactions 1, 5 and 6 are cochords. For convenience, in the rest of this Appendix, we order the reactions as $(1, 6, 5, 4, 7)$ and label them as $(I, II, III, IV, V)$. The chemostatting and internal stoichiometric matrices then write:

$$\mathbb{S}_Y = \begin{pmatrix} 0 & 0 \\ 1 & 0 \\ 0 & 1 \end{pmatrix} \qquad \mathbb{S} = \begin{pmatrix} 1 & -1 & 0 & -1 & 0 \\ 0 & 1 & -1 & 0 & -1 \\ 0 & 0 & 1 & 1 & 1 \end{pmatrix}. \tag{F1}$$

The row-reducing matrix $\mathbb{G}$ is found as described in Sec. A 2 and $-\mathbb{GS}$ takes the form of Eq. (22) with

$$\mathbb{G} = \begin{pmatrix} -1 & -1 & -1 \\ 0 & -1 & -1 \\ 0 & 0 & -1 \end{pmatrix} \qquad \mathbb{T} = \begin{pmatrix} 0 & 0 \\ 1 & 0 \\ 1 & 1 \end{pmatrix}. \tag{F2}$$

The cycles and cocycles are found using the matrix $\mathbb{T}$ as summarized in Table I, with the rank of $\mathbb{S}$ being $M = 3$. We read from the matrix $\mathbb{T}$ that reaction I is a bridge since it is involved in none of the cycles.

Then, the key matrix $\mathbb{L}_P$ defined in Eq. (70) takes the form

$$\mathbb{L}_P = \begin{pmatrix} \frac{1}{\lambda_{II}} + \frac{1}{\lambda_{III}} + \frac{1}{\lambda_{IV}} & \frac{1}{\lambda_{III}} \\ \frac{1}{\lambda_{III}} & \frac{1}{\lambda_{III}} + \frac{1}{\lambda_{V}} \end{pmatrix}. \tag{F3}$$

The chord currents are obtained from the last lines of Eq. (96) as

$$\begin{pmatrix} J_{IV} \\ J_V \end{pmatrix} = \begin{pmatrix} -\dfrac{\lambda_{IV}(\lambda_{II}+\lambda_{III}+\lambda_V) J_3^{\text{ext}} + \lambda_{IV}(\lambda_{III}+\lambda_V) J_2^{\text{ext}}}{\lambda_{II}(\lambda_{III}+\lambda_{IV}+\lambda_V) + \lambda_{IV}(\lambda_{III}+\lambda_V)} \\ \dfrac{\lambda_{IV}\lambda_V J_2^{\text{ext}} - \lambda_{II}\lambda_V J_3^{\text{ext}}}{\lambda_{II}(\lambda_{III}+\lambda_{IV}+\lambda_V) + \lambda_{IV}(\lambda_{III}+\lambda_V)} \end{pmatrix}. \tag{F4}$$

We check explicitly that, as shown on general grounds in Sec. V B 5 and Appendix E, they are monotonic functions of each of the individual $\lambda_I...\lambda_V$ when the others are fixed (with each of these parameters being positive). As expected, they are functions only of ratios of $\lambda$'s, and taking limits of the $\lambda$'s to 0 and to $+\infty$ yields well-defined and finite chord currents. Using these properties, one obtains the results displayed in Fig. 12a for the possible values taken by the chord currents in (F4), for the specific choice $(J_2^{\text{ext}}, J_3^{\text{ext}}) = (0.2, 0.3)$. The cochord currents are obtained in a similar manner from the first lines of Eq. (96) and one obtains the results displayed in Fig. 12b.

---